\documentclass[%
 reprint,
 groupedaddress,
 amsmath,amssymb,
 aps,
 pre,
 floatfix,
 polish
]{revtex4-2}

\usepackage{graphicx} 
\graphicspath{{figures/}} 
\usepackage{dcolumn} 
\usepackage{bm} 
\usepackage[T1]{fontenc}
\PassOptionsToPackage{hyphens}{url}
\usepackage[breaklinks=true,colorlinks=true,linkcolor=blue,citecolor=red,urlcolor=magenta]{hyperref}
\usepackage{subfigure}

\usepackage{color, xcolor} 
\usepackage{enumitem} 

\usepackage{lipsum} 

\usepackage[utf8]{inputenc} 
\usepackage[english]{babel} 
\usepackage{cprotect} 

\usepackage{mathtools} 
\usepackage{amsthm} 
\usepackage{mathrsfs} 
\usepackage{amsfonts} 
\usepackage{extarrows} 
\usepackage{accents} 
\usepackage{braket} 
\usepackage{cancel} 
\usepackage{comment}

\usepackage{amsmath} 
\usepackage{dsfont} 

\usepackage{cleveref} 
\usepackage{nameref} 

\crefformat{equation}{eq.\,(#2#1#3)}
\Crefformat{equation}{Eq.\,(#2#1#3)}
\crefformat{figure}{fig.\,#2#1#3}
\Crefformat{figure}{Fig.\,#2#1#3}
\crefformat{table}{tab.\,#2#1#3}
\Crefformat{table}{Tab.\,#2#1#3 }

\crefmultiformat{equation}{eqs.\,(#2#1#3)}%
{ and~(#2#1#3)}{, (#2#1#3)}{ and~(#2#1#3)}
\Crefmultiformat{equation}{Eqs.\,(#2#1#3)}%
{ and~(#2#1#3)}{, (#2#1#3)}{ and~(#2#1#3)}
\crefmultiformat{figure}{figs.\,#2#1#3}%
{ and~#2#1#3}{, #2#1#3}{ and~#2#1#3}
\Crefmultiformat{figure}{Figs.\,#2#1#3}%
{ and~#2#1#3}{, #2#1#3}{ and~#2#1#3}
\crefmultiformat{table}{tabs.\,#2#1#3}%
{ and~#2#1#3}{, #2#1#3}{ and~#2#1#3}
\Crefmultiformat{table}{Tabs.\,#2#1#3}%
{ and~#2#1#3}{, #2#1#3}{ and~#2#1#3}

\newcommand{\R}{\mathbb{R}}

\newcommand{\indi}{\text{\boldmath$\delta$}}

\begin{document}

\title{Stochastic Thermodynamics of Social Imitation beyond Energetics}


\author{Luis Irisarri}
\thanks{These two authors contributed equally to this work}

\author{Lucas Trigal}%
\thanks{These two authors contributed equally to this work}

\author{Raúl Toral}

\author{Gonzalo Manzano}
\email{gonzalo.manzano@ifisc.uib-csic.es}

\affiliation{
 Instituto de Física Interdisciplinar y Sistemas Complejos, IFISC (CSIC-UIB),
Campus Universitat Illes Balears, E-07122 Palma de Mallorca, Spain
}%

\begin{abstract}
The development of stochastic thermodynamics during the last decades prompted the discovery of novel nonequilibrium relations refining our understanding of the second law in small fluctuating systems and its connection with information theory. 
A fundamental open question is whether these powerful tools can illuminate other areas of complex systems, such as social phenomena, where energy plays no fundamental role. 
Here we develop a framework that derives a ``second law" for social systems. Similarly to Landauer's principle, it constrains spontaneous changes in agent attributes (opinions, cultural traits, etc.) and their informational entropy.
We illustrate this framework to toy agent-based models of social imitation with non-trivial phase diagrams. We demonstrate how cornerstone results---fluctuation theorems, kinetic and thermodynamic uncertainty relations, and second-law-like inequalities---emerge naturally in this context, even across symmetry-breaking transitions. These results reveal fundamental trade-offs in opinion currents arising from the competition between herding and anti-conformity. Moreover, they provide inference tools to extract model parameters from observations of stochastic changes in agents.
\end{abstract}

\maketitle

\section{\label{sec:introduction}Introduction}

Thermodynamics is arguably one of the most robust and successful theories of modern physics. It survived two scientific revolutions and historically contributed to the development of other areas such as chemistry, engineering or ecology, while its applications continue to provide new insights in many adjacent fields~\cite{Kondepudi,Callen,Nielsen20,Roegen}. From cosmology and black holes~\cite{Wald05}, down to the microscopic and quantum realms~\cite{Goold16,Vinjanampathy16}, thermodynamics has been extended to cover a broad range of scales and nonequilibrium situations. In particular, the last decades have witnessed the development of stochastic thermodynamics~\cite{Sekimoto10,Seifert12}, which has emerged as a powerful framework for describing small fluctuating systems out of equilibrium~\cite{Ciliberto17}. Stochastic thermodynamics enables us to refine our understanding of thermodynamic laws and their statistical nature~\cite{Jarzynski11}, with a strong link with information theory~\cite{Parrondo15,Lutz15}. Recent developments of the theory include universal trade-off relations such as the thermodynamic uncertainty relations~\cite{Barato15,Todd16,Horowitz20} and their kinetic counterparts~\cite{Terlizzi18,Yan19,Vo22}; speed limit theorems~\cite{Shiraishi18,Falasco20,LP22,Vu23a,Vu23b}, as well as martingale fluctuation relations for entropy production extrema and stopping times~\cite{Neri17,Chetrite18,gambling21,survival22,roldanMartingalesPhysicistsTreatise2023}.

Concurrently, a great amount of research has been devoted to the study of social phenomena from the perspective of complex systems~\cite{Weidlich91,Castellano09,Opinion_new_review,Lazer20}, namely, systems composed of many interacting components that exhibit emergent behaviour. Using methods from statistical physics, stochastic dynamics and network theory, agent-based models have provided insights in phenomena such as cultural drift and cultural dissemination~\cite{Axelrod97,Castellano00,Centola07}, opinion dynamics and polarization~\cite{Dandekar13,Juanili14,Granha22,Ojer23,Starnini25}, or language adoption, variation and change~\cite{Abrams03,Castello06}, just to mention some examples.

A fair question that is gaining increasing interest is whether we can apply concepts and tools from stochastic thermodynamics to gain insight in other areas in complex systems~\cite{Rao16,noaEntropyProductionTool2019, dasilvaAnalysisEarlierTimes2020, Freitas21,Korbel21,Rao22,manzanoThermodynamicsComputationsAbsolute2024,Sorkin24,Wolpert24,Falasco25}. { Moreover, one may wonder if that would be possible even in situations that lack a thermodynamic foundation in terms of energy, heat or work, such as sociophysics~\cite{BahrMicro,BahrMacro, crochikEntropyProductionMajorityvote2005}.  In particular, obtaining a meaningful framework that allows for the application of the powerful toolbox developed in stochastic thermodynamics during the last decades (nonequilibrium fluctuation relations, thermodynamic and kinetic uncertainty relations, second-law inequalities, etc) would represent a valuable contribution to enrich the study of social system dynamics from the statistical physics viewpoint.}  

In this context, recent works attempted to apply the framework of stochastic thermodynamics in social dynamics models, such as the majority-vote model and some variations~\cite{tomeStochasticThermodynamicsOpinion2023, hawthorneNonequilibriumThermodynamicsMajority2023,oliveiraEntropyProductionCooperative2024}. { These works \emph{assume} local detailed balance relations by postulating an effective energy function and introducing thermal reservoirs with different temperatures. However, the weak connection of these entities with their thermodynamic counterparts, together with the arbitrary restrictions introduced by enforcing relations between them, have substantially limited the scope and application of this approach.}


In this work, we take a different { route}, that allows us to analyze social phenomena from the viewpoint of stochastic thermodynamics without the need of postulating any energy or temperature. { We show that the consistent modeling of the mechanisms responsible for the change in the agents attributes guarantees the \emph{derivation} of a generalized local detailed balance relation from which the underlying structure of stochastic thermodynamics can be recovered in the social context. 
Focusing on toy models,} our motivation is to provide a neat example of how the framework of stochastic thermodynamics can be applied to gain new insights in sociophysics without the need of extra ad hoc postulates. In this manner, we pave the way for using stochastic thermodynamics as a framework to study complex systems in general, without the need for a particular physical interpretation of the model.

In contrast to previous works, we consider a family of suitable imitation models that are microscopically reversible~\cite{Boltzmann}, that is, that every elementary process in the model (e.g. a change of cultural trait or in the opinion of an agent) is accompanied by its reverse process, which is not ruled out from the model definition. We analyze the models using concepts and tools from stochastic thermodynamics without enforcing any energetic interpretation, focusing only on the mathematical structure of the model. As a result we find footprints of the second law, which imposes tight constraints on the plausible evolution of the system observables both at the ensemble level and at the level of fluctuations. Similarly to Landauer's principle linking information and heat~\cite{Parrondo15, Lutz15}, the universal relations that we obtain and analyze here link information-theoretical quantities such as entropy changes with social attributes (opinion or traits) currents. They include second law-like inequalities, thermodynamic and kinetic uncertainty relations~\cite{Horowitz20} and fluctuation theorems that may be used for inference purposes.
Moreover, we show that the model exhibits both first and second-order phase transitions with different hallmarks and discuss associated thermodynamic features of symmetry breaking~\cite{Roldan2014Jun}. { Although a central goal of this work is to present the framework in a sociophysics context and illustrate some of their main tools in a relatively simple, solvable model example, we also explicitly show how the framework generalizes when considering arbitrary network structure, heterogeneity, and other phenomena often playing important roles in social dynamics.}

{
The paper is structured as follows: In Sec.~\ref{sec:Model} we introduce a general social-imitation model under the assumption of microscopic reversibility. In Sec.~\ref{sec:Stationary State} we analyze the long-time dynamics of the model and characterize its phase diagram. In Sec.~\ref{sec:Opinion Thermodynamics} we present our main results, establishing a thermodynamic framework for social systems and illustrating it for the social-imitation model introduced in Sec.~\ref{sec:Model}. In Sec.~\ref{sec:generalizations} we discuss how the framework extends beyond the binary well-mixed setting, including multi-opinion models and complex interaction structures. Finally, in Sec.~\ref{sec:Conclusions} we summarize the main conclusions of our work and discuss possible future directions. The Appendices contain technical details of the calculations and further explanations of the methods employed throughout.}

\section{Model}\label{sec:Model}
\begin{figure}[t]
\includegraphics[width=0.85\columnwidth]{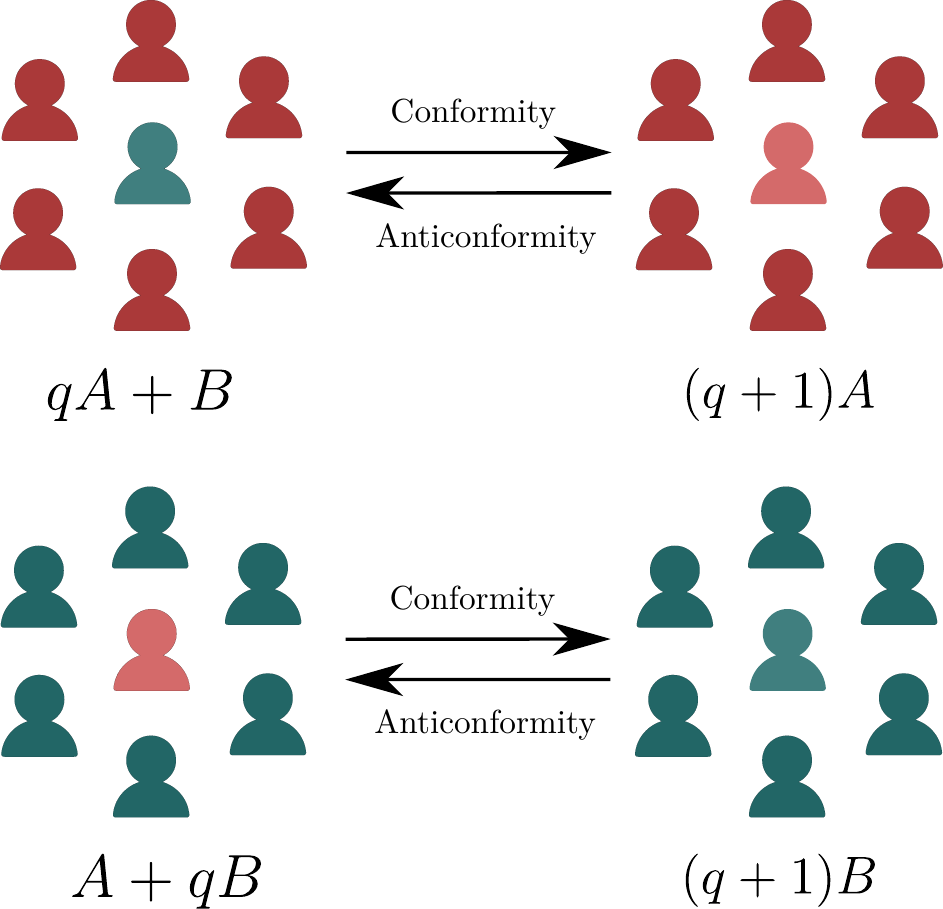}
\caption{\label{fig:social mechanisms} Illustration of herding and anticonformity mechanisms in the two reactions \eqref{eq:Model_Reaction_1} (top) and \eqref{eq:Model_Reaction_2} (bottom).
The herding mechanism occurs when the $q$ agents are of opposite opinion, leading the selected agent to conform and change its opinion. Conversely, the anticonformity mechanism occurs when the $q$ agents share the same opinion as the selected agent, prompting it to differentiate and change its opinion. The example shown corresponds to the case of sampling without repetition with $q = 6$: a chosen agent (lighter color) is influenced by 6 distinct neighbor agents (darker color) to switch its initial state. Note that the interpretation of the parameter $q$ depends on the sampling scheme: without repetition, $q$ represents the exact number of distinct neighbors sampled; with repetition, $q$ denotes the total number of interactions.
}
\end{figure}

We consider a system of $N$ agents, each of which can be in one of two states, $A$ or $B$, corresponding to a (binary) agent's attribute. Typically this attribute is referred to as an opinion, but it could represent a generic social or cultural trait (use of symbols, social norms, values, traditions, language, among others). The agents interact with each other in an all-to-all network topology, meaning that each agent can interact with any other in the system. The interactions between agents are governed by two social mechanisms: imitation (herding) and differentiation (anticonformity) from others. In the herding mechanism, a randomly selected agent changes its attribute (e.g. opinion) when confronted with $q $ other agents holding the opposite opinion. In the anticonformity mechanism, the selected agent changes opinion when confronted with $q$ other agents holding the same opinion. In both cases, the $q \in \{ 1,2,\dots \} $ agents are chosen uniformly at random from the rest of the population (i.e., excluding the selected agent) allowing or not repetition. See \Cref{fig:social mechanisms} for a depiction of the social mechanisms.

The attributes dynamics can be described by a continuous-time Markov process, which we represent schematically through the following two reactions describing possible opinion changes occurring at random times:
\begin{subequations}
\label{eq:Model Reactions}
\begin{align}
 \label{eq:Model_Reaction_1}
 q A + B \xrightleftharpoons[a_1]{h_1} (q + 1) A ,\\
 \label{eq:Model_Reaction_2}
 A + q B \xrightleftharpoons[a_2]{h_2} (q + 1) B,
\end{align}
\end{subequations}
where $h_r, a_r \in [0, \infty)$ for $r \in \{ 1, 2 \}$ are the \textit{reaction rates} parameters of the model { characterizing the intrinsic velocities at which these changes socially occur.} The first reaction describes both herding and anticonformity changes in opinion $A$ and the second reaction describes analogous herding and anticonformity changes in opinion $B$. We are interested in the aggregated variables, $n_{A}(t), n_{B}(t) \in \{ 0, 1, \dots, N \}$, representing the total number of agents in state $A$ and $B$, respectively, at time $t$. Assuming a fixed number of agents, $n_{A}(t) + n_{B}(t) = N$, allows us to reduce the number of relevant aggregated variables to the first one, $n(t) \equiv n_{A}(t) = N - n_B(t)$. { From a social perspective, this variable represents the strength of opinion (or attribute) $A$ in the society.}

This model is a generalisation of the so-called non-linear or $q$-voter model~\cite{castellanoNonlinear$q$voterModel2009} where the possibility of change through anticonformity is added to the traditional herding mechanism~\cite{Holley1975,Clifford1973}. The analysis of such differentiation mechanism have remained more elusive~\cite{nyczkaPhaseTransitions$q$voter2012, nyczka2013} with respect to other noise mechanisms~\cite{Kirman93,Granovsky95}, although it may have implications for polarization in societies~\cite{siedleckiInterplayConformityAnticonformity2016}. Moreover, while in the original $q$-voter model the parameter $q$ takes integer values, here we will consider the more general case $q \in \R_+ $, so that the probability of opinion change varies non-linearly and continuously with the proportion of neighbors holding the opposite opinion. These so-called group interactions are also commonly used in the literature of nonlinear voter models~\cite{peraltaAnalyticalNumericalStudy2018a,ramirezOrderingDynamicsNonlinear2024,Mendes2025}.

As any stochastic process, our model is fully characterized by the probability distribution of its random variables, that is, the probability of finding $n$ agents in state $A$ at time $t$, denoted by $P_n(t)$. It follows the following master equation \cite{Kampen92,Toral2014StochasticNM}:
\begin{equation}
 \label{eq:Master Equation}
 \frac{dP_n(t)}{dt} =\sum_{r} \sum_{m} \left[W_{n, m}^{(r)} P_{m}(t)-W_{m,n}^{(r)}P_n \right],
\end{equation}
$\forall n$, where $W_{m, n}^{(r)} \equiv W^{(r)}(n \to m)$ is the transition rate from state $n$ to state $m$ due to reaction $r$. The total transition rate from state $n$ to state $m$ is then given by the sum of the contributions of this particular jump due to each reaction: $W_{m, n} \equiv \sum_{r} W_{m, n}^{(r)}$.

We remark that the separation of the rates into the different mechanisms producing the changes (the two reactions) is essential to guarantee \textit{micro-reversibility} in the model. Micro-reversibility (or microscopic reversibility) was first introduced by Boltzmann in the context of the kinetic theory of gases~\cite{Boltzmann}, and refers to the decomposition of the microscopic dynamical evolution of a system into elementary processes, each of which possesses a corresponding time-reversed process. It is at the core of stochastic thermodynamics~\cite{Esposito10,Manzano18}, leading to the principle of local detailed balance when transitions are triggered by the exchanges of energy and matter with (equilibrium) thermal reservoirs~\cite{Bauer15, Maes21, Falasco21}. 
{ Socially, micro-reversibility means that if an agent can be persuaded to adopt opinion $A$ by means of some mechanism (captured by a reaction), the agent can also be persuaded to return to its previous opinion $B$ with non-zero probability. }


According to the scheme represented by Eq.~\eqref{eq:Model Reactions}, the only non-null transition rates can be generally written as:
\begin{subequations}\label{eq:transition rates}
\begin{align}
 W_{n+1, n}^{(1)} &= h_1\, (N - n)\, g(n), \label{eq:W1plus} \\
 W_{n-1, n}^{(1)} &= a_1 \, n \, g(n-1), \label{eq:W1minus}\\
 W_{n+1, n}^{(2)} &= a_2 \, (N - n) \, g(N-n-1), \label{eq:W2plus} \\
 W_{n-1, n}^{(2)} &= h_2 \, n \, g(N-n), \label{eq:W2minus}
\end{align}
\end{subequations}
where $g(n) \equiv g(n; q)$ encapsulates the nonlinear group interaction effects. This function represents the dependence of the probability of change on the density of agents, chosen at random from the population, that hold the opposite opinion to the selected agent. 

The particular form of $ g(n) $ depends on the sampling scheme and the particular social influence mechanism producing it. If sampling is done allowing repetition among the $N-1$ neighbors (excluding the selected agent), then $ g(n) = [n/(N-1)]^{q} $. While if repetitions are not allowed $g(n) = (n)_{q} / (N - 1)_q $, where $ (n)_{q}= \Gamma(n)/\Gamma(n - q)$ denotes the \textit{falling factorial} in terms of the Gamma function, (when $q$ is an integer $(n)_{q} = \binom{n}{q} $ becomes the binomial coefficient). In the $N \gg 1$ limit $g(n)$ for both types of sampling coincide. 
In threshold models~\cite{Nowak2020,Nowak2022}, a state change occurs when, among the $q$ agents sampled at random, at least $q_0$ hold the opposite opinion, leading to a more complicated nonlinear $g(n)$. Other nonlinear forms of $g(n)$—as in the $\epsilon$-voter model~\cite{Castello_2006}, majority-rule models~\cite{Galam2002,Oliveira1992}, or group voter models with social temperature~\cite{Krause2012}—can be incorporated straightforwardly within our formulation. Moreover, we remark that in our framework the parameters $h_i$ and $a_i$ for $i=1,2$ above might be time-dependent, hence leading to rates that may explicitly depend on time. 

It will be useful to rewrite the transition rates in terms of symmetric, $\Gamma^{(r)}_{m, n}=\Gamma^{(r)}_{n, m}$, and antisymmetric, $A^{(r)}_{m, n}=-A^{(r)}_{n, m}$, contributions as $W^{(r)}_{m, n} = \Gamma^{(r)}_{m, n} e^{A^{(r)}_{m, n}/2}$ with:
\begin{align}
 \label{eq:Affinity definition}
\Gamma^{(r)}_{m, n} &\equiv \sqrt{W^{(r)}_{m,n} W^{(r)}_{n, m}} ~;~~ A^{(r)}_{m, n} \equiv \ln \frac{W^{(r)}_{m, n}}{W^{(r)}_{n, m}}.
\end{align}
The symmetric contribution provides a notion of the \textit{traffic} between states $n$ and $m$, i.e. of the activity of the transition independent of its bias.  \textcolor{black}{In social terms, it reflects how fast the opinion (or social attribute) strengths may stochastically oscillate between values $n$ and $m$ due to each reaction $r$.}  
In our model it is $\Gamma_{n+1, n}^{(r)} = \sqrt{a_r h_r} \sqrt{(N-n)(n+1)} f^{(r)}(n)$ with $f^{(1)}(n) = g(n)$ and $f^{(2)}(n)= g(N-n-1)$. 

The antisymmetric term $A_{m, n}$ is known as the \textit{affinity} of the transition from state $n$ to state $m$, which measures how biased the transition is{, that is, the tendency in the society to change the strength of a social attribute (opinion $A$) from $n$ to $m$ with respect to the opposite change from $m$ to $n$}. Given the specific form of the transition rates in Eqs.~\eqref{eq:transition rates}, irrespective of the form of the function $g(n)$, it reads:
\begin{equation} \label{eq:Affinity}
 A_{n+ 1, n}^{(r)} = S_{n + 1}^{\mathrm{int}} - S_{n}^{\mathrm{int}} + \mu_r,
\end{equation}
where $\mu_r \equiv \sigma_r \ln (h_r / a_r)$ with $\sigma_{1} = +1 \,(\sigma_2 = -1)$ is a dimensionless parameter that measures the relative strength of the herding and anticonformity mechanisms towards opinion $A$ (positive when opinion $A$ is favored in each reaction), and:
\begin{equation}\label{eq:Internal Entropy}
 S^{\mathrm{int}}_n = \ln \binom{N}{n}.
\end{equation}
is the internal (Boltzmann) entropy of the aggregated opinion state $n$, namely 
the logarithm of the probability of an internal configuration with $n$ agents in opinion $A$ out of $N$ total agents \footnote{The introduction of the Boltzmann constant $k$ in the definition of entropy is unnecessary when dealing with systems which do not have a clear thermodynamic interpretation and hence we take $k=1$ throughout.}. The internal entropy reaches its maximum at $n = N/2$ and its minimum at the extremes or consensus states ($n = 0$ and $n = N$). 

{ In the social context, Eq.~\eqref{eq:Affinity} allows us to distinguish between two effects driving changes in the society attributes (or opinions). The first one, $S_{n + 1}^{\mathrm{int}} - S_{n}^{\mathrm{int}}$ is purely entropic, namely, it has a neat statistical origin, being equal for all reactions. It accounts for the difference in the number of social configurations leading to a resulting opinion strength $n$ in the society. The second one, $\mu_r$, is instead independent of the actual value of the opinion or social attribute strength, but intrinsic to the model reactions $r$, capturing the relevant social psychology features behind the model. In particular, it measures the relative strength of the imitation mechanism with respect to the differentiation one, that is, how much often any individual of the society may be persuaded to change its opinion when finding others with different opinions, rather than by finding other with its same opinion.}

Interestingly, the identification of the entropy terms in Eq.~\eqref{eq:Affinity}, allow us to interpret the (otherwise arbitrary) decomposition in Eq.~\eqref{eq:Affinity definition} as a generalized local detailed balance condition for the transition rates:
\begin{equation} \label{eq:local}
 \frac{W_{m, n}^{(r)}}{W_{n, m}^{(r)}} = e^{S^{\rm int}_{m} - S^{\rm int}_n + (m-n)\mu_r},
\end{equation}
with $m = n \pm 1$ since we are dealing with a one-step jump process, and $\mu_r$ plays the analogous role of a chemical potential difference at constant temperature associated to a chemostated reaction $r$. The above equation implies that transitions towards higher system entropy states are exponentially favored by all reactions, while the parameters $\mu_r$ enforce a fixed intrinsic bias associated to each reaction $r$. In this sense, the exponent in the r.h.s. of the above equation might be interpreted as a stochastic free entropy change (or change in Massieu potential)~\cite{Callen,Guryanova16,Rao18}, where, however, any reference to energy is absent \footnote{Technically speaking, when the two reactions act simultaneously, the differences in generalized chemical potentials make the second term in the exponent to induce a non-conservative force, no longer derivable from a potential~\cite{Rao18}.}.

The \textit{probability current} (or opinion{/attribute} current in the context of our model) between states $n$ and $m$ induced by the reaction $r$ are 
\begin{equation} \label{eq:Jmn}
J_{m,n}^{(r)}(t)=W_{m, n}^{(r)} P_n(t) - W_{n, m}^{(r)} P_m(t).
\end{equation}
The total current for transitions between states $n$ and $m$ (independently of which reaction produces it) is $J_{m,n}(t) =\sum_r J_{m,n}^{(r)}(t)$. In simple words, the current measures the ``imbalance'' in the probability of jumping between { values} $n$ and $m$ { of the social attribute}. If the current is positive $J_{m,n}^{(r)}>0$, the probability that reaction $r$ triggers a jump from $n$ to $m$ is larger than the one of the opposite jump from $m$ to $n$, and vice versa for $J_{m,n}^{(r)}<0$. Probability currents are an essential tool in stochastic thermodynamics which are at the basis of the definition of heat and particle currents, as well as entropy flows and the second law~\cite{Horowitz20}.

Finally, the \textit{dynamical activity} between states $n$ and $m$ induced by the reaction $r$ is
\begin{equation} \label{eq:K}
K_{m,n}^{(r)}(t)=W_{m, n}^{(r)} P_n(t) + W_{n, m}^{(r)} P_m(t).
\end{equation}
The total dynamical activity of the transition is then given by $K_{m,n}(t) = \sum_r K_{m,n}^{(r)}(t)$. The dynamical activity measures the total number of jumps per unit time between states $n$ and $m$. { This is a measure of the total level of fluctuations arising in the opinion strength, i.e. the volatility of opinions, and is conceptually similar to some notions of social temperature~\cite{BahrMicro,BahrMacro}, but well-defined in a general nonequilibrium context}. Together with the traffic, $\Gamma_{m,n}^{(r)}$, introduced above, and contrary to the affinities and currents, the dynamical activity captures genuine time-symmetric or \emph{frenetic} aspects that become important sufficiently far from equilibrium~\cite{Maes20}.

The model possesses the 
parameters: $a_1,a_2,h_1,h_2$. For clarity in the interpretation of the model and in the analysis of the system's critical phenomena we find it convenient to introduce the following reduced dimensionless parameters:
\begin{align} \label{eq:parameters}
 \lambda \equiv \sqrt{\frac{h_1 h_2}{a_1 a_2}},
 &&
 \chi \equiv\sqrt{\frac{h_1 a_2}{h_2 a_1}},
 &&
 \theta \equiv \sqrt{\frac{a_1 h_1}{a_2 h_2}}.
\end{align}
Here $\lambda$ compares the intrinsic strengths of the herding and anticonformity mechanisms and thus indicates which one dominates. Specifically, $\lambda>1$ signals an intrinsic bias toward herding, $\lambda<1$ a bias toward anticonformity, and $\lambda=1$ no intrinsic preference. Parameter $\chi$ quantifies the asymmetry between the two opinions: When $\chi>1$, the rates intrinsically favor opinion $A$; when $\chi<1$, they favor opinion $B$; and for $\chi=1$ the opinions are equally favored. Parameter $\theta$ sets the relative weight of the two reactions: $\theta>1$ means reaction 1 occurs more often than reaction 2, $\theta<1$ the opposite, and $\theta=1$ equal weighting. Finally, we can also introduce a parameter controlling the global time-scale of the opinion dynamics as $\omega \equiv \sqrt[4]{h_1 h_2 a_1 a_2}$, which has units of a rate~\footnote{We will generically set the time-scales of the dynamics by setting $\omega = 1$} .

\section{Stationarity, Equilibrium, and Phase Transitions}\label{sec:Stationary State}
Using the probability currents, the master equation~\eqref{eq:Master Equation} can be written as a continuity equation
\begin{equation} \label{eq:master2}
\frac{dP_n(t)}{dt}+J_n(t)=0,\quad\forall n,
\end{equation}
where $J_n(t) = \sum_r J_n^{(r)}(t) =\sum_r \sum_{m\ne n}J^{(r)}_{m,n}(t)$ is the total escape probability current \textit{out} of state $n$, with $J_{n}^{(r)}(t)$ the corresponding escape probability current induced by reaction $r$, and $J_{m,n}^{(r)}(t)$ is given by Eq.~\eqref{eq:Jmn}. Whenever the dynamics of the continuous-time Markov process~\eqref{eq:Master Equation} is defined by an irreducible time-independent rate matrix (i.e., then the reaction rates of the model are fixed), $\mathbb{W} = \{W_{m,n}\}$, it will reach asymptotically a time-independent stationary distribution, $P_n^{\mathrm{st}}$. Consequently, from Eq.~\eqref{eq:master2}, the escape probability currents also vanish asymptotically, $J_n^{\mathrm{st}}=0,\,\forall n$. 

Moreover, from the theory of stochastic processes~\cite{Kampen92}, it is well known that for one-step jump processes in a bounded domain $n \in \{ 0, 1, \dots ,N \}$, the stationary distribution also satisfies the (stronger) condition of \textit{global detailed balance}, $J_{m,n}^{\mathrm{st}}=0~\forall n,m$, that is,
\begin{equation}\label{eq:GDB}
 W_{n, m} P_m^{\mathrm{st}} = W_{m, n} P_n^{\mathrm{st}}\quad\forall n,m,
\end{equation}
which rules out the possibility of cycles in the steady state. In this case, $P_n^{\mathrm{st}}$ can be calculated through a recursive relation leading to:
\begin{equation}\label{eq:Stationary Distribution Analytical Solution}
\begin{split}
 P^{\mathrm{st}}_n &= P^{\mathrm{st}}_0 \prod_{k=0}^{n-1} \frac{W_{k+1, k}}{W_{k, k+1}} = P^{\mathrm{st}}_0 \, \chi^n \binom{N}{n}
 \\ 
 & \times \prod_{k=0}^{n-1} \left[ \frac{\lambda \theta g(k) + g(N-k-1)}{\theta g(k) + \lambda g(N-k-1)} \right],
\end{split}
\end{equation}
$\forall n > 0$, where $P^{\mathrm{st}}_0$ is fixed by the normalization condition $\sum_{n=0}^N P^{\mathrm{st}}_n=1$. In some cases, Eq.~\eqref{eq:Stationary Distribution Analytical Solution} yields a closed-form expression for $P^{\mathrm{st}}_n$; otherwise, $P^{\mathrm{st}}_n$ can be computed numerically.

Importantly, even though the global detailed-balance condition in Eq.~\eqref{eq:GDB} holds for $P_n^{\rm st}$, one should not identify this stationary distribution with thermodynamic equilibrium, but with a non-equilibrium steady state (NESS). Indeed, while the aggregate currents vanish, $J_{n,m}^{\rm st}=0$, the reaction-resolved currents generally do not: $J_{n,m}^{(r)}\neq 0$ in general, with $J_{n,m}^{(1)}=-\,J_{n,m}^{(2)}$. Thus, although the distribution is stationary, the herding and anticonformity mechanism can remain active and sustain large internal opinion currents.

On the other hand, we say that a system reaches equilibrium if there exists a time-independent probability distribution $P_n^{\mathrm{eq}}$ such that all reservoir currents vanish,
$J_n^{\mathrm{eq},(r)}=\sum_{m\neq n}\bigl[W_{m,n}^{(r)}P_n^{\mathrm{eq}}-W_{n,m}^{(r)}P_m^{\mathrm{eq}}\bigr]=0$~$\forall\, n,r$,
or, equivalently, if the stationary distribution satisfies local detailed balance [cf. Eq.~\eqref{eq:local}]: 
\begin{equation}\label{eq:LDB}
W_{n,m}^{(r)}P_m^{\mathrm{eq}}=W_{m,n}^{(r)}P_n^{\mathrm{eq}} ~~~~ \forall\, n,m,r.
\end{equation}
As shown in App.\ \ref{sec:Equilibrium State apdx}, such an equilibrium state can be reached only when the rates satisfy $h_1h_2=a_1a_2$, i.e., $\lambda=1$. In that case, $P_n^{\mathrm{st}}$ in Eq.~\eqref{eq:Stationary Distribution Analytical Solution} reduces to
\begin{equation}\label{eq:Equilibrium_Distribution}
P_n^{\mathrm{eq}}=\binom{N}{n}\,p^{\,n}(1-p)^{N-n},
\end{equation}
which is a binomial distribution with parameters $N$ and $p=\chi/(\chi+1)\in[0,1]$ (so $p=1/2$ when $\chi=1$). Remarkably, this result is independent of $\theta$ and of the nonlinearity parameter $q$. Moreover, it holds for any choice of $g(n)$, in line with the universal character of equilibrium.

Following a common line of inquiry in physics approaches to social dynamics, we investigate whether the model exhibits abrupt changes in behavior---termed generally as \textit{phase transitions}~\footnote{When the number of individuals is finite, the use of the term \textit{phase transition} is clearly an abuse of notation, as a truly symmetry-breaking phase transition can only occur in the limit $N \to \infty$, where the distribution becomes sharply peaked, approaching a sum of delta functions centered at its maxima~\cite{Toral:2011}.}---as its parameters are varied. A phase transition, in this context, refers to a qualitative change in the existence and location of the maxima of the stationary probability distribution $P_n^{\rm st}$, which gives the probability of observing the system with $n$ individuals holding a particular opinion. These maxima thus identify the most probable states, effectively selecting them as the relevant configurations of the system. A maximum at $n = N/2$ corresponds to a \textit{polarized} (or disensus) state, where the population is approximately evenly divided between the two opinions, and no clear majority emerges. In contrast, maxima located near the extremes, $n \approx 0$ or $n \approx N$, indicate \textit{consensus} states, where nearly the entire population shares the same opinion. These extreme configurations reflect a high degree of collective order.

To advance our analysis of the stationary state, we specialize to the case of sampling with repetition, namely $g(n)=[{n}/{(N-1)}]^q$.
A comprehensive analysis of the stationary distribution with the model parameters, reveals that varying the value of the parameter $q$, only produces a qualitative change in its behavior for $q \leq 1$. Specifically, in contrast to $q > 1 $, the stationary distribution is always unimodal if $q \leq 1$ and no transition from unimodal to bimodal can occur. On the other hand, increasing $q$ systematically sharpens the distribution and advances the onset of the unimodal-bimodal transition to lower values of the model parameters. Consequently, without loss of generality, we will adopt $q = 2 $ when a specific value is required. 

\begin{figure}[t]
\includegraphics[width=1\columnwidth]{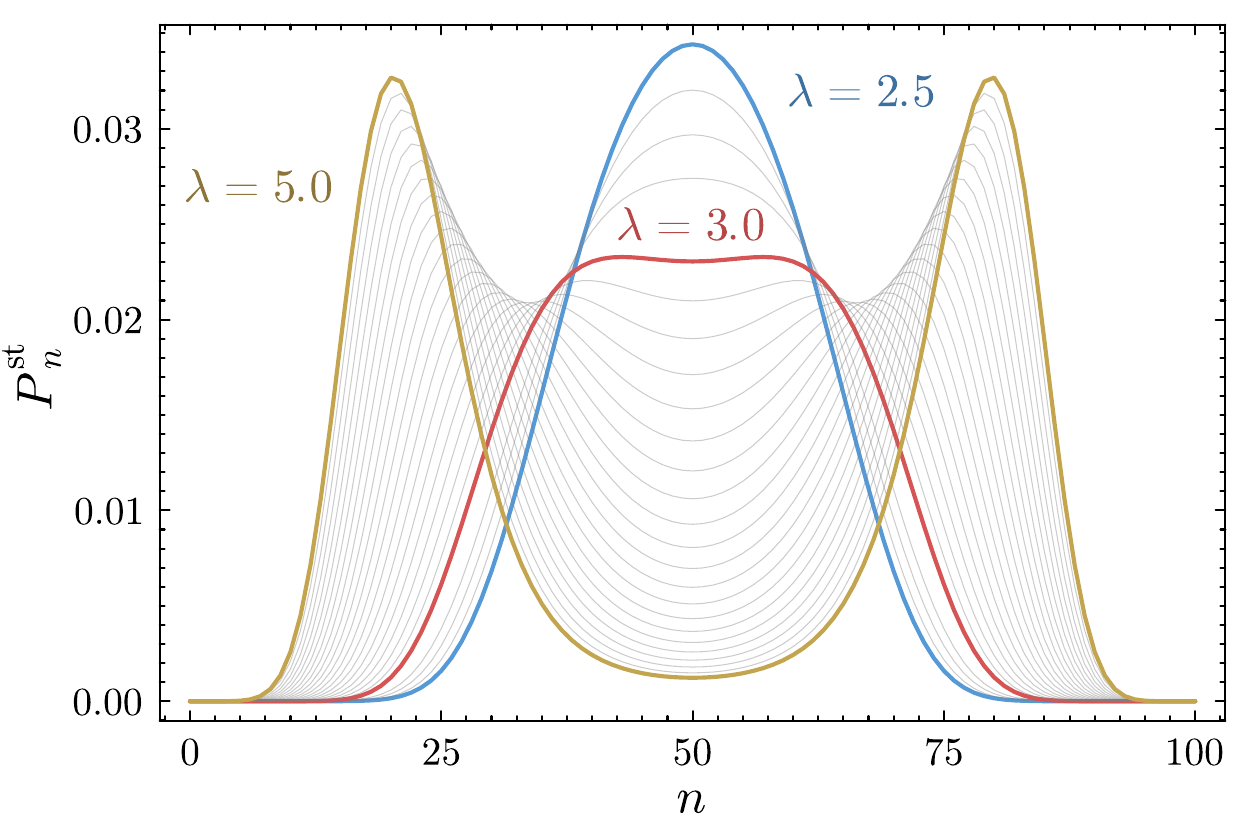}
\caption{\label{fig:stationary probability distribution} 
Stationary probability distribution coming from Eq.~\eqref{eq:Stationary Distribution Analytical Solution} taking $g(n)=\left[{n}/{(n-1)}\right]^q$ for $25$ different values of $\lambda$ equally distributed in the interval $\lambda\in[2.5,5.0]$. We observe how the system transits from a unimodal distribution centered at $N/2$ ($\lambda < 3 $) to a bimodal distribution ($\lambda > 3 $) with the peaks shifting towards the extremes $n \in \{0, N\}$ as $\lambda$ increases. The transition occurs at $ \lambda_{\mathrm{c}} = 3$ as given by \Cref{eq:critical point asymmetric}. \textit{Parameters}: $N = 100, q = 2, \chi = \theta = 1$.}
\end{figure}

For illustrative purposes, let us begin by analyzing the symmetric case $\chi = \theta = 1$ (or $h_1 = h_2$ and $a_1 = a_2$), where no intrinsic bias toward either opinion is introduced and both reactions proceed at equal rates. In Fig.~\ref{fig:stationary probability distribution}, we show the stationary probability distribution Eq.~\eqref{eq:Stationary Distribution Analytical Solution} for several values of the parameter $ \lambda $. We observe that the system transitions (continuously) from a unimodal distribution centered at $ N/2 $ to a bimodal distribution, with the peaks shifting toward the extremes $ n \in \{0, N\} $ as $ \lambda $ increases. In other words, the system undergoes a continuous, second--order transition from a disordered (disensus) state to an ordered (consensus) one around $ \lambda \simeq 3 $. In the limit $ \lambda \to \infty $, i.e., when the herding rate largely exceeds the anticonformity rate, the system reaches a fully consensus state, independently of $ N $, where $ P_{0}^{\mathrm{st}} = P_{N}^{\mathrm{st}} = 1/2 $ and zero otherwise.

\begin{figure}[t]
\includegraphics[width=1\columnwidth]{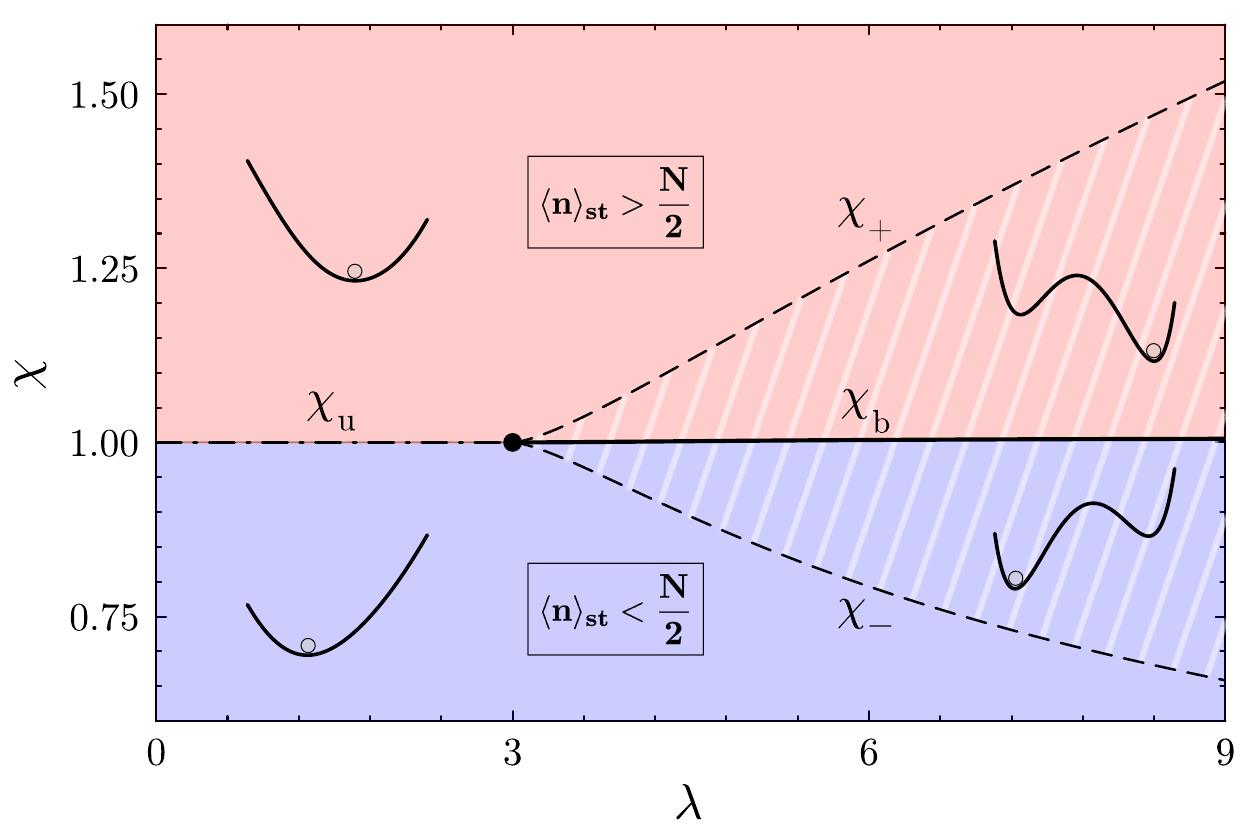}
\caption{\label{fig:phase diagram} 
Phase diagram of the model defined in Eqs.~\eqref{eq:transition rates}, in the $(\lambda, \chi)$ parameter space for $g(n) = [n/(N-1)]^q$, with $q=2$ and symmetric reactions ($\theta=1$). Regions where opinion $A$ is predominant ($\langle n \rangle_{\mathrm{st}} > N/2$) are shown in red, while those favoring opinion $B$ ($\langle n \rangle_{\mathrm{st}} < N/2$) are shown in blue. Each region is further subdivided into unimodal and bimodal zones, delimited by the critical curves $\chi_\pm(\lambda)$ (dashed lines), which converge to the critical point (black dot) at $(\lambda_{\mathrm{c}}, \chi_{\mathrm{c}}) = (3, 1)$ as predicted by Eq.~\eqref{eq:critical point asymmetric}. The line $\chi = 1$ corresponds to the symmetric case illustrated in Fig.~\ref{fig:stationary probability distribution}, which contains the unimodal $\chi_{\rm u}$ (dash-dotted) and bimodal $\chi_{\rm b}(\lambda)$ (solid) transition lines. The insets show the corresponding Fokker--Planck potential $v(x)$ (see Appendix~\ref{sec:Stationary State apdx}) at selected points in the phase diagram, illustrating how the absolute minimum (marked by a gray circle) determines the most probable stationary state.}
\end{figure}

We now consider the general asymmetric case with arbitrary values of $\chi$ and $\theta$. A critical point $ (\lambda_{\mathrm{c}}, \chi_{\mathrm{c}}) $ exists in the phase space, from which two critical lines $ \chi_{\pm}(\lambda) $ bifurcate, as illustrated in \Cref{fig:phase diagram} for $\theta=1$ (for other values of $\theta$ see App.~\ref{sec:Stationary State apdx}). For $ \lambda > \lambda_{\mathrm{c}} $ and $ \chi_{-} < \chi < \chi_{+} $, the system exhibits metastability: the stationary distribution is bimodal with peaks (generally) of unequal height. Outside this metastability region, the distribution is unimodal. 
The phase diagram can be further characterized in the limit of large populations $ N \gg 1 $ following a Fokker-Planck analysis as shown in App.~\ref{sec:Stationary State apdx}. This approach becomes exact in the macroscopic limit $N \to \infty$~\footnote{We will use this approximation for most of our analytical inquiry, for which we avoid carrying the symbols $\simeq$ to ease the notation.}. A standard analysis leads to the analytical expressions for the critical point:
\begin{align}\label{eq:critical point asymmetric}
 \lambda_{\mathrm{c}} = \frac{q+1}{q-1}, ~~~&~~~ \chi_{\mathrm{c}} = \theta^{-1/q}, 
\end{align}
confirming that the system only exhibits a phase transition for $q > 1$. We also notice that the equilibrium condition $\lambda_{\mathrm{eq}} = 1 < \lambda_{\mathrm{c}}$, implies that in equilibrium the system remains permanently in the unimodal phase, and no phase transition can occur for all values of $q>1$. 

Next, we focus on the opinion predominance in the general phase diagram, which differs from the unimodal to the metastable regions.
In the unimodal region, $\chi\notin(\chi_{-},\chi_{+})$ the distribution transits continuously from a predominant consensus at $ B $ $ (n < N/2) $ to a predominant consensus at $ A $ $ (n > N/2) $ when crossing the line 
 $\chi_{\rm u} (\lambda) \equiv ({\lambda + \theta})/({1+\lambda \theta}), \, \forall q.$ 
For equal weights in the reactions $ (\theta = 1) $ as in \Cref{fig:phase diagram}, we obtain $ \chi_{\rm u} = \chi_{\rm c} = 1$. On the other hand, in the metastable region, $\chi\in(\chi_{-},\chi_{+})$, the distribution discontinuously transits,  when crossing the line $ \chi_{\rm b}(\lambda) $, from a bimodal distribution biased towards $ B $, i.e, the peak at $ n < N / 2 $ is higher than the one at $n > N/2$, to the opposite situation where the peak at $ n > N / 2 $ is higher.
In general, this first--order transition line has to be computed numerically by equating the heights of the two peaks of the bimodal distribution, see App.~\ref{sec:Stationary State apdx} for details, although for the case of \Cref{fig:phase diagram} $ (\theta = 1) $, it is possible to prove that $ \chi_{\rm b}(\lambda) = 1 $, for $\lambda\ge\lambda_c$.

To sum up, the system exhibits both first and second-order phase transitions. More precisely, crossing the line $\chi_{\rm b}(\lambda)$ within the metastable region produces a first-order phase transition between two consensus states favoring one opinion or the other, while a second-order, symmetry breaking transition, occurs for any trajectory in parameter space that crosses the critical point $(\lambda_c,\chi_c)$, where one of the two equivalent minima is chosen dynamically.

\section{Imitation thermodynamics}\label{sec:Opinion Thermodynamics}

At this point we develop a framework for stochastic thermodynamics in social systems such as the family of imitation models introduced above. The framework involves two complementary levels of description: the ensemble level, i.e. using the system probability distributions, probability currents, and dynamical activities as introduced above; and the level of trajectories, namely, of single realizations of the stochastic process dictating the changes in the attributes (opinion, cultural traits, etc.) in the population $N$ of agents over time~\footnote{For simplicity we often refer to the agent's attribute as an \textit{opinion}, but we keep in mind that the attribute can correspond to a generic social or cultural trait of the agents}. In order to provide a well-established foundation for the framework, we will start by introducing the stochastic thermodynamics of social systems at the trajectory level. In this level we derive universal nonequilibrium fluctuation relations that serve as the basis to obtain emergent universal inequalities at the ensemble level, reminiscent of the second law of thermodynamics.

\subsection{Thermodynamics along trajectories and fluctuation relations}
\label{TaTaFR}
We introduce a trajectory of the stochastic variable $n(t)$ during a time interval $[0, \tau]$ as the sequence $\mathbf{n}_{[0,\tau]}\equiv \{ n(t): 0\leq t \leq \tau \}$. Such a trajectory provides information about the evolution of the number of agents with opinion (or trait) $A$ over time. However, it does not contain all the relevant information for describing the full thermodynamics of imitation models, since the reaction leading to the corresponding change in $n(t)$ is missing in $\mathbf{n}_{[0,\tau]}$. For a full account of the relevant information we instead use a description based on the transitions or ``jumps'' in the opinions occurring during the interval $[0, \tau]$. More precisely, we introduce a trajectory as  $\gamma_{[0, \tau]} = \{n_0, (k_1, r_1, t_1), ... (k_i, r_i, t_i), ... ,(k_J, r_J, t_J)\}$ where $n_0$ is the initial value of $n(t)$ at time $t=0$, and the numbers $(k_i, r_i, t_i)$ denote the times $t_i \in \mathds{R}^+$ at which a change in the opinion of an agent is verified, the corresponding jump $k_i = \{+ , -\}$ adding or subtracting an agent with opinion $A$, and the reaction producing it, $r_i = \{1, 2 \}$, and $J$ is the total number of jumps along the trajectory~\footnote{An arbitrary number of reactions or jumps involving more than one agent change in opinion can be naturally incorporated in the framework.}. Such description is typically employed to describe the thermodynamics of jump trajectories in monitored quantum systems~\cite{Manzano22}, but it has been also shown useful in the classical domain~\cite{Hiura21,Vu23b}, e.g. when only a limited set of visible transitions are available~\cite{Martinez19, VdMeer22,Harunari22}.

The probability of observing the trajectory $\gamma_{[0, \tau]}$, with $J$ jumps along $[0,\tau]$ reads:
\begin{align} \label{eq:prob}
\mathds{P}(\gamma_{[0, \tau]}) & = P_{n_0}(0)~ \mathcal{D}(t_1,0) ~W_{n_1, n_0}^{(r_1)} \mathcal{D}(t_2,t_1) \cdots ~W_{n_{j}, n_{j-1}}^{(r_j)} \nonumber \\ &\cdots \,\mathcal{D}(t_J, t_{J-1}) ~ W_{n_J, n_{J-1}}^{(r_J)}~ \mathcal{D}(\tau, t_J) \, dt_1 \cdots dt_J,
\end{align}
where $P_{n_0}(0)$ is the initial probability to start with $n_0$ agents in opinion $A$, the quantities $\mathcal{D}(t_i, t_j) = e^{- \int_{t_j}^{t_i} dt \sum_{r} \sum_{m} W^{(r)}_{m, n_i} }$ stand for the probability of a ``dwell'' time between $t_j$ and $t_i$ where the system stays in $n_i$ and no jumps occur, and $W_{n_j, n_i}^{(r_j)} dt_j$ are the probabilities for a jump from $n_i$ to $n_j = n_i +k_i$ due to reaction $r_j$ during the (infinitesimal) interval $[t_j, t_j + dt_j]$. We also denoted for convenience $n_0, n_1, ..., n_J$ the sequence of values taken by the variable $n(t)$ after each jump ($n(\tau)= n_J$). Moreover, we notice that the above expression for the probability of a trajectory is valid when the rates $W_{n_j, n_i}^{(r)}$ are time-dependent as a consequence of the modification of the system parameters $h_r$ and $a_r$ during time. We denote in this case $\Lambda = \{h_1(t), h_2(t), a_1(t), a_2(t) ~;~ 0 \leq t \leq \tau\}$ the sequence of values that the model rates take over time, which is usually called a ``driving protocol" in stochastic thermodynamics.

\begin{figure}[t]
\includegraphics[width=1\columnwidth]{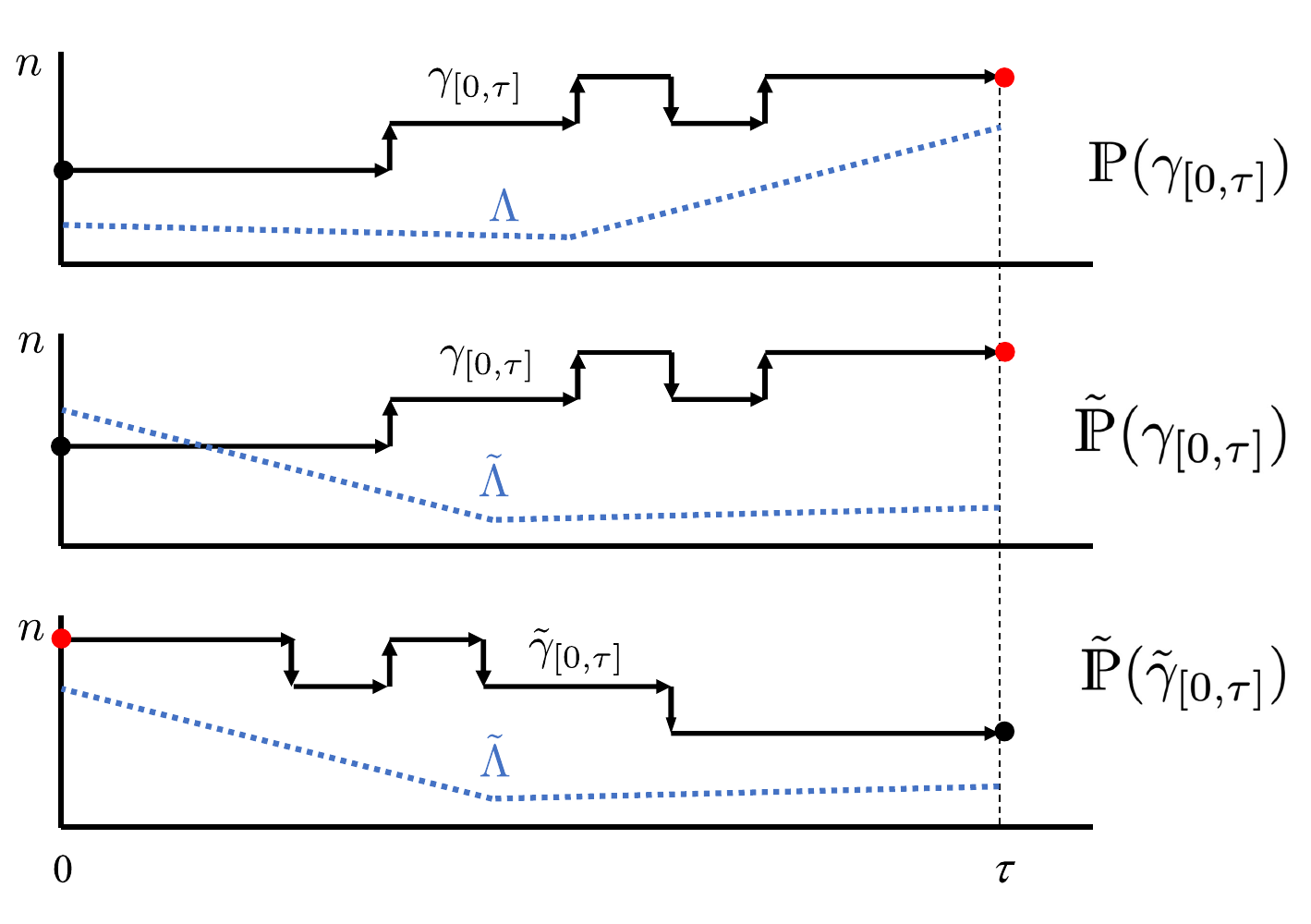}
 \caption{\label{fig:traj}
 Illustration of a forward trajectory, $\gamma_{[0,\tau]}$, with $3$ jumps up ($k_i = +$ for $i=1,2, 4$) and $1$ jump down ($k_3 = -$), together with a driving protocol $\Lambda$ (top plot); and with the inverted protocol of the backward process, $\tilde{\Lambda}$ (middle plot). The corresponding time-reversed trajectory, $\tilde{\gamma}_{[0,\tau]}$, with the inverse jumps ($\tilde{k}_i = -$ for $i=1,2,4$ and $\tilde{k}_3 = +$) in the backward process $\tilde{\Lambda}$ is illustrated in the bottom plot. For obtaining Eq.~\eqref{eq:DFT} we compare the probabilities of trajectories in top and bottom plots.
 }
\end{figure}

One of the cornerstones of stochastic thermodynamics is the so-called detailed fluctuation theorem~\cite{Crooks99,Seifert12,vandenbroeckEnsembleTrajectoryThermodynamics2015}, which establishes a link between the irreversibility of a process with their energetics measured from the heat exchanged with the surrounding medium or the work needed to implement it, along single trajectories. Here we show that an analogous relation can be obtained in our context. In order to obtain it, we introduce the notion of a backward process, consisting in a stochastic process analogous to the one introduced above, but with the particularity of being implemented with the inverted protocol $\tilde \Lambda = \{h_1(\tau - t), h_2(\tau - t), a_1(\tau - t), a_2(\tau - t) ~;~ 0 \leq t \leq \tau\}$, for which the reaction rates follow the time-reversed sequence of values. We denote the probability of trajectories (e.g. $\gamma_{[0,\tau]}$) in the backward process by $\tilde{\mathds{P}}(\gamma_{[0,\tau]})$, to differentiate it from $\mathds{P}(\gamma_{[0,\tau]})$ (see Fig.~\ref{fig:traj}). 

Using the properties of the trajectory probabilities [Eq.~\eqref{eq:prob}] and the local detailed balance relation, Eq.~\eqref{eq:local}, we derive in App~\ref{eq:Corollaries of the Fluctuation Theorems} a version of the detailed fluctuation theorem in stochastic thermodynamics of social systems:
\begin{equation} 
\begin{split} \label{eq:DFT}
 S_{\rm tot}(\gamma_{[0,\tau]}) \equiv \ln \left( \frac{\mathds{P}(\gamma_{[0,\tau]})}{\tilde{\mathds{P}}(\tilde{\gamma}_{[0,\tau]})} \right) = \Delta S_{\rm sys} + \sum_r \mu_r I_r,
\end{split}
\end{equation}
where $\tilde{\gamma}_{[0,\tau]} = \{n_\tau, (\tilde{k}_J, r_J, \tau - t_J), ..., (\tilde{k}_1, r_1, \tau-t_1) \}$ is the time-reversed trajectory associated to $\gamma_{[0,\tau]}$, where the jumps $\tilde{k}_i = \{+, -\}$ are inverted, i.e. to a jump up in the forward trajectory ($k_i= +$), it corresponds a jump down ($\tilde{k}_i = -$) in the time-reversed trajectory, and vice-versa, see Fig.~\ref{fig:traj}. 

In the right-hand side of Eq.~\eqref{eq:DFT} we identify two terms: the first one is the difference in system entropy $ \Delta S_{\rm sys} \equiv \ln P_{n_0}(0) - \ln P_{n_\tau}(\tau) + S_{n_\tau}^{\mathrm{int}} - S_{n_0}^{\mathrm{int}}$ coming from the change in surprisal (or Shannon information), $-\ln P_{n}(t)$, from initial to final system configurations~\cite{Seifert05}, and the difference in internal entropy of the aggregated opinion states at the beginning and at the end of the trajectory [see Eq.~\eqref{eq:Internal Entropy}]; the second one accounts for the accumulated changes in opinion produced by each reaction, $I_r({\gamma_{[0,\tau]}}) \equiv N_+^{(r)} - N_-^{(r)}$, with $N_+^{(r)} (N_-^{(r)})$ the total number of jumps $+ (-)$ from reaction $r$ during the trajectory, multiplied by the (generalized) chemical potentials $\mu_r = \sigma_r \ln (h_r/a_r)$, which quantify the intrinsic bias in each transition towards opinion $A$.

The detailed fluctuation theorem in Eq.~\eqref{eq:DFT} represents a footprint of the second law in our system at the level of fluctuations. It provides a powerful equality that relates an information-theoretical measure of irreversibility (the relative likelihood of trajectories in forward and backward processes) on the l.h.s., with the changes in entropy of the social system, $\Delta S_{\rm sys}$ and the opinion fluxes $I_r$ in the r.h.s., the latter playing the role of energy or particle fluxes in standard stochastic thermodynamics. In this sense, the quantity $S_{\rm tot}$ appearing in Eq.~\eqref{eq:DFT} can be interpreted as the stochastic entropy production in opinion thermodynamics. It quantifies irreversibility in the social dynamics, but instead of relating it to energy dissipation as in physical and chemical systems, it establishes a link to the possible changes in the attributes (e.g. opinion) of the agents in the social system.

Taking the average over all possible trajectories, Eq.~\eqref{eq:DFT} leads to an integral version of the fluctuation theorem:
\begin{equation} \label{eq:IFT}
\langle e^{-S_{\rm tot}} \rangle = \sum_{\gamma_{[0,\tau]}} \tilde{\mathds{P}}(\tilde{\gamma}_{[0,\tau]}) = 1,
\end{equation}
with $\langle f(\gamma_{[0,\tau]}) \rangle = \sum_{\gamma_{[0,\tau]}} \mathds{P}({\gamma}_{[0,\tau]}) f(\gamma_{[0,\tau]})$. The second equality follows from the fact that $\tilde{\mathds{P}}$ is a normalized distribution and $\gamma_{[0,\tau]}$ contains the same stochastic variables as $\tilde{\gamma}_{[0,\tau]}$. 
As in standard stochastic thermodynamics, the integral fluctuation theorem (IFT) \eqref{eq:IFT} puts strict constraints on the allowed statistics of $S_{\rm tot}$~\cite{Jarzynski11}. For example, it implies that negative values of $S_{\rm tot}$ are severely reduced as $\mathrm{Prob}(S_{\rm tot} < - \alpha) \leq e^{-\alpha}$ for $\alpha \in \mathds{R}_+$ (see App~\ref{eq:Corollaries of the Fluctuation Theorems}). 


\subsection{Second-law inequality and ensemble thermodynamics}

The fluctuation relations obtained above have also crucial implications for some thermodynamic quantities at the ensemble level, that is, when taking averages over the set of all possible trajectories. Applying Jensen's inequality to the integral fluctuation theorem in Eq.~\eqref{eq:IFT} (see App.~\ref{eq:Corollaries of the Fluctuation Theorems}), we obtain
\begin{equation} \label{eq:second-law}
 \langle \Delta S_{\rm sys} \rangle + \sum_r \mu_r \langle I_r \rangle \geq 0,
\end{equation}
which corresponds to the second-law inequality for social imitation thermodynamics. The average changes in system entropy are related to Shannon entropy changes as
\begin{equation}
\begin{split}
\langle \Delta S_{\rm sys} \rangle =&~ S[P(\tau)] - S[P(0)] \\
&+ \sum_n [P_n(\tau) - P_n(0)] S_n^{\rm int},
\end{split}
\end{equation}
with $S[P] = - \sum_n P_n \ln P_n$ the Shannon entropy, and the second term arises from the internal entropy changes. This double contribution of the system entropy naturally appears in systems with coarse-grained mesostates~\cite{Herpich2020Jun,Falasco25}, such as in many models of molecular motors~\cite{Seifert19}. On the other side, the average accumulated changes in opinion during the interval, $\langle I_r \rangle$, are related to the probability currents as:
\begin{equation}\label{eq:<dotIr>}
\langle \dot{I}_r \rangle \equiv \lim_{\tau \rightarrow \infty} \frac{\langle I_r \rangle}{\tau} = \sum_{m < n} \sum_n J_{m,n}^{(r)}(t),
\end{equation}
for each reaction $r$, with $J_{m,n}^{(r)}(t)$ in Eq.~\eqref{eq:Jmn} (see App. \ref{Cpr}). Moreover, the average entropy production in the system can be written as:
\begin{equation} \label{eq:EP}
\begin{split}
\langle S_{\rm tot} \rangle &=~ \sum_{\gamma_{[0,\tau]}} \mathds{P}(\gamma_{[0,\tau]}) \ln \left( \frac{\mathds{P}(\gamma_{[0,\tau]})}{\tilde{\mathds{P}}(\tilde{\gamma}_{[0,\tau]})} \right) \\ &= D[\mathds{P}(\gamma_{[0,\tau]}) || \tilde{\mathds{P}}(\tilde{\gamma}_{[0,\tau]})],
\end{split}
\end{equation}
where $D(P||Q) \geq 0$ is the Kullback-Leibler divergence, or relative entropy~\cite{Cover2006}. It is a non-negative measure of statistical distinguishability for any two probability distributions $P$ and $Q$ with same support, and becomes zero if and only if $P=Q$. The identification of the entropy production with a Kullback-Leibler divergence~\cite{Maes03,Kawai07,Gomez-Marin08} has been largely celebrated and employed for inference purposes in living systems~\cite{Roldan10} or computational machines~\cite{manzanoThermodynamicsComputationsAbsolute2024}. Taking the derivative of the entropy production, Eq.~\eqref{eq:EP}, we obtain the entropy production rate:
\begin{equation} \label{eq:EPrate}
 \braket{\dot{S}_{\mathrm{tot}}} = \sum_{r} \sum_{n,m} W^{(r)}_{n, m}(t) P_{m}(t) \ln \left[ \frac{W^{(r)}_{n, m}(t) P_{m}(t)}{W^{(r)}_{m, n}(t) P_n (t)} \right], 
\end{equation}
which is in accordance with the standard definition of the entropy production rate for jump processes in stochastic thermodynamics~\cite{Schnakenberg1976Oct,vandenbroeckEnsembleTrajectoryThermodynamics2015}, with $\braket{\dot{S}_{\mathrm{tot}}}\geq 0$ (see App.~\ref{App:PS_tot}).

\begin{figure*}[t]
 \centering
 \includegraphics[width=\textwidth]{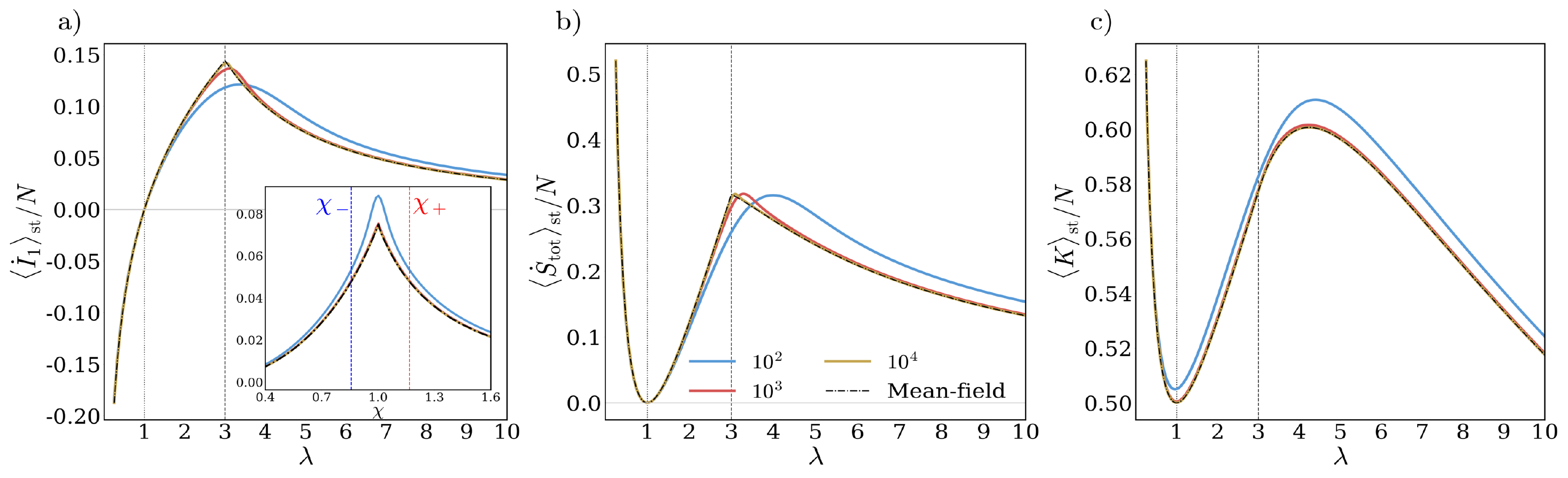}\label{fig:j1}
 
\caption{\label{fig:Ensemble Analysis} 
Ensemble thermodynamic analysis for $q = 2$ and $\theta = 1$. Panels (a--c) show intensive stationary quantities as functions of $\lambda$ with $\chi = 1$: (a) probability current $\langle \dot{I}_1\rangle_{\mathrm{st}}/N$, (b) entropy production rate $\langle \dot{S}_{\mathrm{tot}}\rangle_{\mathrm{st}}/N$ from Eq. (\ref{eq:Stotst}), and (c) dynamical activity $\langle K\rangle_{\mathrm{st}}/N$. These curves are calculated with the stationary distribution Eq. (\ref{eq:Stationary Distribution Analytical Solution}). Exact results for system sizes $N\in\{10^2,10^3,10^4\}$ are shown alongside the mean-field solution (black dash-dotted line, see App.~\ref{sec:Mean-field analytical results}). Vertical dotted and dashed lines mark the equilibrium point ($\lambda_{\mathrm{eq}}=1$) and critical point ($\lambda_{\mathrm{c}}=3$), respectively. Inset in panel (a): dependence of $\langle \dot{I}_1\rangle_{\mathrm{st}}/N$ on $\chi$ at fixed $\lambda=5$, indicating $\chi_{-}$ (blue) and $\chi_{+}$ (red).}
\end{figure*}

Similarly to Landauer's principle linking information and heat~\cite{Lutz15,Parrondo15}, the second-law inequality \eqref{eq:second-law} implies that any change in the configuration of opinions in the system, as measured by the change in entropy $\langle \Delta S_{\rm sys} \rangle$, needs to be compensated by (average) opinion currents $\langle I_r \rangle$. More than that, the biases { in the intrinsic mechanisms leading to opinion changes} $\mu_r$ associated to each reaction determine the possible spontaneous evolution of the system in terms of the distribution of opinions (or traits), and the associated spontaneous currents.

In the long-time run the system reaches the stationary distribution $P_n^{\rm st}$. In that situation the changes in system entropy vanish and the second-law inequality~\eqref{eq:second-law} retains only its second term proportional to the average currents. The entropy production rate becomes:
\begin{equation} \label{eq:Stotst}
 \langle \dot{S}_{\rm tot} \rangle_{\rm st} = \sum_r \mu_r \langle \dot{I}_r \rangle_{\rm st} = \left( \mu_1 - \mu_2 \right) \langle \dot{I}_1 \rangle_{\rm st} \geq 0,
\end{equation}
where the second equality follows from the global detailed balance condition in Eq.~\eqref{eq:GDB}, that implies compensated stationary opinion currents from each reaction, $\langle \dot{I}_1 \rangle_{\rm st} = - \langle \dot{I}_2 \rangle_{\rm st}$, and we used subscripts in the brackets $\langle \cdot \rangle_{\rm st}$ to denote averages in stationary conditions. The last inequality above is a consequence of the non-negativity of Eq.~\eqref{eq:EPrate}.

The above inequality implies that the difference { of the potentials in each reaction act as a themodynamic force, determining} the sign of the stationary opinion currents, where we remark that $\mu_1 - \mu_2 = 2 \ln \lambda$ only depends on the global bias of the model towards herding or anticonformity (and not on the parameters $\chi$ or $\theta$ measuring the intrinsic asymmetry in the opinions or reactions). Therefore, whenever $\mu_1 > \mu_2$ (that is $\lambda > 1$) we have both $\langle \dot{I}_1 \rangle_{\rm st} \geq 0$ and $\langle \dot{I}_2 \rangle_{\rm st} \leq 0$, meaning that both reactions enforce average currents towards herding [see Eqs.~\ref{eq:Model Reactions}] which becomes the dominant mechanism. On the other hand, if $\mu_1 < \mu_2$ (that is $\lambda < 1$), both currents are inverted $\langle \dot{I}_1 \rangle_{\rm st} \leq 0$ and $\langle \dot{I}_2 \rangle_{\rm st} \geq 0$, meaning that the anticonformity mechanism becomes dominant in both reactions. The point where the generalized chemical potentials become equal, $\mu_1 = \mu_2$ ($\lambda = 1$) corresponds to equilibrium and both currents exactly vanish, in accordance with our analysis in Sec.~\ref{sec:Stationary State}.

Different steady-state regimes in the model as a function of $\lambda$ are illustrated in Fig.~\ref{fig:Ensemble Analysis} for increasing system sizes and in the macroscopic limit. Focusing on the symmetric opinions case ($\chi = \theta = 1$) with nonlinearity $q=2$, we show, as a function of $\lambda$, the average opinion current $\langle \dot{I}_1 \rangle_{\rm st}$ [Fig.~\ref{fig:Ensemble Analysis}a], the entropy production rate $\langle \dot{S}_{\rm tot} \rangle_{\rm st}$ [Fig.~\ref{fig:Ensemble Analysis}b], and the total dynamical activity $\langle K \rangle_{\rm st}$ [Fig.~\ref{fig:Ensemble Analysis}c]. The latter is defined as $\langle K \rangle_{\rm st} \equiv \sum_{m<n} \sum_n \sum_r {K_{n, m}^{(r)}}$ [c.f. Eq.~\eqref{eq:K}] and provides the average number of changes in opinion (no matter their sign or the reaction producing them) per unit time. Exact curves for finite $N$ are compared with analytical expressions of the three quantities obtained using the mean-field approximation (macroscopic limit) and given in App~\ref{sec:Mean-field analytical results}, showing an excellent convergence for large values of $N$. 

Focusing on the region $0 < \lambda \leq \lambda_{\mathrm{c}}$ in Fig~\ref{fig:Ensemble Analysis}, where the opinions in the system are polarized , we can observe the two main stationary regimes where anti-conformity dominates $(\lambda < 1)$, leading to $\langle \dot{I}_1 \rangle_{\rm st} < 0$, and where herding is favored $(\lambda > 1)$, with $\langle  \dot{I}_1 \rangle_{\rm st} > 0$, separated by the equilibrium point $(\lambda = 1)$. At equilibrium the currents and the entropy production vanish, indicating zero bias in the system. However, the dynamical activity is minimal but not zero, unveiling the presence of (symmetric) fluctuations in the opinions. We notice that in the limit $\lambda \rightarrow 0$ of infinite bias towards anti-conformity, the three quantities diverge, and the system enters in a frenetic state ($\langle K \rangle_{\rm st} \to \infty$) of maximal polarization with fast changes in opinion back and forth from $A$ to $B$. This behavior contrasts with the opposite case where the bias towards herding is enforced. Indeed as the system crosses the critical point $\lambda_{\rm c} = 3$ (black dashed line) an abrupt change in all three quantities is observed as a consequence of the reorganization of the system into a consensus state. More precisely, both the opinion current and the entropy production reach a local maximum at the critical point, while the dynamical activity suffers a change in curvature. As a consequence, for $\lambda > \lambda_{\rm c}$ we observe how the persistence of the herding behavior in the consensus state lead to smaller currents (it becomes more and more difficult to find other agents with a different opinion) and consequently the irreversibility also diminishes. This situation leads to a ``frozen" consensus state in the limit $\lambda \rightarrow \infty$ with all agents having the same opinion. There, reversibility is recovered with zero opinion currents and entropy production but, contrary to the $\lambda=1$ case, the dynamical activity also vanishes. This indicates the complete absence of fluctuations in the dominant opinion and justifies calling it a ``frozen" state. Finally, in the inset of Fig~\ref{fig:Ensemble Analysis}a we also show the behavior of the currents when varying the intrinsic asymmetry of the opinions $\chi$, for fixed bias $\lambda$. As observed there, the currents are maximum in the symmetric case (when the first-order phase transition takes place) and decrease for increasing asymmetry in both directions.

\subsection{Thermodynamic uncertainty relation}\label{sec:TUR}

A celebrated result in stochastic thermodynamics over recent years has been the so-called thermodynamic uncertainty relation (TUR)~\cite{Barato15,Todd16}, which establishes a simple but powerful relationship between dissipation (as measured by the entropy production), and the precision of currents out of equilibrium. Although different extensions of the TUR have been developed over the last years~\cite{Horowitz20} to include time-dependent driving~\cite{Proesmans19,Timur20} or even quantum effects~\cite{Guarnieri19,Carollo19}, here we focus on its original form, valid for Markovian systems in nonequilibrium stationary states (NESS): 
\begin{equation} \label{eq:TUR}
\frac{\dot{\sigma}^2_\text{st}(I)}{\langle \dot{I}\rangle_{\rm st}^2} \geq \frac{2}{\langle \dot{S}_{\rm tot}\rangle_{\rm st} },
\end{equation}
here $I$ represents an arbitrary stochastic current of the system (typically a heat current or a particle current) and $\dot{\sigma}_\text{st}^2(I)\equiv \lim_{\tau\to\infty}(\langle I^2\rangle_{\rm st} - \langle I \rangle_{\rm st}^2)/\tau$ is the current dispersion rate, which can be obtained from the scaled cumulant generating function using large deviation theory~\cite{Barato15,Todd16} (see App.~\ref{sec:Cal Flu Met}). 

The TUR in Eq.~\eqref{eq:TUR} implies that reducing the relative dispersion of the currents, i.e. making smaller the left hand side, can only come at the cost of increasing the entropy production in the right hand side. Therefore, the TUR unravels a fundamental trade-off between precision and dissipation: having precise currents in a stochastic system is not for free, but it requires a large irreversibility (and dissipation). This means that a coherent behavior with reduced fluctuations, as often observed in the context of complex systems, can only be achieved far from equilibrium, when entropy production becomes large.
\begin{figure}
 \centering
 \includegraphics[width=0.95\linewidth]{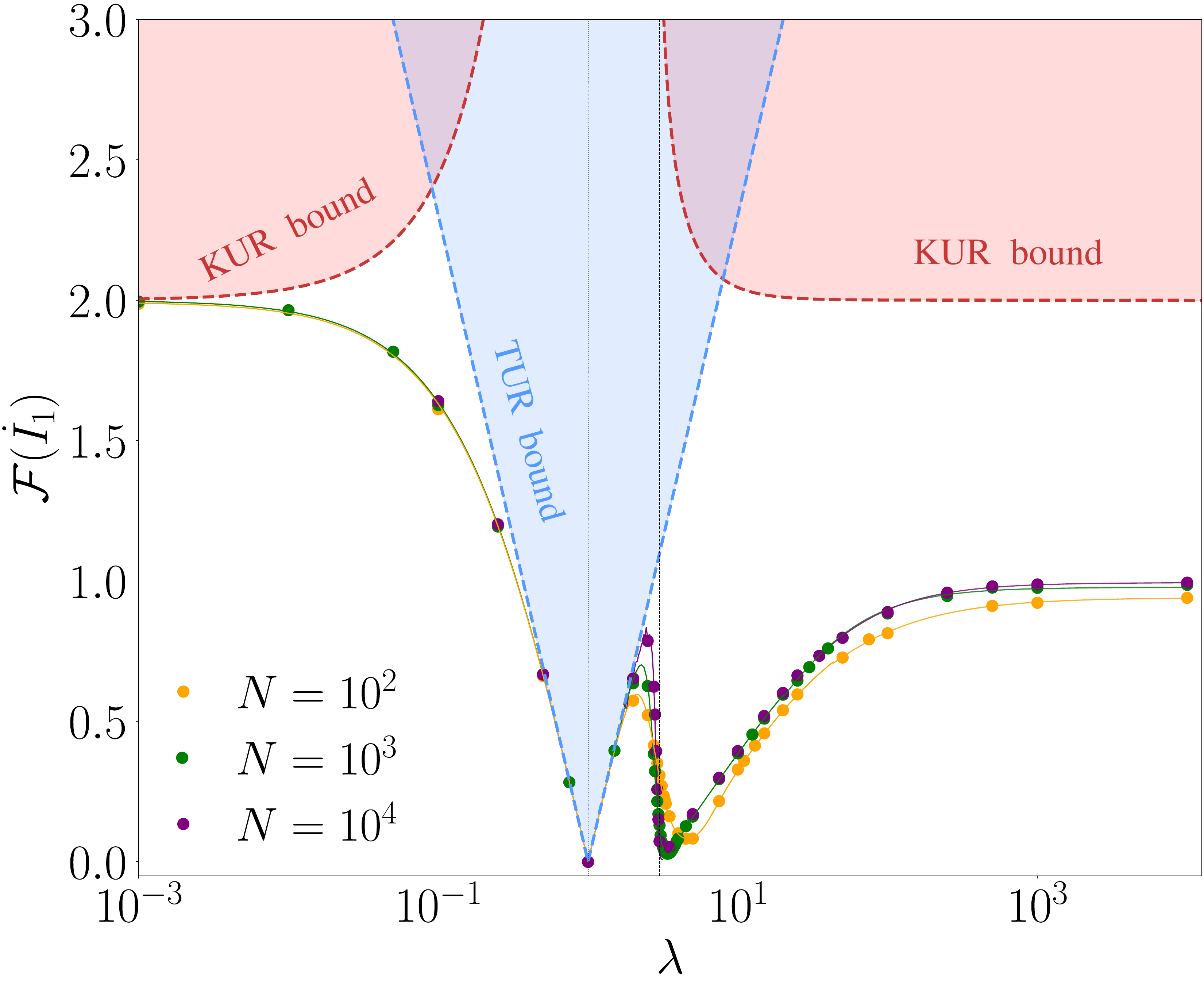}
 \caption{Fano factor $\mathcal{F}(\dot{I}_1)$ as a function of $\lambda$ for the symmetric case ($\chi = \theta = 1$) with $q=2$. Solid lines show exact results computed via Full Counting Statistics for system sizes $N\in\{10^2,10^3,10^4\}$, while circles represent numerical simulations obtained with the Gillespie Method. Shaded regions indicate bounds from the thermodynamic uncertainty relation (TUR), $|\ln \lambda|$ (blue), and kinetic uncertainty relation (KUR), $\langle K \rangle_{\rm st}/|\langle \dot{I}_1 \rangle_{\rm st}|$ (red). Vertical dotted and dashed lines mark the equilibrium point ($\lambda_{\rm eq}=1$) and critical point ($\lambda_{\rm c}=3$), respectively. Parameter: $\omega~\tau = 250$.}
 \label{fig:TUR}
\end{figure}
Applying the TUR in Eq.~\eqref{eq:TUR} to our social model by particularizing it to the opinion current $I_1$, and using the expression for the entropy production in Eq.~\eqref{eq:Stotst}, we obtain, after some rearranging of the terms:
\begin{equation} \label{eq:TURopinion}
\mathcal{F}(\dot{I}_1) \equiv \frac{|\langle \dot{I}_1\rangle_{\rm st}|}{\dot{\sigma}_{\rm st}^2({I}_1)} \leq \frac{|\mu_1 - \mu_2|}{2} = |\ln \lambda|,
\end{equation}
where we introduced the Fano-factor of the current, $\mathcal{F}(\dot{I}_1)$. The Fano factor is a signal-to-noise ratio that measures the variability of the opinion changes with respect to the average opinion current. The above inequality poses an upper limit on the accuracy of the opinion currents that depends only on the generalized chemical potential difference between the two reactions (controlling the herding vs. anti-conformity bias). Finer opinion currents require higher biases in the reaction rates, which can be aligned towards herding ($\mu_1 \gg \mu_2$) or anti-conformity ($\mu_1 \ll \mu_2$).

Far from equilibrium, however, the TUR typically becomes far from tight~\cite{Horowitz20}, meaning that the entropy production is not the only quantity of interest that limits the precision of the currents. Indeed time-symmetric quantities, such as the dynamical activity or the traffic, may also play an important role to understand non-equilibrium behavior~\cite{Maes20}. In this context, another relevant uncertainty relation, called the Kinetic Uncertainty Relation (KUR), has been recently derived~\cite{Terlizzi18}: 
\begin{equation} \label{eq:KUR}
\frac{{\dot{\sigma}_{\rm st}^2({I})}}{\langle \dot{I}\rangle_{\rm st}^2} \geq \frac{1}{\langle{K}\rangle_{\rm st} },
\end{equation}
that sets a limit to the precision of the currents from the dynamical activity [c.f. Eq.~\eqref{eq:TUR}]. The KUR is a universal bound valid for generic Markovian processes in the NESS and can provide a powerful complement to the TUR~\cite{Hiura21}. In terms of the Fano factor of the opinion current $\dot{I}_1$, the KUR can be rewritten as $\mathcal{F}(\dot{I}_1) \leq \langle K \rangle_{\rm st}/\langle \dot{I}_1 \rangle_{\rm st}$, providing us an alternative upper bound to the signal-to-noise ratio achievable by the opinion changes in the system, not based on the asymmetry under time-reversal like Eq.~\eqref{eq:TURopinion}{, but on the total volatility of opinions. Somehow paradoxically, Eq.~\eqref{eq:KUR} implies that large and precise opinion currents need a high volatility of opinions in the society, in the sense of fast fluctuations in the opinions of the agents}.

In order to compute current averages and fluctuations, here and in the following we employ both numerical simulations of stochastic trajectories using the Gillespie algorithm, and spectral methods such as Full Counting Statistics and Large Deviation Theory (valid for large trajectory times)~\cite{Esposito2007Apr,Flindt2008Apr,Walldorf2020May,Landi2024Apr,Ledermann1954Jan}. These methods allow us to obtain all the moments of the current distribution in the NESS for any finite $N$, as detailed in Appendix \ref{sec:Cal Flu Met}.

In Fig.~\ref{fig:TUR} we illustrate the current Fano factor $\mathcal{F}(\dot{I}_1)$ for large system sizes ($N= 10^3$ and $N=10^4$) together with the TUR and KUR bounds, for symmetric opinions ($\chi = \theta = 1$) when varying the bias $\lambda$. As expected, the TUR is saturated at the equilibrium point $\lambda = 1$ (dotted vertical line), where the currents become exactly zero, and it is tight around it. When increasing the bias towards anti-conformity ($\lambda < 1$), the Fano factor increases indicating larger average currents with respect to the fluctuations, but it departs from the limit established by the TUR. There increasing the entropy production has a marginal impact in obtaining more precise currents while the bound established by the KUR becomes tighter. Far from equilibrium, when $\lambda \ll 1$, the KUR is saturated, spotting the role of the dynamical activity as a limiting factor for the accuracy of the currents. On the other hand, for biases towards herding ($\lambda > 1$) the behavior of the Fano factor is quite different due to the presence of the phase transition. The reorganization of the system into a consensus state at $\lambda_{\rm c} = 3$ (vertical dashed line) produces a sudden drop in the Fano factor due to the amplification of fluctuations close to the critical point. Then, as $\lambda$ becomes larger, the precision of the currents increases again and saturates in the large $\lambda$ limit, as the consensus states become frozen. In this limit, contrary to the $\lambda \rightarrow 0$ regime, neither the TUR nor the KUR are saturated. Interestingly, however, the numerical value for the saturation of the Fano factor turns out to be very close to half of the theoretical KUR bound, which suggests that this effect might be related to the symmetry breaking transition in the consensus phase, $\lambda > \lambda_{\rm c}$.

\subsection{Current fluctuations and inference} \label{sec:inference}

The detailed and integral fluctuation relations for entropy production in Eqs.~\eqref{eq:DFT} and \eqref{eq:IFT} are generically valid for arbitrary out-of-equilibrium processes, including those with time-dependent parameters that may describe situations far from the stationary regime. However, stronger versions of these fluctuation relations for the opinion currents can be obtained in the NESS thanks to the addition of the time-translational symmetry characteristic of stationary distributions. In particular, in this case the path probability of trajectories in the time-reversed process, $\tilde{\mathds{P}}(\gamma_{[0,\tau]})$, becomes equivalent to the one in the forward process, $\mathds{P}(\gamma_{[0,\tau]})$ in Eq.~\eqref{eq:prob}. This is a consequence of the fact that the protocol of any process taking place in the NESS is fixed and hence invariant under time-reversal, $\Lambda = \tilde \Lambda$; and the initial states in both the forward and time-reversed process are sampled from the (same) stationary distribution $P_n^{\rm st}$.
\begin{figure}[t]
 \centering
 \includegraphics[width=\linewidth]{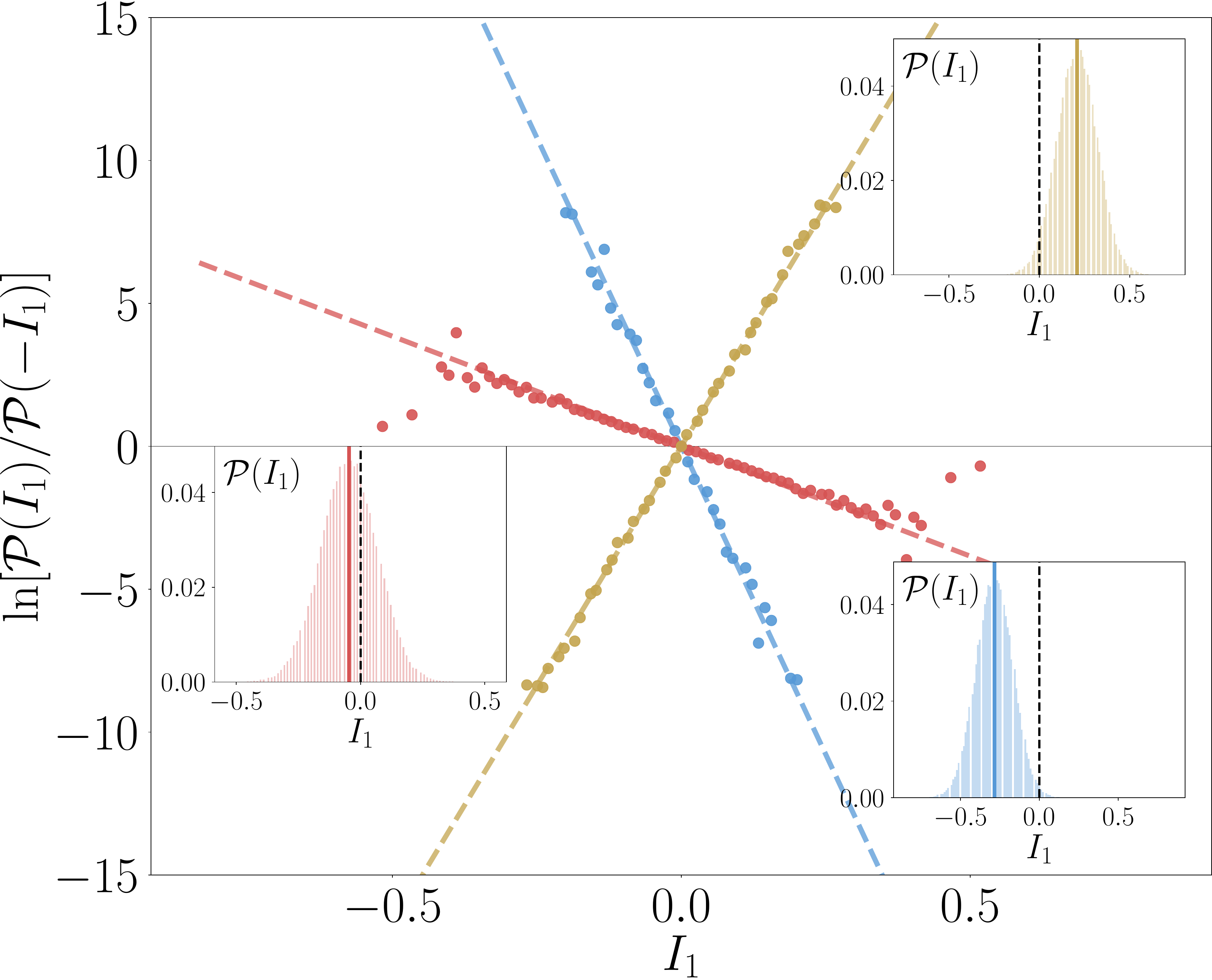}
 \caption{Inference of model parameters from current fluctuations via Eq.~\eqref{eq:Inference}. Linear regressions (dashed lines) yield estimates of the chemical potential difference $\mu_1 - \mu_2=2\ln\lambda$ from the slopes: $-0.575$ ($\lambda=0.75$, blue) with $r^2=0.9952$; $-0.102$ ($\lambda=0.95$, red) with regression coefficient $r^2=0.9941$; and $0.446$ ($\lambda=1.25$, gold) with $r^2=0.9970$. Insets: Corresponding stationary current distributions $\mathcal{P}(I_1)$ obtained from simulations generated with the Gillespie method. Parameters: $\theta=\chi=1$, $q=2$, $N=100$, $\omega~\tau=250$.} 
 \label{fig:inference}
 
\end{figure}
Under the above condition, it is useful to introduce the probability distribution of the integrated opinion current $\mathcal{P}(I)$, where $I$ is a continuous stochastic variable. For the opinion current in the first reaction $I_1$ in the NESS, it is defined as:
\begin{equation}
 \mathcal{P}(I) = \sum_{\gamma_{[0,\tau]}} \mathds{P}(\gamma_{[0,\tau]}) \indi[I - I_1(\gamma_{[0,\tau]})],
\end{equation}
we recall that $I_1(\gamma_{[0,\tau]}) = N_+^{(1)}(\gamma_{[0,\tau]})- N_-^{(1)}(\gamma_{[0,\tau]})$ comes from the stochastic sequence of opinion jumps during a time interval $[0,\tau]$ and $\indi$ above denotes the indicator function, $\indi(x) = 1$ if $x=0$ and $\indi(x) = 0$ otherwise. Analogously for the time-reversed process we have the corresponding probability distribution $\tilde{\mathcal{P}}(I_1) = \sum_{\gamma_{[0,\tau]}} \mathds{P}(\tilde{\gamma}_{[0,\tau]}) \indi[I - I_1(\tilde{\gamma}_{[0,\tau]})]$. 

It is worth noticing that the current $I_1$ associated to the time-reversed trajectory verifies $I_1(\tilde{\gamma}_{[0,\tau]}) = N_-^{(1)}(\gamma_{[0,\tau]}) - N_+^{(1)}(\gamma_{[0,\tau]}) = - I_1 (\gamma_{[0,\tau]})$, since jumps up in the number of agents with opinion $A$ in the forward trajectory translate in jumps down in the time-reversed trajectory and vice versa. As a consequence, the probability of observing a given current value in the time-reversed process is related to the one of observing the opposite sign current in the forward one, that is $\tilde{\mathcal{P}}(I_1) = \mathcal{P}(-I_1)$, which in combination with the detailed fluctuation theorem in Eq.~\eqref{eq:DFT} leads to the following detailed fluctuation theorem for the currents:
\begin{equation}\label{eq:Inference}
\frac{\mathcal{P}(I_1)}{\mathcal{P}(-I_1)}=e^{(\mu_1 - \mu_2)~ I_1},
\end{equation}
valid for the stationary state (for a proof see App.~\ref{apdx:stationary state details}). We note that the above fluctuation relation is stronger than Eq.~\eqref{eq:DFT} since it put constraints between the two tails (corresponding to positive and negative values of $I_1$) of the \emph{same} probability distribution $\mathcal{P}(I_1)$, with no explicit reference to the time-reversed process. It also immediately leads to the integral fluctuation relation, $\langle e^{-(\mu_1 - \mu_2) I_1} \rangle = 1$. { Socially, Eq.~\eqref{eq:Inference} implies that, during an arbitrary interval of time $\tau$, the probability of net opinion changes $I_1$ towards $A$ due to herding becomes exponentially greater (or lower) than the opposite changes $-I_1$ towards $B$ due to anti-conformity, as controlled by the difference in the reaction potentials $\mu_1 - \mu_2$.}

The above strong detailed fluctuation theorem for the current $I_1$, is specially well-suited for inference applications~\cite{Andrieux06,Hayasi10}. In particular, it can be used to efficiently estimating the generalized chemical potential difference in the model $\mu_1 - \mu_2 \,(\,= 2 \ln \lambda)$, measuring the bias towards herding or anti-conformity from a linear regression. 

To be more precise, let us imagine a practical situation where opinion changes triggered by the different reactions can be counted over time. The opinion changes due to reaction $1$ (reaction $2$) could be distinguished from those promoted by the other reaction because an increase in the number of agents with opinion $A$ (opinion $B$) occurs through a herding event. Observing repeatedly the system during several intervals of time of fixed duration, would allow us to sample values of $I_1$. Collecting these values and pairing them in positive and negative twins, one can compute and compare their frequencies of occurrence and the (logarithm) of their ratio, $\ln \left[ \mathcal{P}(I_1)/\mathcal{P}(-I_1) \right]$. As illustrated in Fig.~\ref{fig:inference}, when representing such values with respect to the magnitude of $I_1$, as a consequence of the FT in Eq.~\eqref{eq:Inference}, they appear as straight lines, whose slope corresponds to the bias in generalized chemical potentials, $\mu_1 - \mu_2$. 

\subsection{Thermodynamics of spontaneous symmetry-breaking}\label{sec:Ergodicity Breaking}

Going beyond the NESS, and given the richness of the phase-diagrams typically appearing in opinion dynamics models (as illustrated in Fig.~\ref{fig:phase diagram}; see also Ref.~\cite{Llabres25}), a particularly interesting case where our framework can be applied consists in processes undergoing spontaneous symmetry-breaking through the finite-time variation of the model parameters. More precisely, we are interested in the case where the parameter $\lambda$ controlling the second-order phase transition from a polarized society to a consensus state is varied in time. We therefore consider a driving protocol $\Lambda = \{ \lambda(t): 0\leq t \leq \tau \}$ with other parameters remaining constant ($\chi=\theta =1$ for simplicity), that makes the system transit across the critical point, that is with initial and final values $\lambda(0) < \lambda_{\rm c}$, and $\lambda(\tau) > \lambda_{\rm c}$. 

As mentioned in Sec.~\ref{sec:Stationary State}, the phase transition entails ergodicity breaking in the macroscopic limit due to the existence of degenerate (bistable) stationary solutions in one of its phases. Accordingly, we work in the large-$N$ regime, taking $N$ sufficiently large that the escape probability from the basin of attraction of a given minimum is negligible on the timescales of interest. When crossing the critical point $\lambda_{\rm c}$ from polarization to consensus, the phase space splits into two disconnected regions ($n<N/2$ and $n>N/2$), and the dynamics becomes confined to one of the two consensus states, with transitions between them effectively forbidden. Analogous situations have been analyzed theoretically~\cite{Parrondo01} and observed experimentally in stochastic thermodynamics using a colloidal particle trapped in a tunable double-well potential~\cite{Roldan2014Jun}, where the particle spontaneously localizes in one well as the wells are progressively separated. In such cases, a refined version of the second law in Eq.~\eqref{eq:second-law} can be obtained by conditioning the dynamics to the consensus state $i = \{A, B \}$ spontaneously chosen by the system during the transition: 
\begin{equation}\label{eq:symmetry-breaking}
 \braket{\Delta S_{\mathrm{sys}}}_i + \sum_{r} \mu_r \braket{I_r}_i \geq \log p_i,
\end{equation}
where $p_i$ is the probability that opinion $i$ is chosen (here $p_i = 1/2$ in the symmetric case) and the averages $\langle X \rangle_i = \sum_{\gamma_{[0,\tau]} \in C_i} \mathds{P}(\gamma_{[0,\tau]}) X (\gamma_{[0,\tau]})$ are performed over the set of trajectories $C_i$ ending in consensus $i$ at the final time $\tau$. In particular, $\langle \Delta S_{\rm sys} \rangle_i = S[P^{(i)}(\tau)] - S[P(0)] + \sum_n [P_n^{(i)}(\tau) - P_n(0)] S_n^{\rm int}$, with $P^{(i)}_n(\tau)$ the (normalized) distribution of the system over the corresponding half of the phase space.
\begin{figure}[t]
 \centering
 \includegraphics[width=\linewidth]{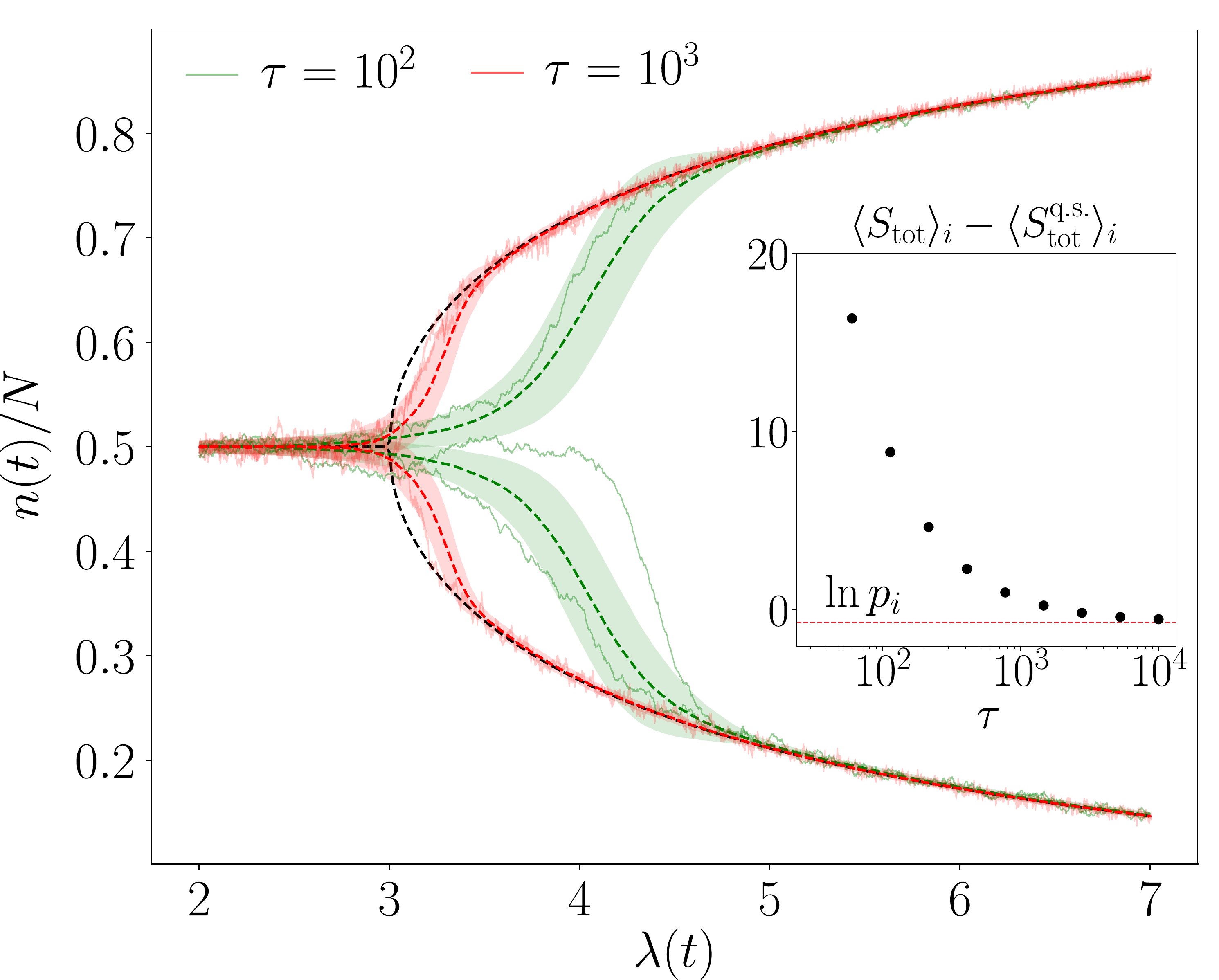}
 \caption{Trajectories of $n(t)$ during a linear ramp protocol, $\lambda(t) = \lambda_0 + (\lambda_\tau - \lambda_0)(t/\tau)$ with $\lambda_0 = 2$ and $\lambda_\tau = 7$. Individual trajectories obtained with Gillespie method (solid lines) are shown for two protocol durations: $\tau = 10^2$ (fast, green) and $\tau = 10^3$ (slow, red). Dashed lines show ensemble averages $\langle n \rangle$ conditioned on the final consensus state, with shaded regions indicating the variance. Black dashed curves represent the mean-field stationary solutions $n_\pm(\lambda)$. Fast protocols exhibit larger fluctuations and delayed symmetry breaking, whereas slow protocols remain closer to the quasi-static branches throughout. Inset: Difference between the average entropy production for finite-time protocols and the quasi-static limit, $\langle S_{\mathrm{tot}} \rangle - \langle S_{\mathrm{tot}}^{\mathrm{q.s.}} \rangle$, as a function of $\tau$ (black circles), showing convergence to the symmetry-breaking bound $\ln p_i = -\ln 2$ (horizontal line). Parameters: $N = 2 \times 10^4$, $q = 2$, $\theta = \chi = 1$.}
 \label{fig:breaking}
\end{figure}

Inequality~\eqref{eq:symmetry-breaking} indicates that knowledge of the consensus state chosen by the system during the dynamics can lead to an apparent reduction of total entropy $ \langle S_{\rm tot} \rangle_i = \braket{\Delta S_{\mathrm{sys}}}_i + \sum_{r} \mu_r \braket{I_r}_i$ (negative l.h.s.) since the r.h.s. is also negative ($p_i \leq 1$)~\footnote{Note however that the standard second law in Eq.~\eqref{eq:second-law} with averages over the entire phase space is always verified in any case.}. This ``daemonic" effect is in close analogy to Szilard's information engine~\cite{Parrondo01}, where a single-particle gas gets trapped in one of the two half of a piston chamber after introducing a partition in the middle of it~\cite{Lutz15,Parrondo15}. As a consequence, the above inequality allows spontaneous processes where the system entropy can be e.g. reduced $\braket{\Delta S_{\mathrm{sys}}}_i < 0$, without the need of a strict compensation from the opinion currents $\braket{I_r}_i$ in the second term.

In physical systems, saturation of inequalities akin to \eqref{eq:symmetry-breaking} occurs for slow isothermal processes in the quasistatic limit, when the system remains in an instantaneous equilibrium state throughout the evolution~\cite{Roldan2014Jun}. In opinion thermodynamics, however, the coexistence of distinct and competing mechanisms for opinion change (our two reactions) implies a nonzero irreversibility even under quasistatic driving, and therefore inequality~\eqref{eq:symmetry-breaking} is not generically saturated. In the quasistatic limit considered here, the system stays arbitrarily close to the instantaneous stationary state where entropy production is minimal, and the total entropy production reduces to the change in the stationary system entropy plus the contribution from the quasistatic probability currents accumulated along the protocol $\lambda(t)$. The l.h.s. of Eq.~\eqref{eq:symmetry-breaking} then becomes:
\begin{equation}
\langle S_{\rm tot}^{ \rm q.s.} \rangle_i = \braket{\Delta S_{\mathrm{sys}}}_i + \int_{0}^\tau \langle \dot{I}_1^{\rm q.s.} \rangle_i \, 2 \log (\lambda) \, \dot{\lambda} \, dt,
\end{equation}
where we used that the quasistatic currents satisfy $\langle \dot{I}_1^{\rm q.s.} \rangle_i = - \langle \dot{I}_2^{\rm q.s.} \rangle_i$ by global detailed balance in the instantaneous stationary state [c.f. Eq.~\eqref{eq:GDB}] and that $\mu_1 - \mu_2 = 2 \ln (\lambda)$.

In Fig.~\ref{fig:breaking} we show sample trajectories from numerical simulations of the number of agents $n(t)$ with opinion $A$ under a driving protocol $\Lambda$ consisting in a linear ramp, $\lambda(t)= \lambda_0 + (\lambda_\tau - \lambda_0)(t/\tau)$, with fixed values $\lambda_0 = 2$ and $\lambda_\tau = 7$, for two different values of $\tau$ (green an red colors), corresponding to different velocities. Corresponding averages over trajectories in the sets $C_i$ (i.e., ending in consensus $i = A, B$), are depicted as dashed lines, and their variances are represented by the shaded areas around them. The spontaneous symmetry breaking transition occurs when the critical point ($\lambda_{\rm c} = 3$) is crossed and the system gets trapped in one of the two consensus states. For slow velocities ($\tau = 10^3$) fluctuations are small and the trapping occurs quite close to the critical point (red curves). The average over each branch $\langle n \rangle_{i}$ is in this case close to the corresponding mean-field solution, $n_{\pm}(\lambda)/N = \frac{1}{2} \pm \frac{1}{2}\sqrt{\frac{\lambda-3}{\lambda+1}}$ during all the protocol. We remark that the mean field values are exactly reproduced only in the quasi-static limit. Faster velocities ($\tau = 10^2$) lead instead to a bigger delay in the trapping (green curves) with larger variance. In any case, as the values of $\lambda$ become large the ergodicity is clearly broken, even at finite $N$, as the fluctuations become small around the two bistable mean-field solutions, and the probability to escape from the corresponding consensus state becomes negligible. The inequality \eqref{eq:symmetry-breaking} is tested in the inset of Fig.~\ref{fig:breaking}. We show the convergence of $\langle S_{\rm tot}\rangle - \langle S^{\rm q.s.}_{\rm tot}\rangle$, to the symmetry breaking bound $\ln p_i = - \ln 2$ ($p_i = 1/2$) from Eq.~\eqref{eq:symmetry-breaking} in the quasi-static regime, here reached for values around $\tau=10^4$.

{

\section{Generalizations}
\label{sec:generalizations}

The framework developed here is not restricted to agents with binary state variables or all-to-all interactions as in the model introduced in Sec.~\ref{sec:Model} and used to illustrate the main tools of the framework. Its essential ingredients, microscopic reversibility and the generalized local detailed balance relation for the elementary transition reactions, are enough to extend the formalism to broader classes of imitation dynamics including multiple attributes, network structure or specific features such as aging or zealotry. 


\subsection{Multiple social attributes}

A first natural extension concerns imitation models with more than two social attributes or opinions. For $M>2$ possible opinions, the mesoscopic state is no longer described by a single occupation number, but by the occupation vector $
\mathbf n=(n_A,n_B,\dots)$ which satisfies $\sum_{i} n_i=N$,
where $n_i$ denotes the number of agents holding opinion $i=\{ A,B,\dots\}$. Due to this normalization constraint, only $M-1$ occupation numbers are independent. Equivalently, the mesoscopic dynamics takes place on an $(M-1)$-dimensional space embedded in the $M$-dimensional occupation space. The corresponding internal entropy is the logarithm of the multiplicity of the mesostate,
\begin{equation}
S_{\rm int}(\mathbf{n})
=
\ln\frac{N!}{\prod_{i} n_i!}.
\end{equation}

Let $A$ and $B$ denote two arbitrary distinct opinions among the $M$ possible ones, and let $\mathbf e_A$ denote the unit vector whose $A$th component is $1$ and all others are zero. Then an elementary change of opinion $A\to B$ transforms the occupation vector as
$\mathbf n\longrightarrow \mathbf n'=\mathbf n-\mathbf e_A+\mathbf e_B $.
In this work we restrict ourselves to reactions involving one-step opinion changes. We therefore assume that the exchange between any pair of distinct opinions may be mediated by several bidirectional reactions. The label $r$ denotes one such reaction. Each reaction connects a specific pair of opinions, although this pair is kept implicit in the notation to avoid overloading it. 

Each reaction $r$ is independent of the others and is characterized, in analogy with Eqs.~\eqref{eq:transition rates}, by two kinetic parameters $h_r$ and $a_r$, which fix the intrinsic bias of the forward and backward elementary updates. In the binary model these parameters correspond to the herding and anti-conformity rates. In addition, each reaction has a configuration-dependent function $g_r(\mathbf n)$, which encodes nonlinear group-interaction effects. We consider here the case in which the same interaction factor $g_r$ modulates both directions of any reversible elementary transition.

For a given reaction  $r$, connecting opinions A and B,  we denote by
$W^{(r)}_{\mathbf n-\mathbf e_A+\mathbf e_B,\mathbf n}$
the transition rate for the elementary jump $A\to B$, and by
$W^{(r)}_{\mathbf n,\mathbf n-\mathbf e_A+\mathbf e_B}$
the corresponding reverse rate $B\to A$. The one-step rates can then be written as 
\begin{subequations}\label{eq:transition-rates-multi-general}
\begin{align}
W_{\mathbf n-\mathbf e_A+\mathbf e_B,\mathbf n}^{(r)}
&= h_r\, n_A\, g_r(\mathbf n), \\
W_{\mathbf n,\mathbf n-\mathbf e_A+\mathbf e_B}^{(r)}
&= a_r\, (n_B+1)\, g_r(\mathbf n).
\end{align}
\end{subequations}
Here, the factor $n_A$ counts the number of agents available to perform the transition $A\to B$, whereas the factor $(n_B+1)$ counts the number of agents holding opinion $B$ in the post-jump configuration $\mathbf n'$. The use of the same factor $g_r(\mathbf n)$ in the two rates expresses the assumption that the nonlinear interaction term is paired under time reversal within each reaction.

With this choice, the generalized local detailed balance relation reads
\begin{equation}
\frac{
W^{(r)}_{\mathbf{n}-\mathbf e_A+\mathbf e_B,\mathbf n}
}{
W^{(r)}_{\mathbf n,\mathbf{n}-\mathbf e_A+\mathbf e_B}
}
=
e^{S_{\rm int}(\mathbf{n}-\mathbf e_A+\mathbf e_B)-S_{\rm int}(\mathbf n)
+\mu_r},
\label{eq:LDB_multi}
\end{equation}
which generalizes Eq.~\eqref{eq:local}. The first two terms on the right-hand side account for the combinatorial change in the multiplicity of the mesostate, while $\mu_r=\ln h_r/a_r$ measures the intrinsic bias associated with reaction $r$. The cancellation of $g_r$ in the ratio is a direct consequence of using the same nonlinear interaction factor for the forward and backward elementary updates.

A trajectory $\gamma_{[0,\tau]}$ is specified by the initial occupation vector $\mathbf n_0$ and by the ordered sequence of elementary jumps occurring during the time interval $[0,\tau]$, each one labels the reaction that produces it. 
For such a reaction-resolved trajectory, $I_r(\gamma_{[0,\tau]})$ is the net number of forward minus backward updates induced by reaction $r$ along the trajectory, with the forward orientation fixed for each reaction. The detailed fluctuation theorem then takes the same form as in Eq.~\eqref{eq:DFT},
but since each reaction connects a specific pair of opinions, the sum over $r$ implicitly includes all pairs of opinions and all reactions mediating exchanges between them. Analogously the system entropy change in that case would read 
$\Delta S_{\rm sys} = \ln P_{\mathbf n_0}(0)-\ln P_{\mathbf n_\tau}(\tau) + S_{\rm int}(\mathbf n_\tau)-S_{\rm int}(\mathbf n_0)$. As a consequence, all the trajectory- and ensemble-level relations discussed in Sec.~\ref{sec:Opinion Thermodynamics} follow in the same way as in the binary case. A qualitative novelty of multi-opinion models is that they may sustain genuine cyclic currents in opinion space, such as $A\to B\to C\to A$, thereby allowing for richer nonequilibrium steady states than in the two-opinion setting.

\subsection{Complex network topologies}

The all-to-all setting is special because the dynamics closes exactly at the mesoscopic level. In that case, all microscopic configurations with the same occupation vector $\mathbf n$ are statistically equivalent: agents carrying the same opinion have identical interaction environments, and the total transition rate out of a mesostate depends only on $\mathbf n$. For more general interaction topologies this exact closure is typically lost, since the position of each opinion on the graph becomes relevant. Nevertheless, the formulation of stochastic thermodynamics remains valid at the microscopic level.

Consider an imitation model with $M$ opinions on a (fixed) interaction graph $G=(V,E)$, defined by a set $V$ of nodes (vertices) and a set $E$ of links (edges) that connect pairs of vertices. The possible node states are $s_\ell\in\{A,B,\dots\}$ and we indicate a microscopic configuration as $\alpha=(s_1,\dots,s_N)$.
Let $\alpha'$ denote the configuration obtained from $\alpha$ after changing, for example, the state of node $\ell$ from opinion $A$ to opinion $B$ by reaction $r$. At the microscopic level, a trajectory is specified by the initial configuration $\alpha_0$ and by the ordered sequence of node updates, each one labeled by the reaction, and the jump time.

For a reversible one-step reaction $r$ mediating the transition 
$A\leftrightarrow B$, we write the microscopic rates as
\begin{subequations}\label{eq:transition-rates-network-general}
\begin{align}
W^{(r)}_{\alpha',\alpha}
&=
h_r\, g_r^{(\ell)}(\alpha),\\
W^{(r)}_{\alpha,\alpha'}
&=
a_r\, g_r^{(\ell)}(\alpha').
\end{align}
\end{subequations}
The network structure is encoded in a node-dependent activity factor $g_r^{(\ell)}(\alpha)$, which summarizes the relevant structural and configurational information entering the update of node $\ell$ through reaction $r$. This factor may depend, for instance, on the number or fraction of agents of each opinion in a given interaction set, the degree of the node, weighted-network information, community-level properties, or motif-level quantities. We assume that $g_r^{(\ell)}(\alpha)$ does not depend on the state of the updating node itself, $s_\ell$. Hence, if $\alpha'$ differs from $\alpha$ only on the value of $s_\ell$, then
$g_r^{(\ell)}(\alpha')=g_r^{(\ell)}(\alpha)$. This assumption ensures that the network structure modulates the activity 
of a reversible reaction, while its bias is entirely encoded in the ratio $h_r/a_r$. Threshold rules, $q$-voter-like nonlinearities, majority-type mechanisms, heterogeneous degree effects, weighted-network rules, or community-based interaction mechanisms can all be incorporated through different choices 
of the activity factor $g_r^{(\ell)}(\alpha)$.

With this choice, each microscopic elementary update satisfies the local detailed balance relation
\begin{equation}
\frac{
W^{(r)}_{\alpha',\alpha}
}{
W^{(r)}_{\alpha,\alpha'}
}
=
e^{\mu_r},
\qquad
{\rm with} \,\,\,\, \mu_r=\ln\frac{h_r}{a_r}.
\label{eq:LDB_micro_network}
\end{equation}
Importantly, in this case no internal-entropy contribution appears in Eq.~\eqref{eq:LDB_micro_network}, since no coarse-graining or grouping of states has been performed. The entropy production along reaction-resolved trajectories takes then the simpler form
\begin{align}
S_{\rm tot}(\gamma_{[0,\tau]})
&= \ln \left( \frac{\mathds{P}(\gamma_{[0,\tau]})}{\tilde{\mathds{P}}(\tilde{\gamma}_{[0,\tau]})} \right)  
\\ &= \ln P_{\alpha_0}(0)-\ln P_{\alpha_\tau}(\tau) + 
\sum_r \mu_r\, I_r(\gamma_{[0,\tau]}), \nonumber
\label{eq:DFT_network}
\end{align}
which can be easily check to verify the IFT in Eq.~\eqref{eq:IFT} and the associated second-law-like inequalities. Similarly the other main tools of the framework can be derived at this description level.

The main effect of a complex network is therefore structural rather than conceptual. In general, many microscopic configurations share the same occupation vector $\mathbf n$ while having different local environments and different escape rates. For example, two configurations with the same number of agents holding each opinion may evolve differently if opinion $A$ is mostly located on highly connected nodes in one case and on peripheral nodes in the other. As a result, the occupation vector $\mathbf n$ is generally not a Markovian state variable by itself, and no exact closed master equation for $\mathbf n$ alone can be derived. Exact mesoscopic closure is recovered in the all-to-all case, and more generally in highly symmetric or lumpable interaction structures~\cite{Herpich2020Jun}. In such cases, all microscopic configurations belonging to the same mesostate have identical coarse-grained transition rates. Outside this special situation, approximate mesoscopic descriptions remain possible by retaining additional topological information. Examples include degree-class occupation variables~\cite{Gleeson2013,Peralta2020}, pair approximations~\cite{Peralta_2018,Sood2005}, motif-based densities~\cite{Cui2022,DEMIREL201468}, or community-level occupation variables~\cite{e25060838,Oestereich2019}. When the mesostate includes such topological information, the corresponding internal entropy reappears naturally as the logarithm of the number of microscopic configurations compatible with the chosen coarse-grained variables, including the constraints imposed by the graph topology.

\subsection{Aging, zealotry and variable population}

Finally, phenomena such as aging, zealotry or those related with a variable population of agents, can also be incorporated within the present framework. Aging generated by explicitly time-dependent rates is already covered by the driven setting discussed above. Aging mechanisms that depends on the time elapsed since the last change of state~\cite{Stark2008,Artime2018} could be incorporated by enlarging the state space to include such internal variables. Zealots~\cite{Mobilia2003,Khalil2018}, nodes that never change opinions, can be incorporated in the node-dependent activity factor $g_r^{(\ell)}(\alpha)$ . A variable population can be implemented in the all-to-all interaction setting by considering the number of agents $N$ as a stochastic variable. Birth and death processes, such as for example $\varnothing \longleftrightarrow A$, can then be incorporated as additional reversible reactions with their own affinities and currents. This allows to tackle phenomena such as migration 

Overall, the binary well-mixed model studied here should be regarded as the minimal analytically tractable representative of a much broader class of socially interacting stochastic systems. What changes across these generalizations is mainly the dimensionality of the state space, the structure of the admissible currents, and the difficulty of constructing an exact coarse-grained description. What remains unchanged is the central message of this work: whenever the elementary transition reactions admit a generalized local detailed balance structure, stochastic thermodynamics provides a well-defined notion of entropy production linking social attributes with information, with cornerstone relations such as fluctuation theorems, thermodynamic and kinetic uncertainty relations, and second-law-like bounds. 

\section{Conclusions and Discussion}
\label{sec:Conclusions}

We have developed a stochastic-thermodynamic framework for social imitation dynamics that does not rely on any explicit notion of energy or temperature. Instead, it links irreversibility to changes in social attributes, typically interpreted as opinions, and to information-theoretical quantities characterizing the prevalence of that opinions at both the trajectory and ensemble levels. In particular, we derived a trajectory entropy production (irreversibility) [Eq.~\eqref{eq:DFT}] satisfying universal fluctuation relations such as the IFT [Eq.~\eqref{eq:IFT}] and second-law-like inequalities [Eqs.~\eqref{eq:second-law} and \eqref{eq:symmetry-breaking}], which admits a Kullback-Leibler representation [Eq.~\eqref{eq:EP}]. This same quantity controls trade-off relations such as the TUR [Eq.~\eqref{eq:TUR}]. Beyond entropy production, the formalism also allows the derivation of strong fluctuation theorems for the attribute currents themselves [Eq.~\eqref{eq:Inference}] with inference applications, and inequalities based on time-symmetric dynamical quantities characterizing the speed of opinion changes, like the KUR~[Eq.~\eqref{eq:KUR}]. In this way, the formalism reveals a unified set of constraints on the magnitude, fluctuations, and precision of the social attributes currents.

To illustrate the framework, we analyzed a minimal imitation model which captures changes in social attributes by imitation and anti-conformity mechanisms, and allows the formulation of the generalized detailed balance condition, Eq.~\eqref{eq:local}, for the reaction rates in Eq.~\eqref{eq:transition rates}.  The model presents a high degree of flexibility in the (possibly) nonlinear influence mechanisms including those in $q$-voter models, threshold models, group and $\epsilon$-voter models or majority-rule models. These models typically present a rich phase diagram in the macroscopic limit with consensus and polarized states, and both first-order and second-order phase transitions between them, as illustrated for the case of $q-$voter influence mechanisms.

Testing the framework in our toy social model, we identified the conditions leading either to equilibrium or to nonequilibrium stationary states. While in equilibrium the stationary distribution shows an universal (binomial) shape only depending on the intrinsic symmetry of the opinions; out of equilibrium the herding and anticonformity mechanisms remain active, sustaining large reaction-resolved opinion currents. There the difference of generalized chemical potentials, $\mu_1-\mu_2$, acts as the relevant social force controlling the dominant bias towards herding or anti-conformity [Eq.~\eqref{eq:Stotst}]. The thermodynamic analysis captures clear signatures of the second-order phase transition, including a local maximum of entropy production at criticality and the emergence of frozen consensus states where both opinion currents and dynamical activity vanish. The TUR and KUR inequalities [Eqs.~\eqref{eq:TUR} and ~\eqref{eq:KUR}] further revealed the presence of fundamental trade-offs in the magnitude and noise of the opinion currents, which we found to be respectively saturated in regimes close and far from equilibrium. 

The strong fluctuation theorem for the opinion currents [Eq.~\eqref{eq:Inference}] showed that the effective force $\mu_1-\mu_2$ puts strict constrains on the fluctuations of opinion changes, and provides a route to infer parameters of the model (such as the relative strength of conformity and anti-conformity attitudes) from the observation stochastic events. This opens the possibility of confronting the theory with empirical data whenever the changes in the opinions can be resolved at the level of the underlying reactions, for instance by distinguishing changes driven by a herding event from those driven by anti-conformity one. Related inference strategies could be explored using the Kullback-Leibler form of the entropy production~\cite{Roldan10,Martinez19}, uncertainty relations~\cite{Seifert19}, or following Refs.~\cite{VdMeer22,Harunari22}. We further examined nonequilibrium driven protocols crossing the critical point where a refined second-law inequality holds [Eq.~\eqref{eq:symmetry-breaking}] reminiscent of information-thermodynamic bounds in feedback-controlled systems~\cite{Lutz15,Parrondo15}. That result suggests the possibility of designing cycles in the social system phase space with an analogous effect to Maxwell's demon. Future works may also investigate hysteretic cycles that cross the first-order phase transition line $\chi_{\rm b}$ in Fig.~\ref{fig:phase diagram}.

The scope of the framework presented here is not limited to the binary (opinions $A$ and $B$) well-mixed setting with two reactions. We showed that the stochastic thermodynamic structure extends naturally to more general agent-based models with an arbitrary number of opinions or attributes, complex interaction graphs, and featuring phenomena such as aging or zealotry. Further directions on which our framework could be generalized include the consideration of evolving networks with a variable number of agents, or heterogeneous agents~\cite{Mendes2025}. In that case, the microscopic state becomes $X=(\alpha,G,N)$, including the number of agents $N$ and the graph structure $G$ as dynamical variables. The same formal logic might be applied: once a generalized local detailed balance relation is specified for the elementary updates, the corresponding fluctuation theorems and nonequilibrium inequalities would follow. Strictly irreversible (unidirectional) transitions that lie outside the standard framework based on microscopic reversibility, would require more advance and specific techniques~\cite{manzanoThermodynamicsComputationsAbsolute2024,Rahav14,Rahav21,Busiello20}.


The present work is meant as an exploratory step showing that stochastic thermodynamics can be meaningfully exported to sociophysics even in the absence of an energetic interpretation. While we focused on a few cornerstone nonequilibrium relations, the same perspective should make it possible to investigate other universal results, including speed limits~\cite{Shiraishi18,Falasco20,LP22,Vu23a,Vu23b} and martingale fluctuation relations for entropy-production extrema and stopping times~\cite{Neri17,Chetrite18,gambling21,survival22,roldanMartingalesPhysicistsTreatise2023}. More broadly, the framework should be applicable to a wide family of agent-based models with microscopically reversible update reactions, including social contagion models~\cite{Bailey}, agent-based models in economics~\cite{Kirman93,Borba2025}, and extensions to complex networks and coarse-grained descriptions beyond the well-mixed case~\cite{Castellano09,Vespignani01,Suchecki05,Lambiotte07,Gleeson11,Carro16}.

}

\begin{acknowledgments}
Partial financial support has been received from Grants PID2021-122256NB-C21/C22, PID2022-140506NB-C21/C22 and PID2024-157493NB-C21/C22 funded by MICIU/AEI/10.13039/501100011033 and by “ERDF/EU”, and the María de Maeztu Program for units of Excellence in R\&D, grant CEX2021-001164-M. L.I. is supported from the CSIC ``Empleo Juvenil del FSE+" program (No. EMPJOV25-IFISC-M2-B-162). L.T. is supported from the CSIC JAE-PRE program (No. JAEPRE23-26). G.M. acknowledges financial support from the ``Ramón y Cajal'' program (No. RYC2021-031121-I) funded by MCIU/AEI/10.13039/501100011033 and European Union NextGenerationEU/PRTR. 
\end{acknowledgments}

\appendix

\section{Equilibrium State}\label{sec:Equilibrium State apdx}

In this appendix we provide a proof for the equilibrium condition $h_1 h_2 = a_1 a_2$ (or equivalently $\lambda = 1$) presented in Sec.~\ref{sec:Stationary State}. We first rewrite the local detailed balance condition in \Cref{eq:LDB} as
\begin{equation}\label{eq:Eq cond 1}
 \frac{W_{n+1, n}^{(r)}}{W_{n, n + 1}^{(r)}} = \frac{P_{n+1}^{\mathrm{eq}}}{P_{n}^{\mathrm{eq}}},
\end{equation}
which is valid for both reactions $\forall r \in \{ 1, 2 \} $ and $n \in \{ 0, 1,\dots, N \} $. Since the r.h.s. of the equation above is independent of $r$, we can equate the l.h.s. rate ratio of the two reactions to find:
\begin{equation}\label{eq:Eq cond 2}
 \frac{W_{n+1, n}^{(1)}}{W_{n, n + 1}^{(1)}} = \frac{W_{n+1, n}^{(2)}}{W_{n, n + 1}^{(2)}}, \quad \forall n \in \{ 0, 1,\dots, N \}.
\end{equation}
Finally, substituting the expression of the rates in \Cref{eq:transition rates} for generic $g(n) $ yields
\begin{equation}\label{eq:Eq cond}
 h_1 h_2 = a_1 a_2,
\end{equation}
or in terms of the reduced dimensionless parameters in Eq.~\eqref{eq:parameters}, $\lambda = 1$.

\section{Stationary State}\label{sec:Stationary State apdx}

In this appendix we analyze the stationary state in the limit $N \gg 1$, where critical phenomena are well defined. More precisely, using a Fokker-Planck approach we fully characterize the phase diagram and state equation of our system.

For $N \gg 1$ it is possible to show that $P_n(t)$ satisfies the Fokker-Planck equation~\cite{vanKampen:2007,Toral2014StochasticNM}
\begin{equation} \label{eq:FP_1D}
\partial_{t} P_n(t) = - \partial_{n}[ F(n) P_n(t)] + \frac{1}{2} \partial_{n}^2 [ D(n) P_n(t)], 
\end{equation}
where $F(n), D(n) $ are the \textit{drift} and \textit{diffusion} functions, respectively, defined as:
\begin{align}
F({n})\equiv &\sum_\ell \ell W(n\to n+\ell)=W_{n+1,{n}}-W_{n-1,{n}},\\
D({n})\equiv &\sum_\ell \ell^2 W(n\to n+\ell)=W_{n+1,{n}}+W_{n-1,{n}}. \nonumber
\end{align}
Introducing the rescaled variable $x \equiv n/N \in [0, 1]$, using the parametrization in \Cref{eq:parameters} and $\omega \equiv \sqrt[4]{a_1 a_2 h_1 h_2}$, the rates \eqref{eq:transition rates} are given by
\begin{equation}\label{eq:transition rates in TD lim}
\begin{split}
 w_+^{(1)}(x) &= \omega \sqrt{\lambda \chi \theta}\, (1 - x)\, g(x), \\
 w_{-}^{(1)}(x) &= \omega \sqrt{\theta / (\lambda \chi)} \, x \, g(x), \\
 w_{+}^{(2)}(x) &= \omega \sqrt{\chi / (\lambda \theta)} \, (1 - x) \, g(1-x), \\
 w_{-}^{(2)}(x) &= \omega \sqrt{\lambda / (\chi \theta)} \, x \, g(1-x), 
\end{split}
\end{equation}
where $w_\pm^{(r)}(x) = W_{n\pm 1, n}^{(r)} / N$. The drift and diffusion functions then read $F(n)=N f(n / N)$ and $D(n)=Nd(n / N)$ with:
\begin{align}
\frac{\sqrt{\lambda\chi\theta}}{\omega} f(x)=&\lambda\chi\theta(1-x)g(x)+\chi(1-x)g(1-x)\\
&-\theta x g(x) -\lambda x g(1 - x),\nonumber\\
\frac{\sqrt{\lambda\chi\theta}}{\omega} d(x)=&\lambda\chi\theta(1-x)g(x)+\chi(1-x)^{q} g(1-x)\\
&+\theta xg(x)+\lambda x g(1-x).\nonumber
\end{align}
The stationary distribution $P_n^{\mathrm{st}}$, obtained by setting $\partial_t P_n = 0$ in Eq.~\eqref{eq:FP_1D}, takes the large-deviation form
\begin{equation} \label{eq:Pst FP}
P_n^{\mathrm{st}}=\mathcal{Z}^{-1}\cdot\exp{\left[-N v\left(\frac{n}{N}\right)\right]},
\end{equation}
where $\mathcal{Z}$ is a normalization constant and $v({x})$ is the potential \footnote{Note that $\omega $ only fixes the relaxation time scale and does not affect the stationary distribution.}:
\begin{equation} \label{eq:potential}
 v({x})=-2\int^{x}\frac{f(z)}{d(z)}dz.
\end{equation}
Extrema of $v(x)$ are found by solving $f(x) = 0$. The most probable stationary states correspond to absolute minima of $v(x)$, or equivalently, to absolute maxima of $P_n^{\mathrm{st}}$. Local minima of $v(x)$ correspond to metastable states.

In the limit $N \gg 1$, we have $g(x) = x^q$ with $q \in \mathbb{R}_+$ for both sampling with and without repetition. Therefore, our analysis holds for both modeling schemes.

\begin{figure}[t]
\includegraphics[width=1\columnwidth]{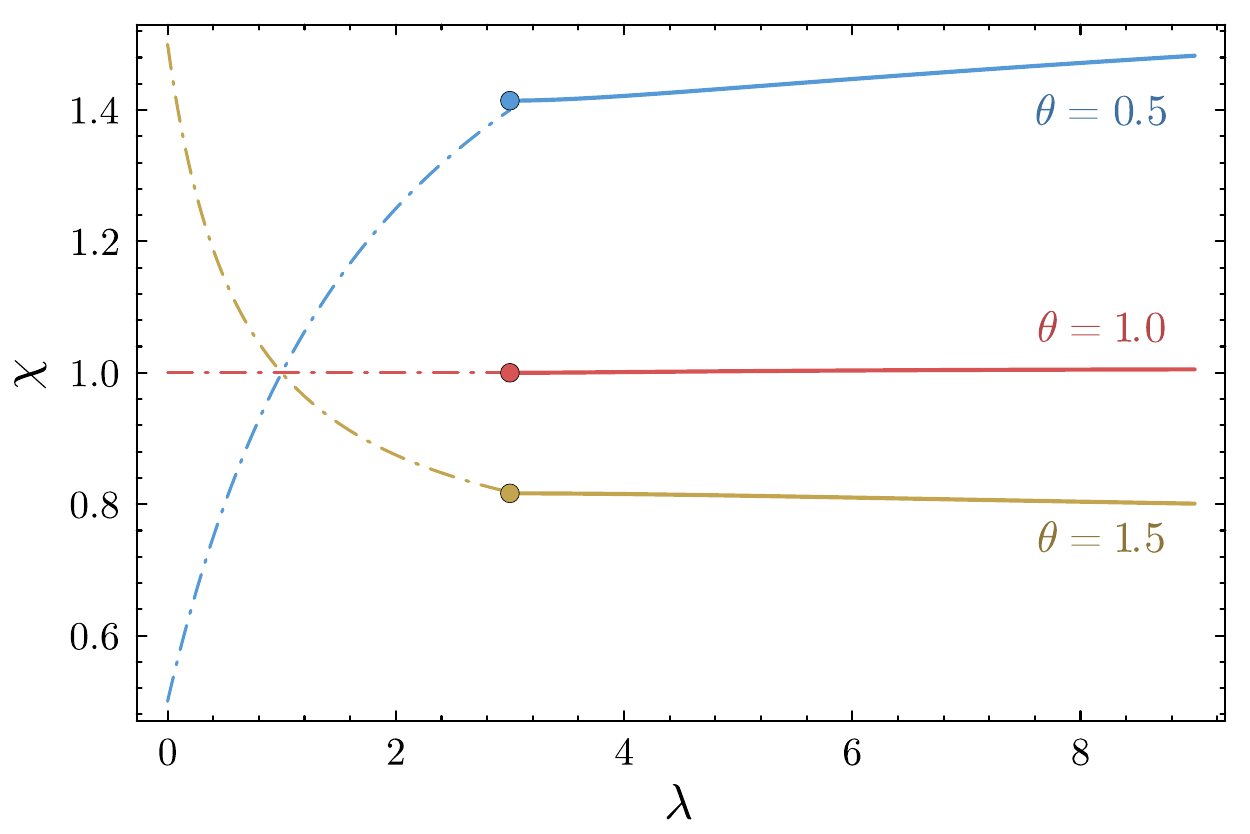}
\caption{\label{fig:phase diagram comparison theta} 
Phase diagrams in the $(\lambda, \chi)$ parameter space for $g(x) = x^q$ with $q=2$, for different values of $\theta \in \{0.5, 1, 1.5\}$. The critical curves $\chi_\pm(\lambda)$ [Eq.~\eqref{eq:chi plus minus}] are omitted for clarity. Critical points are $(3, \sqrt{2})$, $(3, 1)$, $(3, \sqrt{2/3})$ for $\theta = 0.5, 1, 1.5$, respectively, as predicted by Eq.~\eqref{eq:critical point asymmetric}. The unimodal transition line $\chi_{\rm u}(\lambda) = (\lambda + \theta)/(1 + \lambda\theta)$ (dash-dotted) separates opinion-dominant regions, while the bimodal transition line $\chi_{\rm b}(\lambda)$ (solid) marks the first-order transition, computed numerically via $v(x_-) = v(x_+)$.}
\end{figure}

The general behavior is shown in Fig.~\ref{fig:phase diagram comparison theta}. There exists a critical point $(\lambda_{\mathrm{c}}, \chi_{\mathrm{c}})$ with $\lambda_{\mathrm{c}} = (q+1)/(q-1)$ and $\chi_{\mathrm{c}} = \theta^{-1/q}$, from which two critical curves $\chi_{\pm}(\lambda, \theta)$ emerge, delimiting a metastable region. Inside this region, there are three real stationary solutions $\{x_-, x_0, x_+\}$ satisfying $x_- < x_0 < x_+$, where $x_\pm$ are stable states [minima of $v(x)$] and $x_0$ is an unstable state [local maximum of $v(x)$]. Outside the metastable region, only one stable solution $x_{\mathrm{u}}$ exists~\footnote{Since $f(0) = \chi \geq 0$ and $f(1) = -\theta \leq 0$, there is at least one solution $x_{\mathrm{st}} \in [0, 1]$ provided $\chi \neq 0$ and $\theta \neq 0$.}. Figure~\ref{fig:phase diagram comparison theta} shows that $\theta$ shifts the transition lines while preserving the qualitative behavior.

The analytical expressions for the stationary states $x_{\mathrm{st}}$ and the critical curves $\chi_{\pm}(\lambda, \theta)$ can be obtained in closed form only for particular values of $q$. For the general asymmetric case, the stationary states are intrinsically complex and provide little insight, making it more convenient to solve $f(x_{\mathrm{st}}) = 0$ numerically. For the symmetric case, useful relations emerge. Specifically, $x_{\mathrm{u}} = 1/2$ for any $q$. For $q = 2$: 
\begin{equation}\label{eq:xst q2}
  x_{\pm }(\lambda, \chi = 1, \theta = 1) = 
 \frac{1}{2} 
 \left( 1 \pm \sqrt{\frac{\lambda - 3}{1 + \lambda}} \right). 
\end{equation}
The curves delimiting the metastable region can be obtained analytically for $q \in \mathbb{N}$ by setting the discriminant of $f(x) = 0$ to zero. For $q = 2$:
\begin{equation}\label{eq:chi plus minus}
 \chi_\pm(\lambda , \theta )=\sqrt{\frac{\lambda^4+18\lambda^2-27\pm\sqrt{(\lambda^2-1)(\lambda^2-9)^3}}{8\theta\lambda^3}}.
\end{equation}
Both curves start from the critical point $\lambda_{\mathrm{c}}(q = 2) = 3$, generalizing to arbitrary $q$. Numerically, these curves can be found by imposing $f(x_{\mathrm{st}}) = 0$ and $f'(x_\mp) = 0$ for $\chi_{\pm}(\lambda, \theta)$, respectively.

Regarding opinion predominance, in the unimodal region the distribution undergoes a continuous transition from predominant consensus at $B$ ($x < 1/2$) to predominant consensus at $A$ ($x > 1/2$) when crossing $\chi_{\rm u}(\lambda) \equiv (\theta + \lambda)/(1 + \theta\lambda)$, valid for all $q$. In the metastable region, the distribution undergoes a discontinuous transition from a bimodal distribution biased towards $B$ (peak at $x < 1/2$ higher than at $x > 1/2$) to the opposite when crossing $\chi_{\rm b}(\lambda)$. This line is computed numerically via $v(x_-) = v(x_+)$. Note that $\chi_{\rm u}(\lambda_{\mathrm{c}}) \neq \chi_{\mathrm{c}}$ in general, whereas $\chi_{\rm b}(\lambda_{\mathrm{c}}) = \chi_{\mathrm{c}}$.

In summary, the system exhibits both first- and second-order phase transitions. Crossing through the critical point produces a second-order transition, while crossing $\chi_{\rm b}(\lambda)$ within the metastable region produces a first-order transition between consensus states.

\section{Corollaries of the Fluctuation Theorems}\label{eq:Corollaries of the Fluctuation Theorems}

In this appendix we proof the detailed fluctuation theorem in Eq.~\eqref{eq:DFT} and some of its corollaries presented in \Cref{sec:Opinion Thermodynamics}. 

\subsection{Detailed Fluctuation Theorem in  Eq.~\eqref{eq:DFT}}

We start from the l.h.s. of \Cref{eq:DFT}, $S_{\rm tot}= \mathds{P}(\gamma_{[0, \tau]})/\tilde{\mathds{P}}(\tilde{\gamma}_{[0, \tau]})$, and substitute the path probabilities of trajectories in the forward process [\eqref{eq:prob}] and in the backward one:
\begin{align} \label{eq:probback}
\tilde{\mathds{P}}(\tilde{\gamma}_{[0, \tau]})  & = P_{n_\tau}(\tau)~ \mathcal{D}(\tau,t_J) ~ W_{n_{J-1}, n_{J}}^{(r_J)}~... \\ &...~W_{n_0, n_1}^{(r_1)} \mathcal{D}(t_1,0)~ dt_1 ... dt_J, \nonumber
\end{align}
to obtain:
\begin{equation}
    \frac{\mathds{P}(\gamma_{[0, \tau]})}{\tilde{\mathds{P}}(\tilde{\gamma}_{[0, \tau]})} = 
    \frac{P_{n_0}(0)}{P_{n_\tau}(\tau)} \prod_{j=1}^{J} \frac{W_{n_{j}, n_{j-1}}^{(r_j)}(t_j)}{W_{n_{j-1}, n_{j}}^{(r_j)}(t_j)}.
\end{equation}
Applying Eq.~\eqref{eq:local} and taking logarithms of both sides, we obtain: 
\begin{align}\label{eq:tot1}
    S_{\mathrm{tot}}[\gamma_{[0, \tau]}] &= \ln P_{n_0}(0) - \ln P_{n_\tau}(\tau)  \\ 
    & \qquad + \sum_{j=1}^{J} \left[ S_{n_j}^{\mathrm{int}} - S_{n_{j-1}}^{\mathrm{int}} + (n_j - n_{j-1}) \mu_{r_j} \right]. \nonumber
\end{align}
This is a telescoping sum: the intermediate terms $S_{n_j}^{\mathrm{int}}$ cancel pairwise, leaving only the boundary contributions $S_{n_\tau}^{\mathrm{int}} - S_{n_0}^{\mathrm{int}}$, which combine with the logarithmic terms to give the system entropy change, $\Delta S_{\mathrm{sys}} \equiv \ln P_{n_0}(0) - \ln P_{n_\tau}(\tau) + S_{n_\tau}^{\mathrm{int}} - S_{n_0}^{\mathrm{int}}$. Meanwhile, the sum $\sum_{j=0}^{J} (n_j - n_{j-1}) \mu_{r_j}$ telescopes to the net opinion changes along the path, which can be decomposed as $\mu_1 I_1 + \mu_2 I_2$, where $I_r$ denotes the net number of transitions due to reaction $r$. Thus, from Eq.~\eqref{eq:tot1} we recover Eq.~\eqref{eq:DFT}, $S_{\mathrm{tot}}[\gamma_{[0, \tau]}] = \Delta S_{\mathrm{sys}} + \mu_1 I_1 + \mu_2 I_2$.

\subsection{Bound on the Negative Entropy Production Tail}
In this subsection we prove the tail bound on negative entropy production, i.e.
\begin{equation} \label{eq:tail}
\mathrm{Prob}(S_{\mathrm{tot}} \leq -\alpha) \leq \exp(-\alpha),    
\end{equation}
for $\alpha > 0$. This result follows directly from combining the integral fluctuation theorem in Eq.~\eqref{eq:IFT} with Markov's inequality. 

We first recall Markov's inequality: for a non-negative random variable $X$ with probability density $f(x)$ and any $a > 0$, 
\begin{align}
    \langle X \rangle &= \int_{0}^{\infty} x f(x) dx = \int_{0}^{a} x f(x) dx + \int_{a}^{\infty} x f(x) dx \nonumber \\
    & \geq \int_{a}^{\infty} x f(x) dx \geq a \int_{a}^{\infty} f(x) dx = a \, \mathrm{Prob}(X \geq a).
\end{align}
where we denoted $x$ the values of the variable $X$. Rearranging terms, it yields:
\begin{equation}
\mathrm{Prob}(X \geq a) \leq \langle X \rangle / a.
\end{equation}

Now, applying the above inequality to $X = e^{-S_{\mathrm{tot}}}$ (which is non-negative) and $a = e^\alpha$ gives
\begin{equation}
    \mathrm{Prob}(e^{-S_{\mathrm{tot}}} \geq e^\alpha) \leq \frac{\langle e^{-S_{\mathrm{tot}}} \rangle}{e^\alpha}.
\end{equation}
The event $\{e^{-S_{\mathrm{tot}}} \geq e^\alpha\}$ is equivalent to $\{S_{\mathrm{tot}} \leq -\alpha\}$, and by the integral fluctuation theorem in Eq.~\eqref{eq:IFT}, we have $\langle e^{-S_{\mathrm{tot}}} \rangle = 1$ in the numerator of the right hand side. Therefore, from the above inequality we obtain $\mathrm{Prob}(S_{\mathrm{tot}} \leq -\alpha) \leq e^{-\alpha}$ in Eq.~\eqref{eq:tail}, as claimed.

\subsection{Second-law inequality \eqref{eq:second-law}}

The proof follows from combining the integral fluctuation theorem in Eq.~\eqref{eq:IFT} with Jensen's inequality. We first recall Jensen's inequality: for a convex function $\varphi$ and a random variable $X$,
\begin{equation}
    \varphi(\langle X \rangle) \leq \langle \varphi(X) \rangle.
\end{equation}
Applying the above inequality to $X = -S_{\mathrm{tot}}$ and the convex function $\varphi(x) = e^x$, it yields
\begin{equation}
    e^{-\langle S_{\mathrm{tot}} \rangle} \leq \langle e^{-S_{\mathrm{tot}}} \rangle.
\end{equation}
Finally, from the integral fluctuation theorem in Eq.~\eqref{eq:IFT}, the right-hand side of the above inequality becomes the unity: $\langle e^{-S_{\mathrm{tot}}} \rangle = 1$. Therefore,
\begin{equation}
    e^{-\langle S_{\mathrm{tot}} \rangle} \leq 1,
\end{equation}
which implies $\langle S_{\mathrm{tot}} \rangle \geq 0$ by taking logarithms in both sides. Recalling the decomposition $S_{\mathrm{tot}} = \Delta S_{\rm sys} + \sum_r \mu_r I_r$ from \eqref{eq:DFT}, we then obtain
\begin{equation}
    \langle S_{\mathrm{tot}} \rangle = \langle \Delta S_{\rm sys} \rangle + \sum_r \mu_r \langle I_r \rangle \geq 0,
\end{equation}
as stated in Eq.~\eqref{eq:second-law}.
\subsection{Positivity of the entropy production rate}
\label{App:PS_tot}
Starting from the ensemble expression for the instantaneous entropy production rate, Eq.~(\ref{eq:EPrate}),
\begin{equation}
\begin{aligned}
\big\langle \dot S_{\mathrm{tot}}(t) \big\rangle
&=
\sum_{r}\sum_{n,m}
W^{(r)}_{n m}(t)\,P_m(t)
\ln\!\left[
\frac{W^{(r)}_{n m}(t)\,P_m(t)}
     {W^{(r)}_{m n}(t)\,P_n(t)}
\right],
\end{aligned}
\label{eq:Sdot_explicit}
\end{equation}
we now show that it is non–negative at all times.

For a fixed reaction $r$, it is convenient to introduce the shorthand
\begin{equation}
a_{n m} \equiv W^{(r)}_{n m}(t)\,P_m(t),
\qquad
b_{n m} \equiv W^{(r)}_{m n}(t)\,P_n(t),
\label{eq:anm_bnm_def}
\end{equation}
so that the contribution of reaction $r$ to the entropy production rate can be written as
\begin{equation}
\big\langle \dot S_{\mathrm{tot}}^{(r)}(t) \big\rangle
=
\sum_{n,m} a_{n m}
\ln\frac{a_{n m}}{b_{n m}}.
\label{eq:Sdot_r_ab}
\end{equation}
By simple re-indexing of the double sum we observe that
\begin{equation}
\begin{aligned}
\sum_{n,m} a_{n m}
&=
\sum_{n,m} W^{(r)}_{n m}(t)\,P_m(t)
\\
&=
\sum_{n,m} W^{(r)}_{m n}(t)\,P_n(t)
=
\sum_{n,m} b_{n m},
\end{aligned}
\label{eq:sum_a_equals_sum_b}
\end{equation}
that is, the total outflow and inflow associated with reaction $r$ coincide when summed over all pairs $(n,m)$.

We now invoke the log--sum inequality (a particular case of Jensen's inequality): for any families of non–negative numbers $\{a_k\}$ and $\{b_k\}$ such that $\sum_k a_k = \sum_k b_k$, one has
\begin{equation}
\begin{aligned}
\sum_k a_k \ln\frac{a_k}{b_k}
&\;\ge\;
\Bigl(\sum_k a_k\Bigr)
\ln\!\frac{\sum_k a_k}{\sum_k b_k}
= 0.
\end{aligned}
\label{eq:log_sum}
\end{equation}
Applying Eq.~\eqref{eq:log_sum} to the sets $\{a_{n m}\}$ and $\{b_{n m}\}$ defined in Eq.~\eqref{eq:anm_bnm_def}, and using Eq.~\eqref{eq:sum_a_equals_sum_b}, we obtain
\begin{equation}
\big\langle \dot S_{\mathrm{tot}}^{(r)}(t) \big\rangle
=
\sum_{n,m} a_{n m}\ln\frac{a_{n m}}{b_{n m}}
\;\ge\; 0
\qquad
\text{for all } r.
\label{eq:Sdot_r_positive}
\end{equation}
Summing over all reactions finally yields
\begin{equation}
\begin{aligned}
\big\langle \dot S_{\mathrm{tot}}(t) \big\rangle
&=
\sum_{r} \big\langle \dot S_{\mathrm{tot}}^{(r)}(t) \big\rangle
\\
&=
\sum_{r}\sum_{n,m} a_{n m}\ln\frac{a_{n m}}{b_{n m}}
\;\ge\; 0,
\end{aligned}
\label{eq:Sdot_total_positive}
\end{equation}
which establishes the non–negativity of the entropy production rate.

\section{Integrated opinion currents and their rates}
\label{Cpr}
We prove the equivalence, introduced in Eq.~\eqref{eq:<dotIr>}, between the (reaction–resolved) probability currents $J_{m, n}^r$, and the time derivative of the ensemble average of the integrated opinion current $\langle \dot{I}_r \rangle$. We proceed by 
fixing an (arbitrary) order on the state space: for each unordered pair of states $\{m,n\}$ with $m<n$, we declare the ordered transition $(n\to m)$ “positive’’ and its reverse $(m\to n)$ “negative.’’ For a given reaction $r$, let $N^{(r)}_{m,n}(\tau)$ be the counting variable equal to the total number of jumps $n\!\to\! m$ due to reaction $r$, observed up to time $\tau$. We recover the definition of the total number of jumps in \Cref{sec:Opinion Thermodynamics} by summing over the pair of states, that is $N^{(r)}_{+}(\tau)=\sum_{m<n} N^{(r)}_{m,n(\tau)}
$. The integrated current associated to reaction $r$ along a single trajectory $\gamma_{[0,\tau]}$ can be written as
\begin{equation}
I_r\!\left(\gamma_{[0,\tau]}\right)
=\sum_{m<n}\left[N^{(r)}_{m,n}(\tau)-N^{(r)}_{n,m}(\tau)\right],
\label{eq:Ir_def}
\end{equation}
where $I^{(r)}_{m,n}(t)=N^{(r)}_{m,n}(t)-N^{(r)}_{n,m}(t)$ is the transition–resolved integrated current for a fixed ordered transition $(n\to m)$. Its (distribution–valued) time derivative can be represented as a shot–noise measure with Dirac spikes at the jump times. In stochastic–calculus notation this reads
\begin{equation}
\dot I^{(r)}_{m,n}(t)=\frac{1}{dt}\Big[dN^{(r)}_{m,n}(t)-dN^{(r)}_{n,m}(t)\Big],
\label{eq:shotnoise}
\end{equation}
where $dN^{(r)}_{m,n}(t)$ is a Poisson counting increment: it equals $1$ if a jump $n\!\to\! m$ due to $r$ occurs in $[t,t{+}dt)$ and $0$ otherwise. These stochastic variables satisfy $dN^{(r)}_{m,n}(t)\,dN^{(r)}_{m',n'}(t)
=\,\delta_{m m'}\,\delta_{n n'}\,dN^{(r)}_{m,n}(t)$.

Taking expectations in \eqref{eq:shotnoise} gives
\begin{equation}
\big\langle \dot I^{(r)}_{m,n}(t)\big\rangle
=\frac{1}{dt}\left[\left\langle dN^{(r)}_{m,n}(t)\right\rangle
-\left\langle dN^{(r)}_{n,m}(t)\right\rangle\right],
\label{eq:avg_start}
\end{equation}
where, using standard Poisson–increment rules, for each ordered transition $(n\to m)$ and reaction $r$, for the jump $n\!\to\! m$ at time $t$ we have:
\begin{align}
\label{eq:avg_intensity_1}
\left\langle dN^{(r)}_{n,m}(t)\right\rangle
&=P_m(t)\,W^{(r)}_{n,m}(t)\,dt,\\
\left\langle dN^{(r)}_{m,n}(t)\right\rangle
&=P_n(t)\,W^{(r)}_{m,n}(t)\,dt.
\label{eq:avg_intensity_2}
\end{align}
Substituting Eqs.~\eqref{eq:avg_intensity_1}–\eqref{eq:avg_intensity_2} into Eq.~\eqref{eq:avg_start} immediately yields the reaction–resolved probability current across the ordered transition $(n\to m)$:
\begin{equation}
\begin{split}
\big\langle \dot I^{(r)}_{m,n}(t)\big\rangle
&= W^{(r)}_{m,n}(t)\,P_n(t)\;-\;W^{(r)}_{n,m}(t)\,P_m(t) \\
&=:J^{(r)}_{m,n}(t),
\end{split}
\label{eq:transition_current}
\end{equation}
c.f. Eq.~\eqref{eq:Jmn}. Summing \eqref{eq:transition_current} over pairs $m<n$ we obtain, for the integrated current \eqref{eq:Ir_def}:
\begin{equation}
\begin{split}
\langle \dot I_r(t) \rangle &= \frac{d}{dt}\,\big\langle I_r(t)\big\rangle \\
&=\sum_{m<n} \big\langle \dot I^{(r)}_{m,n}(t)\big\rangle
=\sum_{m<n} J^{(r)}_{m,n}(t).
\label{eq:sum_current}
\end{split}
\end{equation}
Finally, in the long–time limit the system reaches a stationary regime, in which the current becomes time–independent. Therefore under stationary regime conditions:
\begin{equation}
\langle I_r(\tau)\rangle=\int_0^\tau \langle \dot I_r(t)\rangle\,dt = \langle \dot I_r\rangle\,\tau \qquad (\tau \to \infty),
\end{equation}
and, equivalently, $\langle \dot I_r\rangle \;=\lim_{\tau\to\infty}\frac{\langle I_r(\tau)\rangle}{\tau}$, which completes the proof.

\section{Mean-field Results}\label{sec:Mean-field analytical results}

In this appendix, we present the mean-field approach used to obtain analytical results for the ensemble thermodynamic quantities in the stationary regime.  
For any function $f(n)$ of the number of agents with a opinion $A$, we approximate $\sum_n P_n(t) f(n) \simeq f(\langle n \rangle)$, which is a reliable approximation for $N \gg 1$, and becomes exact in the macroscopic limit $N \to \infty$. We will use this approximation for most of our analytical inquiry, for which we will avoid carrying the symbols $\simeq$ to ease the notation.

The time-dependent expected number of agents in state $A $, $\langle n(t) \rangle$, can be obtained [rewriting the master equation \eqref{eq:Master Equation}] from the following equation~\cite{Toral2014StochasticNM}
: 
\begin{equation}\label{eq:mean-field equation}
 \frac{d\langle n(t) \rangle}{dt} = \sum_n P_n(t) \left( {W_{n+1, n}} - {W_{n-1, n}}\right) = \braket{\dot{I}(t)},
\end{equation}
which in the stationary state ($ d_t \braket{n}_{\mathrm{st}} = 0 $), corroborates that $\braket{\dot{I}}_{\mathrm{st}} = 0$, or $\braket{\dot{I}_1}_{\mathrm{st}} = - \braket{\dot{I}_2}_{\mathrm{st}}$. Similarly, the average probability currents, and dynamical activities can be written as:
\begin{align}
  \label{eq:Ir as expected values}
 \braket{\dot{I}_{r}} &= 
 \sum_n P_n(t) \left( {W_{n+1,n}^{(r)}} - {W_{n-1,n}^{(r)}} \right), \\
  \label{eq:Kr as expected values}
 \braket{K_{r}} &= 
 \sum_n P_n(t) \left( {W_{n+1,n}^{(r)}} + {W_{n-1,n}^{(r)}} \right).
\end{align}
For our analysis, we consider here the change of variables $x \equiv n / N $ and the re-parametrizations introduced at the end of \Cref{sec:Model}, i.e, \( (h_r, a_r) \to (\lambda, \chi, \theta, \omega) \). With these considerations, applying the mean-field approximation to \Cref{eq:Ir as expected values,eq:Kr as expected values}, they reduce to: 
\begin{align}
\label{eq:MF probability current 1}
    \braket{\dot{I}_1(t)} &= N \omega \sqrt{\lambda \chi \theta} \left[ 1 - \left(1 + \frac{1}{\lambda \chi} \right) \braket{x(t)} \right] g(\braket{x(t)}), \\
    \label{eq:MF probability current 2}
    \braket{I_2(t)} &= N \omega \sqrt{\frac{\chi }{\lambda \theta }} \left[ 1- \left(1 + \frac{\lambda}{\chi} \right) \braket{x(t)} \right] g(1-\braket{x(t)}),\\
    \label{eq:MF dynamical activity 1}
    \braket{K_1(t)} &= N \omega \sqrt{\lambda \chi \theta} \left[ 1 - \left(1 - \frac{1}{\lambda \chi} \right) \braket{x(t)} \right] g(\braket{x(t)}), \\
    \label{eq:MF dynamical activity 2}
    \braket{K_2(t)} &= N \omega \sqrt{\frac{\chi }{\lambda \theta }} \left[ 1- \left(1 - \frac{\lambda}{\chi} \right) \braket{x(t)} \right] g(1-\braket{x(t)}),
\end{align}
which are extensive quantities, i.e, they scale with the number of agents $N$. These are the general non-equilibrium expressions for the mean-field probability currents and dynamical activities of the two-reaction system. However, we notice that even though these expressions seem compact, the complex dependecies on the parameters $(\lambda, \chi, \theta, \omega)$ remain hidden in $\braket{x(t)}$,
for which the solution of \eqref{eq:mean-field equation} is needed. 

In the equilibrium state, achieved at $\lambda = 1 $ [see Eq.~\eqref{eq:Eq cond}], we can use \cref{eq:MF probability current 1,eq:MF probability current 2} to show that both probability currents vanish, $\braket{\dot{I}_r}_{\mathrm{eq}} = 0 $ for $r=1,2$. As a consequence the entropy production rate [see Eq.~\eqref{eq:Stotst}] also becomes zero, $\braket{\dot{S}_{\mathrm{tot}}}_{\mathrm{eq}} = 0 $, as expected for an equilibrium state. On the other hand, the dynamical activities [\cref{eq:MF dynamical activity 1,eq:MF dynamical activity 2}] associated to each reaction are non-zero and given by $\braket{{K}_1}_{\mathrm{eq}} = 2 N \omega \sqrt{\theta \chi^{2q + 1}} / (1 + \chi)^{q + 1} $, $\braket{{K}_2}_{\mathrm{eq}} = 2 \chi^q \braket{{K}_1}_{\mathrm{eq}}$. This confirms that in the equilibrium state, the dynamics of the system is not frozen. Also, we remark that the dynamical activity does capture a $q$ dependence even for the equilibrium state, contrary to the equilibrium probability distribution \eqref{eq:Equilibrium_Distribution}.

In the nonequilibrium stationary state the ensemble quantities capture the critical behavior summarized in App.~\ref{sec:Stationary State apdx}. In fact, the symmetric case ($\chi=\theta=1$) suffices to illustrate the main features of interest. For this situation, we were able to find analytical mean-field expressions for $g(x)=x^q$ in Eqs.~\eqref{eq:transition rates}, which models non-linear social influence both sampling with and without repetition in the thermodynamic limit $N \to \infty $. For $\lambda$ below the critical point in Eq.~\eqref{eq:critical point asymmetric}, the stationary mean is $\langle x \rangle_{\mathrm{st}}=1/2$ for all $q$, and we obtain:
\begin{align}
 \label{eq:Stationary Probability Currents 1 below}
 \braket{\dot{I}(\lambda \leq \lambda_{\mathrm{c}})}_{\mathrm{st}}
 & = 
 N \omega ~ \frac{\lambda - 1}{2^{q + 1} \sqrt{\lambda} }
 \\
 \label{eq:Stationary Dynamical Activities}
 \braket{K(\lambda \leq \lambda_{\mathrm{c}})}_{\mathrm{st}}
 & = 
 N \omega ~ \frac{\lambda + 1}{2^{q} \sqrt{\lambda} }
\end{align}
which for $\lambda \rightarrow 0$, tend respectively to $\infty$ and $-\infty$. 
At the critical point, we instead obtain: 
\begin{align}
 \label{eq:Stationary Probability Currents 1 critical}
 \braket{\dot{I}(\lambda = \lambda_{\mathrm{c}},\chi=1)}_{\mathrm{st}}
 &= 
 N~ \frac{\omega}{2^{q}} \frac{1}{\sqrt{q^{2} - 1}},
 \\
 \label{eq:Stationary Dynamical Activities critical}
 \braket{K(\lambda = \lambda_{\mathrm{c}},\chi=1)}_{\mathrm{st}}
 &= 
 N~ \frac{\omega}{2^{q - 1}} \frac{q}{\sqrt{q^{2} - 1}},
\end{align}
which correspond to a maximum and to an inflexion point, respectively. Above the critical point the expression of the currents may differ with \( q \). However their behavior in the limit $\lambda \to \infty$ is universal. In that case, $\braket{x}_{\mathrm{st}} \in \{ 0, 1 \} $ depending on which maxima of the stationary distribution we are in, but both yield the same results; $\braket{\dot{I}_1(\lambda \to \infty)} = \braket{K(\lambda \to \infty)} = 0$, as for finite \( N \). For \( q = 2 \), substituting Eq.~\eqref{eq:xst q2} into \Cref{eq:MF probability current 1,eq:MF dynamical activity 1}, we obtain closed-form expressions for the currents above the critical point:
\begin{align}
    \braket{\dot{I}(\lambda \geq \lambda_{\mathrm{c}})}_{\mathrm{st}}
    &= N~ \frac{\omega}{\sqrt{\lambda} (\lambda + 1)},\\
    \braket{K(\lambda \geq \lambda_{\mathrm{c}})}_{\mathrm{st}} 
    &= N~ \frac{2 \omega}{\sqrt{\lambda}} \frac{\lambda - 1}{\lambda + 1},
\end{align}
and substituting them into the entropy production rate expression, $\langle \dot{S}_{\rm tot} \rangle =(\mu_1 - \mu_2) \langle I_1 \rangle_{\rm st}$, we have:
\begin{equation}\label{eq:dtStot sym}
  \langle \dot{S}_{\mathrm{tot}} \rangle =
    \begin{cases}
    \dfrac{N~ \omega \ln\lambda\left(\lambda-1\right)}{4 \sqrt{\lambda}}, & \lambda \le \lambda_c,\\[8pt]
    \dfrac{N~ \omega 2\,\ln\lambda}{\sqrt{\lambda}\,(\lambda+1)}, & \lambda > \lambda_c.
    \end{cases}
\end{equation}

\section{Full Counting Statistics}\label{sec:Cal Flu Met}
In this appendix we summarize the method of Full Counting Statistics \cite{Esposito2009Dec,Esposito2007Apr,Walldorf2020May} used to analyze the statistical properties of opinion currents. Fluctuations play a central role in stochastic thermodynamics, but obtaining their statistics directly by sampling trajectories is often computationally costly. To circumvent this issue, one can compute the moments of the stationary probability distribution of the opinion current, $\mathcal{P}(I_r)$, using generating-function techniques. Throughout, we focus on the stationary regime (long-time limit).

We consider a stationary trajectory $\gamma_{[0,\tau]}$. The integrated current up to time $\tau$, as defined in Sec.~\ref{TaTaFR}, is the difference between the accumulated number of jumps up and down for each reaction $r$, namely $I_r(\gamma_{[0,\tau]}\big)=N_{+}^{(r)}-N_{-}^{(r)}\!\big.$ The statistics of the stationary current are encoded in the scaled cumulant generating function (SCGF),
\begin{equation}
C_r(\xi_r)\equiv \lim_{t\to\infty}\frac{1}{t}\,\ln \int_{-\infty}^{+\infty}\mathcal{P} \left[I_r(t)\right]\,e^{\xi_r I_r(t)}\,dI_r(t),
\end{equation}
where $\xi_r$ is a counting field for reaction $r$. The SCGF admits the power-series expansion $C_r(\xi_r)=\sum_{k\ge 1}\tfrac{c_k}{k!}\,\xi_r^k$, whose coefficients are the scaled cumulants of the stationary distribution of $I_r$:
\begin{align}
\label{FCS_mean}
c_1 \equiv \langle \dot{I}_r\rangle_{\mathrm{st}} &= \frac{1}{\tau}\left.\frac{\partial C_r(\xi_r)}{\partial \xi_r}\right|_{\xi_r=0},\\
c_2 \equiv \sigma^2_{\mathrm{st}}(I_r) &= \frac{1}{\tau}\left.\frac{\partial^2 C_r(\xi_r)}{\partial \xi_r^2}\right|_{\xi_r=0}.
\label{FCS_var}
\end{align}
Higher-order cumulants follow from $\partial_{\xi_r}^n C_r\big|_{\xi_r=0}$. A complementary characterization is provided by large-deviation theory~\cite{Touchette2009Jul}. In the long-time limit
\begin{equation} 
\mathcal{P} \big(I_r\big)\asymp e^{-t\,\psi(I_r)},
\end{equation}
where $\psi(I_r)$ is the so-called rate function. It can be obtained from the SCGF by a Legendre transform:
\begin{equation}
\psi_r\!\big(I_r\big)=\max_{\xi_r}\,\big[\xi_r I_r - C_r(\xi_r)\big].
\end{equation}
For a given reaction, the rate function satisfies the detailed fluctuation theorem for the opinion current, which can be alternatively stated as $\psi_1(I_1)-\psi_1(-I_1)=(\mu_1-\mu_2)\,I_1$ (see also Sec.~\ref{sec:inference}). Consequently, computing $C_r(\xi_r)$ is central to the analysis of current fluctuations. In this work we obtain it via a spectral (tilted-operator) method (see a detailed review in Ref.~\cite{Landi2024Apr}).

We assume a finite state space (of size $N$) and work in a matrix representation. We can construct the rate matrix (or Liouvillian) operator governing the dynamics of the probability density as an $N \times N$ matrix with elements: 
\[ \label{eq:Liouvilian}
(\mathbb{W})_{n,m}=
\begin{cases}
-\displaystyle\sum_{r}\!\big[W^{(r)}_{n,n-1}+W^{(r)}_{n,n+1}\big], & m=n,\\[8pt]
\displaystyle\sum_{r} W^{(r)}_{n,n-1}, & m=n-1,\\[8pt]
\displaystyle\sum_{r} W^{(r)}_{n,n+1}, & m=n+1,
\end{cases}
\]
and zero otherwise. In terms of the Liouvillian operator, the master equation~\eqref{eq:Master Equation} can be written as $\ket{ \dot{P}(t)} = \mathbb{W} \ket{P(t)}$, with the probability vector $\ket{P}=(P_1, P_2, ..., P_N)^\intercal$, and similarly for $\ket{\dot{P}(t)}$.

The SCGF is generated by the long-time evolution of a tilted rate matrix (or tilted Liouvillian)  acting on the initial probability vector as:
\begin{align}
\label{Def_Cr_SM}
C_r(\xi_r)&=\lim_{t\to\infty}\frac{1}{t}\ln \big\langle e^{\xi_r I_r}\big\rangle \nonumber \\ 
&= \lim_{t\to\infty}\frac{1}{t}\ln \bra{1}\,e^{\mathbb{W}_{\xi_r} t}\,\ket{P(t=0)},
\end{align}
with $\ket{1}$ the unit vector. The tilted rate matrix (or tilted Liouvillian) $\mathbb{W}_{\xi_r}$ is constructed by decomposing the generator into (i) off-diagonal terms, which encode jumps between distinct states, and (ii) diagonal terms, which collect total escape rates and enforce probability conservation. Tilting attaches a counting field to each jump reaction: off-diagonal entries that increase the measured current are multiplied by $e^{+\xi_r}$, whereas those that decrease it are multiplied by $e^{-\xi_r}$; diagonal entries are unchanged. For the model considered here, the tilted operator has tridiagonal form,
\begin{widetext}
\begin{equation}
\mathbb{W}_{\boldsymbol{\xi}}=
\begingroup
\setlength{\arraycolsep}{6pt}\renewcommand{\arraystretch}{1.2}
\begin{pmatrix}
\ddots & \cdots & \cdots & \cdots & \ddots \\[2pt]
\cdots
  & \displaystyle\sum_{r} W^{(r)}_{\,n-1,\,n-2}\,e^{-\xi_r}
  & 0
  & 0
  & \cdots \\[8pt]
\cdots
  & -\displaystyle\sum_{r}\!\big[W^{(r)}_{\,n-1,\,n-2}+W^{(r)}_{\,n-1,\,n}\big]
  & \displaystyle\sum_{r} W^{(r)}_{\,n,\,n-1}\,e^{-\xi_r}
  & 0
  & \cdots \\[8pt]
\cdots
  & \displaystyle\sum_{r} W^{(r)}_{\,n-1,\,n}\,e^{\xi_r}
  & -\displaystyle\sum_{r}\!\big[W^{(r)}_{\,n,\,n-1}+W^{(r)}_{\,n,\,n+1}\big]
  & \displaystyle\sum_{r} W^{(r)}_{\,n+1,\,n}\,e^{-\xi_r}
  & \cdots \\[8pt]
\cdots
  & 0
  & \displaystyle\sum_{r} W^{(r)}_{\,n,\,n+1}\,e^{\xi_r}
  & -\displaystyle\sum_{r}\!\big[W^{(r)}_{\,n+1,\,n}+W^{(r)}_{\,n+1,\,n+2}\big]
  & \cdots \\[8pt]
\cdots
  & 0
  & 0
  & \displaystyle\sum_{r} W^{(r)}_{\,n+1,\,n+2}\,e^{\xi_r}
  & \cdots \\[2pt]
\ddots & \cdots & \cdots & \cdots & \ddots
\end{pmatrix}
\endgroup
\label{eq:Lxi-3col-5rows}
\end{equation}
\end{widetext}
where $\boldsymbol{\xi}=\xi_r$ collects the counting fields for all reactions. Jumps contributing to $N_-^{(r)}$ appear on the superdiagonal, while those contributing to $N_+^{(r)}$ appear on the subdiagonal. In the long-time limit, Eq.~\eqref{Def_Cr_SM} is governed by the dominant eigenvalue of $\mathbb{W}_{\xi_r}$—the one with the largest real part—denoted $\zeta_0(\xi_r)$, with $\zeta_0(0)=0$:
\begin{equation}
C_r(\xi_r)=\zeta_0(\xi_r).
\end{equation}
Once the functional form of $C_r(\xi_r)$ has been determined, $\langle \dot{I}_r \rangle_{\text{st}}$ can be obtained by means of Eq.(\ref{FCS_mean}), and $\sigma^2_{\mathrm{st}}(I_r)$ by means of Eq.(\ref{FCS_var}) The main limitation of this approach is computational: for very large $N$, evaluating the dominant eigenvalue becomes prohibitively expensive. However for the social imitation model analyzed in this work we could obtain results using this method up to $N=10^4$ agents.

\section{Proof of the strong fluctuation theorem for currents}\label{apdx:stationary state details}

In this appendix we present a detailed derivation of the strong fluctuation theorem for integrated opinion currents. As stated in the main text in Sec. \ref{sec:inference}, starting from the path probability of a given trajectory $\gamma_{[0,\tau]}$, denoted $\mathds{P}(\gamma_{[0,\tau]})$, we define the probability to observe a value $I_1$ of the integrated current (for a fixed time window $[0,\tau]$) as
\begin{equation} \label{eq:currentapx}
\mathcal{P}\big(I_1\big)
= \sum_{\gamma_{[0,\tau]}} \mathds{P}(\gamma_{[0,\tau]})\,
\indi\left[I_1 - I_1(\gamma_{[0,\tau]})\right].
\end{equation}
where $\indi(x)=1$ if $x=0$ and $\indi(x)=0$, otherwise.

Analogously, the probability to observe the same current with opposite sign can be defined from the probability of the time-reversed trajectory $\tilde{\gamma}_{[0,\tau]}$,
\begin{equation}
\mathcal{P}\big(-I_1\big)
= \sum_{\tilde{\gamma}_{[0,\tau]}} \tilde{\mathds{P}}(\tilde{\gamma}_{[0,\tau]})\,
\indi\left[-I_1 - I_1(\tilde{\gamma}_{[0,\tau]})\right].
\end{equation}
By the detailed fluctuation theorem of Eq.~\eqref{eq:DFT}, the ratio of forward and backward path probabilities satisfies $\tilde{\mathds{P}}(\tilde{\gamma}_{[0,\tau]}) = e^{\,- S_{\rm tot}(\gamma_{[0,\tau]})} \mathds{P}(\gamma_{[0,\tau]})$, where $S_{\rm tot}(\gamma_{[0,\tau]})$ is the total entropy production along $\gamma_{[0,\tau]}$. Using the one-to-one correspondence between paths and their time reversals, and microreversibility of the current
\(
I_1(\tilde{\gamma}_{[0,\tau]}) = -\,I_1(\gamma_{[0,\tau]})
\),
we can change variables $\tilde{\gamma}_{[0,\tau]}\mapsto \gamma_{[0,\tau]}$ and rewrite the distribution of the negative current as
\begin{equation}
\mathcal{P}\big(-I_1\big)
= \sum_{\gamma_{[0,\tau]}} \mathds{P}(\gamma_{[0,\tau]})\,
e^{-S_{\rm tot}(\gamma_{[0,\tau]})}\,
\indi\left[I_1 - I_1(\gamma_{[0,\tau]})\right],
\end{equation}
where we also used the symmetry of the indicator function. In the stationary regime, the system entropy change over $[0,\tau]$ vanishes, so the total entropy production equals the medium entropy production. For a single integrated current $I_1$ one has the linear relation $S_{\rm tot}(\gamma_{[0,\tau]}) = (\mu_1 - \mu_2)\, I_1(\gamma_{[0,\tau]})$. 
Using this steady-state form, we note that on the support of the indicator functions $I_1(\gamma_{[0,\tau]})=I_1$ is fixed, hence the exponential factor becomes a constant and factors out of the denominator:
\begin{equation}
\mathcal{P}\big(-I_1\big)
= e^{-(\mu_1 - \mu_2) I_1} \sum_{\gamma_{[0,\tau]}} \mathds{P}(\gamma_{[0,\tau]})\,
\,
\indi\left[I_1 - I_1(\gamma_{[0,\tau]})\right],
\end{equation}
c.f. Eq.~\eqref{eq:currentapx}. As a consequence, the ratio of probabilities yields the strong fluctuation theorem stated in Eq.~\eqref{eq:Inference}:
\begin{equation}
\frac{\mathcal{P}(I_1)}{\mathcal{P}(-I_1)} \;=\; e^{(\mu_1 - \mu_2)\, I_1}.
\end{equation}



\clearpage
\bibliography{apssamp}

\providecommand{\noopsort}[1]{}\providecommand{\singleletter}[1]{#1}%
\begin{thebibliography}{153}%
\makeatletter
\providecommand \@ifxundefined [1]{%
 \@ifx{#1\undefined}
}%
\providecommand \@ifnum [1]{%
 \ifnum #1\expandafter \@firstoftwo
 \else \expandafter \@secondoftwo
 \fi
}%
\providecommand \@ifx [1]{%
 \ifx #1\expandafter \@firstoftwo
 \else \expandafter \@secondoftwo
 \fi
}%
\providecommand \natexlab [1]{#1}%
\providecommand \enquote  [1]{``#1''}%
\providecommand \bibnamefont  [1]{#1}%
\providecommand \bibfnamefont [1]{#1}%
\providecommand \citenamefont [1]{#1}%
\providecommand \href@noop [0]{\@secondoftwo}%
\providecommand \href [0]{\begingroup \@sanitize@url \@href}%
\providecommand \@href[1]{\@@startlink{#1}\@@href}%
\providecommand \@@href[1]{\endgroup#1\@@endlink}%
\providecommand \@sanitize@url [0]{\catcode `\\12\catcode `\$12\catcode
  `\&12\catcode `\#12\catcode `\^12\catcode `\_12\catcode `\%12\relax}%
\providecommand \@@startlink[1]{}%
\providecommand \@@endlink[0]{}%
\providecommand \url  [0]{\begingroup\@sanitize@url \@url }%
\providecommand \@url [1]{\endgroup\@href {#1}{\urlprefix }}%
\providecommand \urlprefix  [0]{URL }%
\providecommand \Eprint [0]{\href }%
\providecommand \doibase [0]{https://doi.org/}%
\providecommand \selectlanguage [0]{\@gobble}%
\providecommand \bibinfo  [0]{\@secondoftwo}%
\providecommand \bibfield  [0]{\@secondoftwo}%
\providecommand \translation [1]{[#1]}%
\providecommand \BibitemOpen [0]{}%
\providecommand \bibitemStop [0]{}%
\providecommand \bibitemNoStop [0]{.\EOS\space}%
\providecommand \EOS [0]{\spacefactor3000\relax}%
\providecommand \BibitemShut  [1]{\csname bibitem#1\endcsname}%
\let\auto@bib@innerbib\@empty
\bibitem [{\citenamefont {Kondepudi}\ and\ \citenamefont
  {Prigogine}(2015)}]{Kondepudi}%
  \BibitemOpen
  \bibfield  {author} {\bibinfo {author} {\bibfnamefont {D.}~\bibnamefont
  {Kondepudi}}\ and\ \bibinfo {author} {\bibfnamefont {I.}~\bibnamefont
  {Prigogine}},\ }\href@noop {} {\emph {\bibinfo {title} {{Modern
  Thermodynamics: From Heat Engines to Dissipative Structures}}}},\ \bibinfo
  {edition} {2nd}\ ed.\ (\bibinfo  {publisher} {New Delhi: Wiley india Pvt.
  Ltd},\ \bibinfo {year} {2015})\BibitemShut {NoStop}%
\bibitem [{\citenamefont {Callen}(2015)}]{Callen}%
  \BibitemOpen
  \bibfield  {author} {\bibinfo {author} {\bibfnamefont {H.~B.}\ \bibnamefont
  {Callen}},\ }\href@noop {} {\emph {\bibinfo {title} {Thermodynamics and an
  {{Introduction}} to {{Thermostatistics}}}}},\ \bibinfo {edition} {2nd}\ ed.\
  (\bibinfo  {publisher} {New Delhi: Wiley india Pvt. Ltd},\ \bibinfo {year}
  {2015})\BibitemShut {NoStop}%
\bibitem [{\citenamefont {Nielsen}\ \emph {et~al.}(2020)\citenamefont
  {Nielsen}, \citenamefont {Müller}, \citenamefont {Marques}, \citenamefont
  {Bastianoni},\ and\ \citenamefont {Jørgensen}}]{Nielsen20}%
  \BibitemOpen
  \bibfield  {author} {\bibinfo {author} {\bibfnamefont {S.~N.}\ \bibnamefont
  {Nielsen}}, \bibinfo {author} {\bibfnamefont {F.}~\bibnamefont {Müller}},
  \bibinfo {author} {\bibfnamefont {J.~C.}\ \bibnamefont {Marques}}, \bibinfo
  {author} {\bibfnamefont {S.}~\bibnamefont {Bastianoni}},\ and\ \bibinfo
  {author} {\bibfnamefont {S.~E.}\ \bibnamefont {Jørgensen}},\ }\bibfield
  {title} {\bibinfo {title} {Thermodynamics in ecology—an introductory
  review},\ }\bibfield  {journal} {\bibinfo  {journal} {Entropy}\ }\textbf
  {\bibinfo {volume} {22}},\ \href {https://doi.org/10.3390/e22080820}
  {10.3390/e22080820} (\bibinfo {year} {2020})\BibitemShut {NoStop}%
\bibitem [{\citenamefont {Georgescu-Roegen}(1971)}]{Roegen}%
  \BibitemOpen
  \bibfield  {author} {\bibinfo {author} {\bibfnamefont {N.}~\bibnamefont
  {Georgescu-Roegen}},\ }\href@noop {} {\emph {\bibinfo {title} {The Entropy
  Law and the Economic Process}}}\ (\bibinfo  {publisher} {Harvard University
  Press},\ \bibinfo {address} {Cambridge, MA and London, England},\ \bibinfo
  {year} {1971})\BibitemShut {NoStop}%
\bibitem [{\citenamefont {Wald}(2005)}]{Wald05}%
  \BibitemOpen
  \bibfield  {author} {\bibinfo {author} {\bibfnamefont {R.~M.}\ \bibnamefont
  {Wald}},\ }\bibinfo {title} {The thermodynamics of black holes},\ in\ \href
  {https://doi.org/10.1007/0-387-24992-3_1} {\emph {\bibinfo {booktitle}
  {Lectures on Quantum Gravity}}},\ \bibinfo {editor} {edited by\ \bibinfo
  {editor} {\bibfnamefont {A.}~\bibnamefont {Gomberoff}}\ and\ \bibinfo
  {editor} {\bibfnamefont {D.}~\bibnamefont {Marolf}}}\ (\bibinfo  {publisher}
  {Springer US},\ \bibinfo {address} {Boston, MA},\ \bibinfo {year} {2005})\
  pp.\ \bibinfo {pages} {1--37}\BibitemShut {NoStop}%
\bibitem [{\citenamefont {Goold}\ \emph {et~al.}(2016)\citenamefont {Goold},
  \citenamefont {Huber}, \citenamefont {Riera}, \citenamefont {Rio},\ and\
  \citenamefont {Skrzypczyk}}]{Goold16}%
  \BibitemOpen
  \bibfield  {author} {\bibinfo {author} {\bibfnamefont {J.}~\bibnamefont
  {Goold}}, \bibinfo {author} {\bibfnamefont {M.}~\bibnamefont {Huber}},
  \bibinfo {author} {\bibfnamefont {A.}~\bibnamefont {Riera}}, \bibinfo
  {author} {\bibfnamefont {L.~d.}\ \bibnamefont {Rio}},\ and\ \bibinfo {author}
  {\bibfnamefont {P.}~\bibnamefont {Skrzypczyk}},\ }\bibfield  {title}
  {\bibinfo {title} {The role of quantum information in thermodynamics—a
  topical review},\ }\href {https://doi.org/10.1088/1751-8113/49/14/143001}
  {\bibfield  {journal} {\bibinfo  {journal} {Journal of Physics A:
  Mathematical and Theoretical}\ }\textbf {\bibinfo {volume} {49}},\ \bibinfo
  {pages} {143001} (\bibinfo {year} {2016})}\BibitemShut {NoStop}%
\bibitem [{\citenamefont {Vinjanampathy}\ and\ \citenamefont
  {Anders}(2016)}]{Vinjanampathy16}%
  \BibitemOpen
  \bibfield  {author} {\bibinfo {author} {\bibfnamefont {S.}~\bibnamefont
  {Vinjanampathy}}\ and\ \bibinfo {author} {\bibfnamefont {J.}~\bibnamefont
  {Anders}},\ }\bibfield  {title} {\bibinfo {title} {Quantum thermodynamics},\
  }\href {https://doi.org/10.1080/00107514.2016.1201896} {\bibfield  {journal}
  {\bibinfo  {journal} {Contemporary Physics}\ }\textbf {\bibinfo {volume}
  {57}},\ \bibinfo {pages} {545} (\bibinfo {year} {2016})}\BibitemShut
  {NoStop}%
\bibitem [{\citenamefont {Sekimoto}(2010)}]{Sekimoto10}%
  \BibitemOpen
  \bibfield  {author} {\bibinfo {author} {\bibfnamefont {K.}~\bibnamefont
  {Sekimoto}},\ }\href@noop {} {\emph {\bibinfo {title} {Stochastic
  energetics}}},\ Vol.\ \bibinfo {volume} {799}\ (\bibinfo  {publisher}
  {Springer},\ \bibinfo {year} {2010})\BibitemShut {NoStop}%
\bibitem [{\citenamefont {Seifert}(2012)}]{Seifert12}%
  \BibitemOpen
  \bibfield  {author} {\bibinfo {author} {\bibfnamefont {U.}~\bibnamefont
  {Seifert}},\ }\bibfield  {title} {\bibinfo {title} {Stochastic
  thermodynamics, fluctuation theorems and molecular machines},\ }\href@noop {}
  {\bibfield  {journal} {\bibinfo  {journal} {Rep. Prog. Phys.}\ }\textbf
  {\bibinfo {volume} {75}},\ \bibinfo {pages} {126001} (\bibinfo {year}
  {2012})}\BibitemShut {NoStop}%
\bibitem [{\citenamefont {Ciliberto}(2017)}]{Ciliberto17}%
  \BibitemOpen
  \bibfield  {author} {\bibinfo {author} {\bibfnamefont {S.}~\bibnamefont
  {Ciliberto}},\ }\bibfield  {title} {\bibinfo {title} {Experiments in
  stochastic thermodynamics: Short history and perspectives},\ }\href
  {https://doi.org/10.1103/PhysRevX.7.021051} {\bibfield  {journal} {\bibinfo
  {journal} {Phys. Rev. X}\ }\textbf {\bibinfo {volume} {7}},\ \bibinfo {pages}
  {021051} (\bibinfo {year} {2017})}\BibitemShut {NoStop}%
\bibitem [{\citenamefont {Jarzynski}(2011)}]{Jarzynski11}%
  \BibitemOpen
  \bibfield  {author} {\bibinfo {author} {\bibfnamefont {C.}~\bibnamefont
  {Jarzynski}},\ }\bibfield  {title} {\bibinfo {title} {Equalities and
  inequalities: Irreversibility and the second law of thermodynamics at the
  nanoscale},\ }\href@noop {} {\bibfield  {journal} {\bibinfo  {journal} {Annu.
  Rev. Condens. Matter Phys.}\ }\textbf {\bibinfo {volume} {2}},\ \bibinfo
  {pages} {329} (\bibinfo {year} {2011})}\BibitemShut {NoStop}%
\bibitem [{\citenamefont {Parrondo}\ \emph {et~al.}(2015)\citenamefont
  {Parrondo}, \citenamefont {Horowitz},\ and\ \citenamefont
  {Sagawa}}]{Parrondo15}%
  \BibitemOpen
  \bibfield  {author} {\bibinfo {author} {\bibfnamefont {J.~M.~R.}\
  \bibnamefont {Parrondo}}, \bibinfo {author} {\bibfnamefont {J.~M.}\
  \bibnamefont {Horowitz}},\ and\ \bibinfo {author} {\bibfnamefont
  {T.}~\bibnamefont {Sagawa}},\ }\bibfield  {title} {\bibinfo {title}
  {Thermodynamics of information},\ }\href@noop {} {\bibfield  {journal}
  {\bibinfo  {journal} {Nature Phys.}\ }\textbf {\bibinfo {volume} {11}},\
  \bibinfo {pages} {131} (\bibinfo {year} {2015})}\BibitemShut {NoStop}%
\bibitem [{\citenamefont {Lutz}\ and\ \citenamefont
  {Ciliberto}(2015)}]{Lutz15}%
  \BibitemOpen
  \bibfield  {author} {\bibinfo {author} {\bibfnamefont {E.}~\bibnamefont
  {Lutz}}\ and\ \bibinfo {author} {\bibfnamefont {S.}~\bibnamefont
  {Ciliberto}},\ }\bibfield  {title} {\bibinfo {title} {{Information: From
  Maxwell’s demon to Landauer’s eraser}},\ }\href
  {https://doi.org/10.1063/PT.3.2912} {\bibfield  {journal} {\bibinfo
  {journal} {Physics Today}\ }\textbf {\bibinfo {volume} {68}},\ \bibinfo
  {pages} {30} (\bibinfo {year} {2015})}\BibitemShut {NoStop}%
\bibitem [{\citenamefont {Barato}\ and\ \citenamefont
  {Seifert}(2015)}]{Barato15}%
  \BibitemOpen
  \bibfield  {author} {\bibinfo {author} {\bibfnamefont {A.~C.}\ \bibnamefont
  {Barato}}\ and\ \bibinfo {author} {\bibfnamefont {U.}~\bibnamefont
  {Seifert}},\ }\bibfield  {title} {\bibinfo {title} {Thermodynamic uncertainty
  relation for biomolecular processes},\ }\href
  {https://doi.org/10.1103/PhysRevLett.114.158101} {\bibfield  {journal}
  {\bibinfo  {journal} {Phys. Rev. Lett.}\ }\textbf {\bibinfo {volume} {114}},\
  \bibinfo {pages} {158101} (\bibinfo {year} {2015})}\BibitemShut {NoStop}%
\bibitem [{\citenamefont {Gingrich}\ \emph {et~al.}(2016)\citenamefont
  {Gingrich}, \citenamefont {Horowitz}, \citenamefont {Perunov},\ and\
  \citenamefont {England}}]{Todd16}%
  \BibitemOpen
  \bibfield  {author} {\bibinfo {author} {\bibfnamefont {T.~R.}\ \bibnamefont
  {Gingrich}}, \bibinfo {author} {\bibfnamefont {J.~M.}\ \bibnamefont
  {Horowitz}}, \bibinfo {author} {\bibfnamefont {N.}~\bibnamefont {Perunov}},\
  and\ \bibinfo {author} {\bibfnamefont {J.~L.}\ \bibnamefont {England}},\
  }\bibfield  {title} {\bibinfo {title} {Dissipation bounds all steady-state
  fluctuations},\ }\href {https://doi.org/10.1103/PhysRevLett.116.120601}
  {\bibfield  {journal} {\bibinfo  {journal} {Phys. Rev. Lett.}\ }\textbf
  {\bibinfo {volume} {116}},\ \bibinfo {pages} {120601} (\bibinfo {year}
  {2016})}\BibitemShut {NoStop}%
\bibitem [{\citenamefont {Horowitz}\ and\ \citenamefont
  {Gingrich}(2020)}]{Horowitz20}%
  \BibitemOpen
  \bibfield  {author} {\bibinfo {author} {\bibfnamefont {J.~M.}\ \bibnamefont
  {Horowitz}}\ and\ \bibinfo {author} {\bibfnamefont {T.~R.}\ \bibnamefont
  {Gingrich}},\ }\bibfield  {title} {\bibinfo {title} {Thermodynamic
  uncertainty relations constrain non-equilibrium fluctuations},\ }\href
  {https://doi.org/10.1038/s41567-019-0702-6} {\bibfield  {journal} {\bibinfo
  {journal} {Nat. Phys.}\ }\textbf {\bibinfo {volume} {16}},\ \bibinfo {pages}
  {15} (\bibinfo {year} {2020})}\BibitemShut {NoStop}%
\bibitem [{\citenamefont {Terlizzi}\ and\ \citenamefont
  {Baiesi}(2018)}]{Terlizzi18}%
  \BibitemOpen
  \bibfield  {author} {\bibinfo {author} {\bibfnamefont {I.~D.}\ \bibnamefont
  {Terlizzi}}\ and\ \bibinfo {author} {\bibfnamefont {M.}~\bibnamefont
  {Baiesi}},\ }\bibfield  {title} {\bibinfo {title} {Kinetic uncertainty
  relation},\ }\href {https://doi.org/10.1088/1751-8121/aaee34} {\bibfield
  {journal} {\bibinfo  {journal} {Journal of Physics A: Mathematical and
  Theoretical}\ }\textbf {\bibinfo {volume} {52}},\ \bibinfo {pages} {02LT03}
  (\bibinfo {year} {2018})}\BibitemShut {NoStop}%
\bibitem [{\citenamefont {Yan}\ \emph {et~al.}(2019)\citenamefont {Yan},
  \citenamefont {Hilfinger}, \citenamefont {Vinnicombe},\ and\ \citenamefont
  {Paulsson}}]{Yan19}%
  \BibitemOpen
  \bibfield  {author} {\bibinfo {author} {\bibfnamefont {J.}~\bibnamefont
  {Yan}}, \bibinfo {author} {\bibfnamefont {A.}~\bibnamefont {Hilfinger}},
  \bibinfo {author} {\bibfnamefont {G.}~\bibnamefont {Vinnicombe}},\ and\
  \bibinfo {author} {\bibfnamefont {J.}~\bibnamefont {Paulsson}},\ }\bibfield
  {title} {\bibinfo {title} {Kinetic uncertainty relations for the control of
  stochastic reaction networks},\ }\href
  {https://doi.org/10.1103/PhysRevLett.123.108101} {\bibfield  {journal}
  {\bibinfo  {journal} {Phys. Rev. Lett.}\ }\textbf {\bibinfo {volume} {123}},\
  \bibinfo {pages} {108101} (\bibinfo {year} {2019})}\BibitemShut {NoStop}%
\bibitem [{\citenamefont {Vo}\ \emph {et~al.}(2022)\citenamefont {Vo},
  \citenamefont {Van~Vu},\ and\ \citenamefont {Hasegawa}}]{Vo22}%
  \BibitemOpen
  \bibfield  {author} {\bibinfo {author} {\bibfnamefont {V.~T.}\ \bibnamefont
  {Vo}}, \bibinfo {author} {\bibfnamefont {T.}~\bibnamefont {Van~Vu}},\ and\
  \bibinfo {author} {\bibfnamefont {Y.}~\bibnamefont {Hasegawa}},\ }\bibfield
  {title} {\bibinfo {title} {Unified thermodynamic–kinetic uncertainty
  relation},\ }\href {https://doi.org/10.1088/1751-8121/ac9099} {\bibfield
  {journal} {\bibinfo  {journal} {Journal of Physics A: Mathematical and
  Theoretical}\ }\textbf {\bibinfo {volume} {55}},\ \bibinfo {pages} {405004}
  (\bibinfo {year} {2022})}\BibitemShut {NoStop}%
\bibitem [{\citenamefont {Shiraishi}\ \emph {et~al.}(2018)\citenamefont
  {Shiraishi}, \citenamefont {Funo},\ and\ \citenamefont
  {Saito}}]{Shiraishi18}%
  \BibitemOpen
  \bibfield  {author} {\bibinfo {author} {\bibfnamefont {N.}~\bibnamefont
  {Shiraishi}}, \bibinfo {author} {\bibfnamefont {K.}~\bibnamefont {Funo}},\
  and\ \bibinfo {author} {\bibfnamefont {K.}~\bibnamefont {Saito}},\ }\bibfield
   {title} {\bibinfo {title} {Speed limit for classical stochastic processes},\
  }\href {https://doi.org/10.1103/PhysRevLett.121.070601} {\bibfield  {journal}
  {\bibinfo  {journal} {Phys. Rev. Lett.}\ }\textbf {\bibinfo {volume} {121}},\
  \bibinfo {pages} {070601} (\bibinfo {year} {2018})}\BibitemShut {NoStop}%
\bibitem [{\citenamefont {Falasco}\ and\ \citenamefont
  {Esposito}(2020)}]{Falasco20}%
  \BibitemOpen
  \bibfield  {author} {\bibinfo {author} {\bibfnamefont {G.}~\bibnamefont
  {Falasco}}\ and\ \bibinfo {author} {\bibfnamefont {M.}~\bibnamefont
  {Esposito}},\ }\bibfield  {title} {\bibinfo {title} {Dissipation-time
  uncertainty relation},\ }\href
  {https://doi.org/10.1103/PhysRevLett.125.120604} {\bibfield  {journal}
  {\bibinfo  {journal} {Phys. Rev. Lett.}\ }\textbf {\bibinfo {volume} {125}},\
  \bibinfo {pages} {120604} (\bibinfo {year} {2020})}\BibitemShut {NoStop}%
\bibitem [{\citenamefont {Garc\'{\i}a-Pintos}\ \emph
  {et~al.}(2022)\citenamefont {Garc\'{\i}a-Pintos}, \citenamefont {Nicholson},
  \citenamefont {Green}, \citenamefont {del Campo},\ and\ \citenamefont
  {Gorshkov}}]{LP22}%
  \BibitemOpen
  \bibfield  {author} {\bibinfo {author} {\bibfnamefont {L.~P.}\ \bibnamefont
  {Garc\'{\i}a-Pintos}}, \bibinfo {author} {\bibfnamefont {S.~B.}\ \bibnamefont
  {Nicholson}}, \bibinfo {author} {\bibfnamefont {J.~R.}\ \bibnamefont
  {Green}}, \bibinfo {author} {\bibfnamefont {A.}~\bibnamefont {del Campo}},\
  and\ \bibinfo {author} {\bibfnamefont {A.~V.}\ \bibnamefont {Gorshkov}},\
  }\bibfield  {title} {\bibinfo {title} {Unifying quantum and classical speed
  limits on observables},\ }\href {https://doi.org/10.1103/PhysRevX.12.011038}
  {\bibfield  {journal} {\bibinfo  {journal} {Phys. Rev. X}\ }\textbf {\bibinfo
  {volume} {12}},\ \bibinfo {pages} {011038} (\bibinfo {year}
  {2022})}\BibitemShut {NoStop}%
\bibitem [{\citenamefont {Van~Vu}\ and\ \citenamefont
  {Saito}(2023{\natexlab{a}})}]{Vu23a}%
  \BibitemOpen
  \bibfield  {author} {\bibinfo {author} {\bibfnamefont {T.}~\bibnamefont
  {Van~Vu}}\ and\ \bibinfo {author} {\bibfnamefont {K.}~\bibnamefont {Saito}},\
  }\bibfield  {title} {\bibinfo {title} {Topological speed limit},\ }\href
  {https://doi.org/10.1103/PhysRevLett.130.010402} {\bibfield  {journal}
  {\bibinfo  {journal} {Phys. Rev. Lett.}\ }\textbf {\bibinfo {volume} {130}},\
  \bibinfo {pages} {010402} (\bibinfo {year} {2023}{\natexlab{a}})}\BibitemShut
  {NoStop}%
\bibitem [{\citenamefont {Van~Vu}\ and\ \citenamefont
  {Saito}(2023{\natexlab{b}})}]{Vu23b}%
  \BibitemOpen
  \bibfield  {author} {\bibinfo {author} {\bibfnamefont {T.}~\bibnamefont
  {Van~Vu}}\ and\ \bibinfo {author} {\bibfnamefont {K.}~\bibnamefont {Saito}},\
  }\bibfield  {title} {\bibinfo {title} {Thermodynamic unification of optimal
  transport: thermodynamic uncertainty relation, minimum dissipation, and
  thermodynamic speed limits},\ }\href
  {https://doi.org/10.1103/PhysRevX.13.011013} {\bibfield  {journal} {\bibinfo
  {journal} {Physical Review X}\ }\textbf {\bibinfo {volume} {13}},\ \bibinfo
  {pages} {011013} (\bibinfo {year} {2023}{\natexlab{b}})}\BibitemShut
  {NoStop}%
\bibitem [{\citenamefont {Neri}\ \emph {et~al.}(2017)\citenamefont {Neri},
  \citenamefont {Rold\'an},\ and\ \citenamefont {J\"ulicher}}]{Neri17}%
  \BibitemOpen
  \bibfield  {author} {\bibinfo {author} {\bibfnamefont {I.}~\bibnamefont
  {Neri}}, \bibinfo {author} {\bibfnamefont {E.}~\bibnamefont {Rold\'an}},\
  and\ \bibinfo {author} {\bibfnamefont {F.}~\bibnamefont {J\"ulicher}},\
  }\bibfield  {title} {\bibinfo {title} {Statistics of infima and stopping
  times of entropy production and applications to active molecular processes},\
  }\href {https://doi.org/10.1103/PhysRevX.7.011019} {\bibfield  {journal}
  {\bibinfo  {journal} {Phys. Rev. X}\ }\textbf {\bibinfo {volume} {7}},\
  \bibinfo {pages} {011019} (\bibinfo {year} {2017})}\BibitemShut {NoStop}%
\bibitem [{\citenamefont {Chétrite}\ \emph {et~al.}(2019)\citenamefont
  {Chétrite}, \citenamefont {Gupta}, \citenamefont {Neri},\ and\ \citenamefont
  {Rold\'an}}]{Chetrite18}%
  \BibitemOpen
  \bibfield  {author} {\bibinfo {author} {\bibfnamefont {R.}~\bibnamefont
  {Chétrite}}, \bibinfo {author} {\bibfnamefont {S.}~\bibnamefont {Gupta}},
  \bibinfo {author} {\bibfnamefont {I.}~\bibnamefont {Neri}},\ and\ \bibinfo
  {author} {\bibfnamefont {E.}~\bibnamefont {Rold\'an}},\ }\bibfield  {title}
  {\bibinfo {title} {Martingale theory for housekeeping heat},\ }\href
  {https://doi.org/10.1209/0295-5075/124/60006} {\bibfield  {journal} {\bibinfo
   {journal} {Europhysics Letters}\ }\textbf {\bibinfo {volume} {124}},\
  \bibinfo {pages} {60006} (\bibinfo {year} {2019})}\BibitemShut {NoStop}%
\bibitem [{\citenamefont {Manzano}\ \emph {et~al.}(2021)\citenamefont
  {Manzano}, \citenamefont {Subero}, \citenamefont {Maillet}, \citenamefont
  {Fazio}, \citenamefont {Pekola},\ and\ \citenamefont
  {Rold\'an}}]{gambling21}%
  \BibitemOpen
  \bibfield  {author} {\bibinfo {author} {\bibfnamefont {G.}~\bibnamefont
  {Manzano}}, \bibinfo {author} {\bibfnamefont {D.}~\bibnamefont {Subero}},
  \bibinfo {author} {\bibfnamefont {O.}~\bibnamefont {Maillet}}, \bibinfo
  {author} {\bibfnamefont {R.}~\bibnamefont {Fazio}}, \bibinfo {author}
  {\bibfnamefont {J.~P.}\ \bibnamefont {Pekola}},\ and\ \bibinfo {author}
  {\bibfnamefont {E.}~\bibnamefont {Rold\'an}},\ }\bibfield  {title} {\bibinfo
  {title} {Thermodynamics of gambling demons},\ }\href
  {https://doi.org/10.1103/PhysRevLett.126.080603} {\bibfield  {journal}
  {\bibinfo  {journal} {Phys. Rev. Lett.}\ }\textbf {\bibinfo {volume} {126}},\
  \bibinfo {pages} {080603} (\bibinfo {year} {2021})}\BibitemShut {NoStop}%
\bibitem [{\citenamefont {Manzano}\ and\ \citenamefont
  {Rold\'an}(2022)}]{survival22}%
  \BibitemOpen
  \bibfield  {author} {\bibinfo {author} {\bibfnamefont {G.}~\bibnamefont
  {Manzano}}\ and\ \bibinfo {author} {\bibfnamefont {E.}~\bibnamefont
  {Rold\'an}},\ }\bibfield  {title} {\bibinfo {title} {Survival and extreme
  statistics of work, heat, and entropy production in steady-state heat
  engines},\ }\href {https://doi.org/10.1103/PhysRevE.105.024112} {\bibfield
  {journal} {\bibinfo  {journal} {Phys. Rev. E}\ }\textbf {\bibinfo {volume}
  {105}},\ \bibinfo {pages} {024112} (\bibinfo {year} {2022})}\BibitemShut
  {NoStop}%
\bibitem [{\citenamefont {Édgar Roldán}\ \emph {et~al.}(2023)\citenamefont
  {Édgar Roldán}, \citenamefont {Neri}, \citenamefont {Chetrite},
  \citenamefont {Gupta}, \citenamefont {Pigolotti}, \citenamefont {Jülicher},\
  and\ \citenamefont {Sekimoto}}]{roldanMartingalesPhysicistsTreatise2023}%
  \BibitemOpen
  \bibfield  {author} {\bibinfo {author} {\bibnamefont {Édgar Roldán}},
  \bibinfo {author} {\bibfnamefont {I.}~\bibnamefont {Neri}}, \bibinfo {author}
  {\bibfnamefont {R.}~\bibnamefont {Chetrite}}, \bibinfo {author}
  {\bibfnamefont {S.}~\bibnamefont {Gupta}}, \bibinfo {author} {\bibfnamefont
  {S.}~\bibnamefont {Pigolotti}}, \bibinfo {author} {\bibfnamefont
  {F.}~\bibnamefont {Jülicher}},\ and\ \bibinfo {author} {\bibfnamefont
  {K.}~\bibnamefont {Sekimoto}},\ }\bibfield  {title} {\bibinfo {title}
  {Martingales for physicists: a treatise on stochastic thermodynamics
  and beyond},\ }\href {https://doi.org/10.1080/00018732.2024.2317494}
  {\bibfield  {journal} {\bibinfo  {journal} {Advances in Physics}\ }\textbf
  {\bibinfo {volume} {72}},\ \bibinfo {pages} {1} (\bibinfo {year}
  {2023})}\BibitemShut {NoStop}%
\bibitem [{\citenamefont {Weidlich}(1991)}]{Weidlich91}%
  \BibitemOpen
  \bibfield  {author} {\bibinfo {author} {\bibfnamefont {W.}~\bibnamefont
  {Weidlich}},\ }\bibfield  {title} {\bibinfo {title} {Physics and social
  science — the approach of synergetics},\ }\href
  {https://doi.org/https://doi.org/10.1016/0370-1573(91)90024-G} {\bibfield
  {journal} {\bibinfo  {journal} {Physics Reports}\ }\textbf {\bibinfo {volume}
  {204}},\ \bibinfo {pages} {1} (\bibinfo {year} {1991})}\BibitemShut {NoStop}%
\bibitem [{\citenamefont {Castellano}\ \emph
  {et~al.}(2009{\natexlab{a}})\citenamefont {Castellano}, \citenamefont
  {Fortunato},\ and\ \citenamefont {Loreto}}]{Castellano09}%
  \BibitemOpen
  \bibfield  {author} {\bibinfo {author} {\bibfnamefont {C.}~\bibnamefont
  {Castellano}}, \bibinfo {author} {\bibfnamefont {S.}~\bibnamefont
  {Fortunato}},\ and\ \bibinfo {author} {\bibfnamefont {V.}~\bibnamefont
  {Loreto}},\ }\bibfield  {title} {\bibinfo {title} {Statistical physics of
  social dynamics},\ }\href {https://doi.org/10.1103/RevModPhys.81.591}
  {\bibfield  {journal} {\bibinfo  {journal} {Rev. Mod. Phys.}\ }\textbf
  {\bibinfo {volume} {81}},\ \bibinfo {pages} {591} (\bibinfo {year}
  {2009}{\natexlab{a}})}\BibitemShut {NoStop}%
\bibitem [{\citenamefont {Starnini}\ \emph
  {et~al.}(2025{\natexlab{a}})\citenamefont {Starnini}, \citenamefont
  {Baumann}, \citenamefont {Galla}, \citenamefont {Garcia}, \citenamefont
  {Iñiguez}, \citenamefont {Karsai}, \citenamefont {Lorenz},\ and\
  \citenamefont {Sznajd-Weron}}]{Opinion_new_review}%
  \BibitemOpen
  \bibfield  {author} {\bibinfo {author} {\bibfnamefont {M.}~\bibnamefont
  {Starnini}}, \bibinfo {author} {\bibfnamefont {F.}~\bibnamefont {Baumann}},
  \bibinfo {author} {\bibfnamefont {T.}~\bibnamefont {Galla}}, \bibinfo
  {author} {\bibfnamefont {D.}~\bibnamefont {Garcia}}, \bibinfo {author}
  {\bibfnamefont {G.}~\bibnamefont {Iñiguez}}, \bibinfo {author}
  {\bibfnamefont {M.}~\bibnamefont {Karsai}}, \bibinfo {author} {\bibfnamefont
  {J.}~\bibnamefont {Lorenz}},\ and\ \bibinfo {author} {\bibfnamefont
  {K.}~\bibnamefont {Sznajd-Weron}},\ }\href {https://arxiv.org/abs/2507.11521}
  {\bibinfo {title} {Opinion dynamics: Statistical physics and beyond}}
  (\bibinfo {year} {2025}{\natexlab{a}}),\ \Eprint
  {https://arxiv.org/abs/2507.11521} {arXiv:2507.11521 [physics.soc-ph]}
  \BibitemShut {NoStop}%
\bibitem [{\citenamefont {Lazer}\ \emph {et~al.}(2020)\citenamefont {Lazer},
  \citenamefont {Pentland}, \citenamefont {Watts}, \citenamefont {Aral},
  \citenamefont {Athey}, \citenamefont {Contractor}, \citenamefont {Freelon},
  \citenamefont {Gonzalez-Bailon}, \citenamefont {King}, \citenamefont
  {Margetts}, \citenamefont {Nelson}, \citenamefont {Salganik}, \citenamefont
  {Strohmaier}, \citenamefont {Vespignani},\ and\ \citenamefont
  {Wagner}}]{Lazer20}%
  \BibitemOpen
  \bibfield  {author} {\bibinfo {author} {\bibfnamefont {D.~M.~J.}\
  \bibnamefont {Lazer}}, \bibinfo {author} {\bibfnamefont {A.}~\bibnamefont
  {Pentland}}, \bibinfo {author} {\bibfnamefont {D.~J.}\ \bibnamefont {Watts}},
  \bibinfo {author} {\bibfnamefont {S.}~\bibnamefont {Aral}}, \bibinfo {author}
  {\bibfnamefont {S.}~\bibnamefont {Athey}}, \bibinfo {author} {\bibfnamefont
  {N.}~\bibnamefont {Contractor}}, \bibinfo {author} {\bibfnamefont
  {D.}~\bibnamefont {Freelon}}, \bibinfo {author} {\bibfnamefont
  {S.}~\bibnamefont {Gonzalez-Bailon}}, \bibinfo {author} {\bibfnamefont
  {G.}~\bibnamefont {King}}, \bibinfo {author} {\bibfnamefont {H.}~\bibnamefont
  {Margetts}}, \bibinfo {author} {\bibfnamefont {A.}~\bibnamefont {Nelson}},
  \bibinfo {author} {\bibfnamefont {M.~J.}\ \bibnamefont {Salganik}}, \bibinfo
  {author} {\bibfnamefont {M.}~\bibnamefont {Strohmaier}}, \bibinfo {author}
  {\bibfnamefont {A.}~\bibnamefont {Vespignani}},\ and\ \bibinfo {author}
  {\bibfnamefont {C.}~\bibnamefont {Wagner}},\ }\bibfield  {title} {\bibinfo
  {title} {Computational social science: Obstacles and opportunities},\ }\href
  {https://doi.org/10.1126/science.aaz8170} {\bibfield  {journal} {\bibinfo
  {journal} {Science}\ }\textbf {\bibinfo {volume} {369}},\ \bibinfo {pages}
  {1060} (\bibinfo {year} {2020})}\BibitemShut {NoStop}%
\bibitem [{\citenamefont {Axelrod}(1997)}]{Axelrod97}%
  \BibitemOpen
  \bibfield  {author} {\bibinfo {author} {\bibfnamefont {R.}~\bibnamefont
  {Axelrod}},\ }\href {http://www.jstor.org/stable/j.ctt7s951} {\emph {\bibinfo
  {title} {The Complexity of Cooperation: Agent-Based Models of Competition and
  Collaboration}}}\ (\bibinfo  {publisher} {Princeton University Press},\
  \bibinfo {year} {1997})\BibitemShut {NoStop}%
\bibitem [{\citenamefont {Castellano}\ \emph {et~al.}(2000)\citenamefont
  {Castellano}, \citenamefont {Marsili},\ and\ \citenamefont
  {Vespignani}}]{Castellano00}%
  \BibitemOpen
  \bibfield  {author} {\bibinfo {author} {\bibfnamefont {C.}~\bibnamefont
  {Castellano}}, \bibinfo {author} {\bibfnamefont {M.}~\bibnamefont
  {Marsili}},\ and\ \bibinfo {author} {\bibfnamefont {A.}~\bibnamefont
  {Vespignani}},\ }\bibfield  {title} {\bibinfo {title} {Nonequilibrium phase
  transition in a model for social influence},\ }\href
  {https://doi.org/10.1103/PhysRevLett.85.3536} {\bibfield  {journal} {\bibinfo
   {journal} {Phys. Rev. Lett.}\ }\textbf {\bibinfo {volume} {85}},\ \bibinfo
  {pages} {3536} (\bibinfo {year} {2000})}\BibitemShut {NoStop}%
\bibitem [{\citenamefont {Centola}\ \emph {et~al.}(2007)\citenamefont
  {Centola}, \citenamefont {González-Avella}, \citenamefont {Eguíluz},\ and\
  \citenamefont {Miguel}}]{Centola07}%
  \BibitemOpen
  \bibfield  {author} {\bibinfo {author} {\bibfnamefont {D.}~\bibnamefont
  {Centola}}, \bibinfo {author} {\bibfnamefont {J.~C.}\ \bibnamefont
  {González-Avella}}, \bibinfo {author} {\bibfnamefont {V.~M.}\ \bibnamefont
  {Eguíluz}},\ and\ \bibinfo {author} {\bibfnamefont {M.~S.}\ \bibnamefont
  {Miguel}},\ }\bibfield  {title} {\bibinfo {title} {Homophily, cultural drift,
  and the co-evolution of cultural groups},\ }\href
  {https://doi.org/10.1177/0022002707307632} {\bibfield  {journal} {\bibinfo
  {journal} {Journal of Conflict Resolution}\ }\textbf {\bibinfo {volume}
  {51}},\ \bibinfo {pages} {905} (\bibinfo {year} {2007})}\BibitemShut
  {NoStop}%
\bibitem [{\citenamefont {Dandekar}\ \emph {et~al.}(2013)\citenamefont
  {Dandekar}, \citenamefont {Goel},\ and\ \citenamefont {Lee}}]{Dandekar13}%
  \BibitemOpen
  \bibfield  {author} {\bibinfo {author} {\bibfnamefont {P.}~\bibnamefont
  {Dandekar}}, \bibinfo {author} {\bibfnamefont {A.}~\bibnamefont {Goel}},\
  and\ \bibinfo {author} {\bibfnamefont {D.~T.}\ \bibnamefont {Lee}},\
  }\bibfield  {title} {\bibinfo {title} {Biased assimilation, homophily, and
  the dynamics of polarization},\ }\href
  {https://doi.org/10.1073/pnas.1217220110} {\bibfield  {journal} {\bibinfo
  {journal} {Proceedings of the National Academy of Sciences}\ }\textbf
  {\bibinfo {volume} {110}},\ \bibinfo {pages} {5791} (\bibinfo {year}
  {2013})}\BibitemShut {NoStop}%
\bibitem [{\citenamefont {Fern\'andez-Gracia}\ \emph
  {et~al.}(2014)\citenamefont {Fern\'andez-Gracia}, \citenamefont {Suchecki},
  \citenamefont {Ramasco}, \citenamefont {San~Miguel},\ and\ \citenamefont
  {Egu\'{\i}luz}}]{Juanili14}%
  \BibitemOpen
  \bibfield  {author} {\bibinfo {author} {\bibfnamefont {J.}~\bibnamefont
  {Fern\'andez-Gracia}}, \bibinfo {author} {\bibfnamefont {K.}~\bibnamefont
  {Suchecki}}, \bibinfo {author} {\bibfnamefont {J.~J.}\ \bibnamefont
  {Ramasco}}, \bibinfo {author} {\bibfnamefont {M.}~\bibnamefont
  {San~Miguel}},\ and\ \bibinfo {author} {\bibfnamefont {V.~M.}\ \bibnamefont
  {Egu\'{\i}luz}},\ }\bibfield  {title} {\bibinfo {title} {Is the voter model a
  model for voters?},\ }\href {https://doi.org/10.1103/PhysRevLett.112.158701}
  {\bibfield  {journal} {\bibinfo  {journal} {Phys. Rev. Lett.}\ }\textbf
  {\bibinfo {volume} {112}},\ \bibinfo {pages} {158701} (\bibinfo {year}
  {2014})}\BibitemShut {NoStop}%
\bibitem [{\citenamefont {Granha}\ \emph {et~al.}(2022)\citenamefont {Granha},
  \citenamefont {Vilela}, \citenamefont {Wang}, \citenamefont {Nelson},\ and\
  \citenamefont {Stanley}}]{Granha22}%
  \BibitemOpen
  \bibfield  {author} {\bibinfo {author} {\bibfnamefont {M.~F.~B.}\
  \bibnamefont {Granha}}, \bibinfo {author} {\bibfnamefont {A.~L.~M.}\
  \bibnamefont {Vilela}}, \bibinfo {author} {\bibfnamefont {C.}~\bibnamefont
  {Wang}}, \bibinfo {author} {\bibfnamefont {K.~P.}\ \bibnamefont {Nelson}},\
  and\ \bibinfo {author} {\bibfnamefont {H.~E.}\ \bibnamefont {Stanley}},\
  }\bibfield  {title} {\bibinfo {title} {Opinion dynamics in financial markets
  via random networks},\ }\href {https://doi.org/10.1073/pnas.2201573119}
  {\bibfield  {journal} {\bibinfo  {journal} {Proceedings of the National
  Academy of Sciences}\ }\textbf {\bibinfo {volume} {119}},\ \bibinfo {pages}
  {e2201573119} (\bibinfo {year} {2022})}\BibitemShut {NoStop}%
\bibitem [{\citenamefont {Ojer}\ \emph {et~al.}(2023)\citenamefont {Ojer},
  \citenamefont {Starnini},\ and\ \citenamefont {Pastor-Satorras}}]{Ojer23}%
  \BibitemOpen
  \bibfield  {author} {\bibinfo {author} {\bibfnamefont {J.}~\bibnamefont
  {Ojer}}, \bibinfo {author} {\bibfnamefont {M.}~\bibnamefont {Starnini}},\
  and\ \bibinfo {author} {\bibfnamefont {R.}~\bibnamefont {Pastor-Satorras}},\
  }\bibfield  {title} {\bibinfo {title} {Modeling explosive opinion
  depolarization in interdependent topics},\ }\href
  {https://doi.org/10.1103/PhysRevLett.130.207401} {\bibfield  {journal}
  {\bibinfo  {journal} {Phys. Rev. Lett.}\ }\textbf {\bibinfo {volume} {130}},\
  \bibinfo {pages} {207401} (\bibinfo {year} {2023})}\BibitemShut {NoStop}%
\bibitem [{\citenamefont {Starnini}\ \emph
  {et~al.}(2025{\natexlab{b}})\citenamefont {Starnini}, \citenamefont
  {Baumann}, \citenamefont {Galla}, \citenamefont {Garcia}, \citenamefont
  {Iñiguez}, \citenamefont {Karsai}, \citenamefont {Lorenz},\ and\
  \citenamefont {Sznajd-Weron}}]{Starnini25}%
  \BibitemOpen
  \bibfield  {author} {\bibinfo {author} {\bibfnamefont {M.}~\bibnamefont
  {Starnini}}, \bibinfo {author} {\bibfnamefont {F.}~\bibnamefont {Baumann}},
  \bibinfo {author} {\bibfnamefont {T.}~\bibnamefont {Galla}}, \bibinfo
  {author} {\bibfnamefont {D.}~\bibnamefont {Garcia}}, \bibinfo {author}
  {\bibfnamefont {G.}~\bibnamefont {Iñiguez}}, \bibinfo {author}
  {\bibfnamefont {M.}~\bibnamefont {Karsai}}, \bibinfo {author} {\bibfnamefont
  {J.}~\bibnamefont {Lorenz}},\ and\ \bibinfo {author} {\bibfnamefont
  {K.}~\bibnamefont {Sznajd-Weron}},\ }\href {https://arxiv.org/abs/2507.11521}
  {\bibinfo {title} {Opinion dynamics: Statistical physics and beyond}}
  (\bibinfo {year} {2025}{\natexlab{b}}),\ \Eprint
  {https://arxiv.org/abs/2507.11521} {arXiv:2507.11521 [physics.soc-ph]}
  \BibitemShut {NoStop}%
\bibitem [{\citenamefont {Abrams}\ and\ \citenamefont
  {Strogatz}(2003)}]{Abrams03}%
  \BibitemOpen
  \bibfield  {author} {\bibinfo {author} {\bibfnamefont {D.~M.}\ \bibnamefont
  {Abrams}}\ and\ \bibinfo {author} {\bibfnamefont {S.~H.}\ \bibnamefont
  {Strogatz}},\ }\bibfield  {title} {\bibinfo {title} {Modelling the dynamics
  of language death},\ }\href@noop {} {\bibfield  {journal} {\bibinfo
  {journal} {Nature}\ }\textbf {\bibinfo {volume} {424}},\ \bibinfo {pages}
  {900} (\bibinfo {year} {2003})}\BibitemShut {NoStop}%
\bibitem [{\citenamefont {Castelló}\ \emph {et~al.}(2006)\citenamefont
  {Castelló}, \citenamefont {Eguíluz},\ and\ \citenamefont
  {Miguel}}]{Castello06}%
  \BibitemOpen
  \bibfield  {author} {\bibinfo {author} {\bibfnamefont {X.}~\bibnamefont
  {Castelló}}, \bibinfo {author} {\bibfnamefont {V.~M.}\ \bibnamefont
  {Eguíluz}},\ and\ \bibinfo {author} {\bibfnamefont {M.~S.}\ \bibnamefont
  {Miguel}},\ }\bibfield  {title} {\bibinfo {title} {Ordering dynamics with two
  non-excluding options: bilingualism in language competition},\ }\href
  {https://doi.org/10.1088/1367-2630/8/12/308} {\bibfield  {journal} {\bibinfo
  {journal} {New Journal of Physics}\ }\textbf {\bibinfo {volume} {8}},\
  \bibinfo {pages} {308} (\bibinfo {year} {2006})}\BibitemShut {NoStop}%
\bibitem [{\citenamefont {Rao}\ and\ \citenamefont {Esposito}(2016)}]{Rao16}%
  \BibitemOpen
  \bibfield  {author} {\bibinfo {author} {\bibfnamefont {R.}~\bibnamefont
  {Rao}}\ and\ \bibinfo {author} {\bibfnamefont {M.}~\bibnamefont {Esposito}},\
  }\bibfield  {title} {\bibinfo {title} {Nonequilibrium thermodynamics of
  chemical reaction networks: Wisdom from stochastic thermodynamics},\ }\href
  {https://doi.org/10.1103/PhysRevX.6.041064} {\bibfield  {journal} {\bibinfo
  {journal} {Phys. Rev. X}\ }\textbf {\bibinfo {volume} {6}},\ \bibinfo {pages}
  {041064} (\bibinfo {year} {2016})}\BibitemShut {NoStop}%
\bibitem [{\citenamefont {Noa}\ \emph {et~al.}(2019)\citenamefont {Noa},
  \citenamefont {Harunari}, \citenamefont {{de Oliveira}},\ and\ \citenamefont
  {Fiore}}]{noaEntropyProductionTool2019}%
  \BibitemOpen
  \bibfield  {author} {\bibinfo {author} {\bibfnamefont {C.~E.~F.}\
  \bibnamefont {Noa}}, \bibinfo {author} {\bibfnamefont {P.~E.}\ \bibnamefont
  {Harunari}}, \bibinfo {author} {\bibfnamefont {M.~J.}\ \bibnamefont {{de
  Oliveira}}},\ and\ \bibinfo {author} {\bibfnamefont {C.~E.}\ \bibnamefont
  {Fiore}},\ }\bibfield  {title} {\bibinfo {title} {Entropy production as a
  tool for characterizing nonequilibrium phase transitions},\ }\href
  {https://doi.org/10.1103/PhysRevE.100.012104} {\bibfield  {journal} {\bibinfo
   {journal} {Physical Review E}\ }\textbf {\bibinfo {volume} {100}},\ \bibinfo
  {pages} {012104} (\bibinfo {year} {2019})}\BibitemShut {NoStop}%
\bibitem [{\citenamefont {{da Silva}}\ \emph {et~al.}(2020)\citenamefont {{da
  Silva}}, \citenamefont {{de Oliveira}}, \citenamefont {Tom{\'e}},\ and\
  \citenamefont {{Drugowich de
  Fel{\'i}cio}}}]{dasilvaAnalysisEarlierTimes2020}%
  \BibitemOpen
  \bibfield  {author} {\bibinfo {author} {\bibfnamefont {R.}~\bibnamefont {{da
  Silva}}}, \bibinfo {author} {\bibfnamefont {M.~J.}\ \bibnamefont {{de
  Oliveira}}}, \bibinfo {author} {\bibfnamefont {T.}~\bibnamefont {Tom{\'e}}},\
  and\ \bibinfo {author} {\bibfnamefont {J.~R.}\ \bibnamefont {{Drugowich de
  Fel{\'i}cio}}},\ }\bibfield  {title} {\bibinfo {title} {Analysis of earlier
  times and flux of entropy on the majority voter model with diffusion},\
  }\href {https://doi.org/10.1103/PhysRevE.101.012130} {\bibfield  {journal}
  {\bibinfo  {journal} {Physical Review E}\ }\textbf {\bibinfo {volume}
  {101}},\ \bibinfo {pages} {012130} (\bibinfo {year} {2020})}\BibitemShut
  {NoStop}%
\bibitem [{\citenamefont {Freitas}\ \emph {et~al.}(2021)\citenamefont
  {Freitas}, \citenamefont {Delvenne},\ and\ \citenamefont
  {Esposito}}]{Freitas21}%
  \BibitemOpen
  \bibfield  {author} {\bibinfo {author} {\bibfnamefont {N.}~\bibnamefont
  {Freitas}}, \bibinfo {author} {\bibfnamefont {J.-C.}\ \bibnamefont
  {Delvenne}},\ and\ \bibinfo {author} {\bibfnamefont {M.}~\bibnamefont
  {Esposito}},\ }\bibfield  {title} {\bibinfo {title} {Stochastic
  thermodynamics of nonlinear electronic circuits: A realistic framework for
  computing around $kt$},\ }\href {https://doi.org/10.1103/PhysRevX.11.031064}
  {\bibfield  {journal} {\bibinfo  {journal} {Phys. Rev. X}\ }\textbf {\bibinfo
  {volume} {11}},\ \bibinfo {pages} {031064} (\bibinfo {year}
  {2021})}\BibitemShut {NoStop}%
\bibitem [{\citenamefont {Korbel}\ \emph {et~al.}(2021)\citenamefont {Korbel},
  \citenamefont {Lindner}, \citenamefont {Hanel},\ and\ \citenamefont
  {Thurner}}]{Korbel21}%
  \BibitemOpen
  \bibfield  {author} {\bibinfo {author} {\bibfnamefont {J.}~\bibnamefont
  {Korbel}}, \bibinfo {author} {\bibfnamefont {S.~D.}\ \bibnamefont {Lindner}},
  \bibinfo {author} {\bibfnamefont {R.}~\bibnamefont {Hanel}},\ and\ \bibinfo
  {author} {\bibfnamefont {S.}~\bibnamefont {Thurner}},\ }\bibfield  {title}
  {\bibinfo {title} {Thermodynamics of structure-forming systems},\ }\href
  {https://doi.org/10.1038/s41467-021-21272-7} {\bibfield  {journal} {\bibinfo
  {journal} {Nature Communications}\ }\textbf {\bibinfo {volume} {12}},\
  \bibinfo {pages} {1} (\bibinfo {year} {2021})}\BibitemShut {NoStop}%
\bibitem [{\citenamefont {Rao}\ and\ \citenamefont {Leibler}(2022)}]{Rao22}%
  \BibitemOpen
  \bibfield  {author} {\bibinfo {author} {\bibfnamefont {R.}~\bibnamefont
  {Rao}}\ and\ \bibinfo {author} {\bibfnamefont {S.}~\bibnamefont {Leibler}},\
  }\bibfield  {title} {\bibinfo {title} {Evolutionary dynamics, evolutionary
  forces, and robustness: A nonequilibrium statistical mechanics perspective},\
  }\href {https://doi.org/10.1073/pnas.2112083119} {\bibfield  {journal}
  {\bibinfo  {journal} {Proceedings of the National Academy of Sciences}\
  }\textbf {\bibinfo {volume} {119}},\ \bibinfo {pages} {e2112083119} (\bibinfo
  {year} {2022})}\BibitemShut {NoStop}%
\bibitem [{\citenamefont {Manzano}\ \emph {et~al.}(2024)\citenamefont
  {Manzano}, \citenamefont {Karde{\c s}}, \citenamefont {Rold{\'a}n},\ and\
  \citenamefont {Wolpert}}]{manzanoThermodynamicsComputationsAbsolute2024}%
  \BibitemOpen
  \bibfield  {author} {\bibinfo {author} {\bibfnamefont {G.}~\bibnamefont
  {Manzano}}, \bibinfo {author} {\bibfnamefont {G.}~\bibnamefont {Karde{\c
  s}}}, \bibinfo {author} {\bibfnamefont {{\'E}.}~\bibnamefont {Rold{\'a}n}},\
  and\ \bibinfo {author} {\bibfnamefont {D.~H.}\ \bibnamefont {Wolpert}},\
  }\bibfield  {title} {\bibinfo {title} {Thermodynamics of {{Computations}}
  with {{Absolute Irreversibility}}, {{Unidirectional Transitions}}, and
  {{Stochastic Computation Times}}},\ }\href
  {https://doi.org/10.1103/PhysRevX.14.021026} {\bibfield  {journal} {\bibinfo
  {journal} {Physical Review X}\ }\textbf {\bibinfo {volume} {14}},\ \bibinfo
  {pages} {021026} (\bibinfo {year} {2024})}\BibitemShut {NoStop}%
\bibitem [{\citenamefont {Sorkin}\ \emph {et~al.}(2024)\citenamefont {Sorkin},
  \citenamefont {Diamant}, \citenamefont {Ariel},\ and\ \citenamefont
  {Markovich}}]{Sorkin24}%
  \BibitemOpen
  \bibfield  {author} {\bibinfo {author} {\bibfnamefont {B.}~\bibnamefont
  {Sorkin}}, \bibinfo {author} {\bibfnamefont {H.}~\bibnamefont {Diamant}},
  \bibinfo {author} {\bibfnamefont {G.}~\bibnamefont {Ariel}},\ and\ \bibinfo
  {author} {\bibfnamefont {T.}~\bibnamefont {Markovich}},\ }\bibfield  {title}
  {\bibinfo {title} {{Second Law of Thermodynamics without Einstein
  Relation}},\ }\href {https://doi.org/10.1103/PhysRevLett.133.267101}
  {\bibfield  {journal} {\bibinfo  {journal} {Phys. Rev. Lett.}\ }\textbf
  {\bibinfo {volume} {133}},\ \bibinfo {pages} {267101} (\bibinfo {year}
  {2024})}\BibitemShut {NoStop}%
\bibitem [{\citenamefont {Wolpert}\ \emph {et~al.}(2024)\citenamefont
  {Wolpert}, \citenamefont {Korbel}, \citenamefont {Lynn}, \citenamefont
  {Tasnim}, \citenamefont {Grochow}, \citenamefont {Kardeş}, \citenamefont
  {Aimone}, \citenamefont {Balasubramanian}, \citenamefont {Giuli},
  \citenamefont {Doty}, \citenamefont {Freitas}, \citenamefont {Marsili},
  \citenamefont {Ouldridge}, \citenamefont {Richa}, \citenamefont {Riechers},
  \citenamefont {Rold\'an}, \citenamefont {Rubenstein}, \citenamefont
  {Toroczkai},\ and\ \citenamefont {Paradiso}}]{Wolpert24}%
  \BibitemOpen
  \bibfield  {author} {\bibinfo {author} {\bibfnamefont {D.~H.}\ \bibnamefont
  {Wolpert}}, \bibinfo {author} {\bibfnamefont {J.}~\bibnamefont {Korbel}},
  \bibinfo {author} {\bibfnamefont {C.~W.}\ \bibnamefont {Lynn}}, \bibinfo
  {author} {\bibfnamefont {F.}~\bibnamefont {Tasnim}}, \bibinfo {author}
  {\bibfnamefont {J.~A.}\ \bibnamefont {Grochow}}, \bibinfo {author}
  {\bibfnamefont {G.}~\bibnamefont {Kardeş}}, \bibinfo {author} {\bibfnamefont
  {J.~B.}\ \bibnamefont {Aimone}}, \bibinfo {author} {\bibfnamefont
  {V.}~\bibnamefont {Balasubramanian}}, \bibinfo {author} {\bibfnamefont
  {E.~D.}\ \bibnamefont {Giuli}}, \bibinfo {author} {\bibfnamefont
  {D.}~\bibnamefont {Doty}}, \bibinfo {author} {\bibfnamefont {N.}~\bibnamefont
  {Freitas}}, \bibinfo {author} {\bibfnamefont {M.}~\bibnamefont {Marsili}},
  \bibinfo {author} {\bibfnamefont {T.~E.}\ \bibnamefont {Ouldridge}}, \bibinfo
  {author} {\bibfnamefont {A.~W.}\ \bibnamefont {Richa}}, \bibinfo {author}
  {\bibfnamefont {P.}~\bibnamefont {Riechers}}, \bibinfo {author}
  {\bibfnamefont {E.}~\bibnamefont {Rold\'an}}, \bibinfo {author}
  {\bibfnamefont {B.}~\bibnamefont {Rubenstein}}, \bibinfo {author}
  {\bibfnamefont {Z.}~\bibnamefont {Toroczkai}},\ and\ \bibinfo {author}
  {\bibfnamefont {J.}~\bibnamefont {Paradiso}},\ }\bibfield  {title} {\bibinfo
  {title} {Is stochastic thermodynamics the key to understanding the energy
  costs of computation?},\ }\href {https://doi.org/10.1073/pnas.2321112121}
  {\bibfield  {journal} {\bibinfo  {journal} {Proceedings of the National
  Academy of Sciences}\ }\textbf {\bibinfo {volume} {121}},\ \bibinfo {pages}
  {e2321112121} (\bibinfo {year} {2024})}\BibitemShut {NoStop}%
\bibitem [{\citenamefont {Falasco}\ and\ \citenamefont
  {Esposito}(2025)}]{Falasco25}%
  \BibitemOpen
  \bibfield  {author} {\bibinfo {author} {\bibfnamefont {G.}~\bibnamefont
  {Falasco}}\ and\ \bibinfo {author} {\bibfnamefont {M.}~\bibnamefont
  {Esposito}},\ }\bibfield  {title} {\bibinfo {title} {Macroscopic stochastic
  thermodynamics},\ }\href {https://doi.org/10.1103/RevModPhys.97.015002}
  {\bibfield  {journal} {\bibinfo  {journal} {Rev. Mod. Phys.}\ }\textbf
  {\bibinfo {volume} {97}},\ \bibinfo {pages} {015002} (\bibinfo {year}
  {2025})}\BibitemShut {NoStop}%
\bibitem [{\citenamefont {Bahr}\ and\ \citenamefont
  {Passerini}(1998{\natexlab{a}})}]{BahrMicro}%
  \BibitemOpen
  \bibfield  {author} {\bibinfo {author} {\bibfnamefont {D.~B.}\ \bibnamefont
  {Bahr}}\ and\ \bibinfo {author} {\bibfnamefont {E.}~\bibnamefont
  {Passerini}},\ }\bibfield  {title} {\bibinfo {title} {Statistical mechanics
  of opinion formation and collective behavior: Micro‐sociology},\ }\href
  {https://doi.org/10.1080/0022250X.1998.9990210} {\bibfield  {journal}
  {\bibinfo  {journal} {The Journal of Mathematical Sociology}\ }\textbf
  {\bibinfo {volume} {23}},\ \bibinfo {pages} {1} (\bibinfo {year}
  {1998}{\natexlab{a}})}\BibitemShut {NoStop}%
\bibitem [{\citenamefont {Bahr}\ and\ \citenamefont
  {Passerini}(1998{\natexlab{b}})}]{BahrMacro}%
  \BibitemOpen
  \bibfield  {author} {\bibinfo {author} {\bibfnamefont {D.~B.}\ \bibnamefont
  {Bahr}}\ and\ \bibinfo {author} {\bibfnamefont {E.}~\bibnamefont
  {Passerini}},\ }\bibfield  {title} {\bibinfo {title} {Statistical mechanics
  of collective behavior: Macro‐sociology},\ }\href
  {https://doi.org/10.1080/0022250X.1998.9990211} {\bibfield  {journal}
  {\bibinfo  {journal} {The Journal of Mathematical Sociology}\ }\textbf
  {\bibinfo {volume} {23}},\ \bibinfo {pages} {29} (\bibinfo {year}
  {1998}{\natexlab{b}})}\BibitemShut {NoStop}%
\bibitem [{\citenamefont {Crochik}\ and\ \citenamefont
  {Tom{\'e}}(2005)}]{crochikEntropyProductionMajorityvote2005}%
  \BibitemOpen
  \bibfield  {author} {\bibinfo {author} {\bibfnamefont {L.}~\bibnamefont
  {Crochik}}\ and\ \bibinfo {author} {\bibfnamefont {T.}~\bibnamefont
  {Tom{\'e}}},\ }\bibfield  {title} {\bibinfo {title} {Entropy production in
  the majority-vote model},\ }\href
  {https://doi.org/10.1103/PhysRevE.72.057103} {\bibfield  {journal} {\bibinfo
  {journal} {Physical Review E}\ }\textbf {\bibinfo {volume} {72}},\ \bibinfo
  {pages} {057103} (\bibinfo {year} {2005})}\BibitemShut {NoStop}%
\bibitem [{\citenamefont {Tom{\'e}}\ \emph {et~al.}(2023)\citenamefont
  {Tom{\'e}}, \citenamefont {Fiore},\ and\ \citenamefont {{de
  Oliveira}}}]{tomeStochasticThermodynamicsOpinion2023}%
  \BibitemOpen
  \bibfield  {author} {\bibinfo {author} {\bibfnamefont {T.}~\bibnamefont
  {Tom{\'e}}}, \bibinfo {author} {\bibfnamefont {C.~E.}\ \bibnamefont
  {Fiore}},\ and\ \bibinfo {author} {\bibfnamefont {M.~J.}\ \bibnamefont {{de
  Oliveira}}},\ }\bibfield  {title} {\bibinfo {title} {Stochastic
  thermodynamics of opinion dynamics models},\ }\href
  {https://doi.org/10.1103/PhysRevE.107.064135} {\bibfield  {journal} {\bibinfo
   {journal} {Physical Review E}\ }\textbf {\bibinfo {volume} {107}},\ \bibinfo
  {pages} {064135} (\bibinfo {year} {2023})}\BibitemShut {NoStop}%
\bibitem [{\citenamefont {Hawthorne}\ \emph {et~al.}(2023)\citenamefont
  {Hawthorne}, \citenamefont {Harunari}, \citenamefont {{de Oliveira}},\ and\
  \citenamefont {Fiore}}]{hawthorneNonequilibriumThermodynamicsMajority2023}%
  \BibitemOpen
  \bibfield  {author} {\bibinfo {author} {\bibfnamefont {F.}~\bibnamefont
  {Hawthorne}}, \bibinfo {author} {\bibfnamefont {P.~E.}\ \bibnamefont
  {Harunari}}, \bibinfo {author} {\bibfnamefont {M.~J.}\ \bibnamefont {{de
  Oliveira}}},\ and\ \bibinfo {author} {\bibfnamefont {C.~E.}\ \bibnamefont
  {Fiore}},\ }\bibfield  {title} {\bibinfo {title} {Nonequilibrium
  {{Thermodynamics}} of the {{Majority Vote Model}}},\ }\href
  {https://doi.org/10.3390/e25081230} {\bibfield  {journal} {\bibinfo
  {journal} {Entropy}\ }\textbf {\bibinfo {volume} {25}},\ \bibinfo {pages}
  {1230} (\bibinfo {year} {2023})}\BibitemShut {NoStop}%
\bibitem [{\citenamefont {Oliveira}\ \emph {et~al.}(2024)\citenamefont
  {Oliveira}, \citenamefont {Wang}, \citenamefont {Dong}, \citenamefont {Du},
  \citenamefont {Fiore}, \citenamefont {Vilela},\ and\ \citenamefont
  {Stanley}}]{oliveiraEntropyProductionCooperative2024}%
  \BibitemOpen
  \bibfield  {author} {\bibinfo {author} {\bibfnamefont {I.~V.~G.}\
  \bibnamefont {Oliveira}}, \bibinfo {author} {\bibfnamefont {C.}~\bibnamefont
  {Wang}}, \bibinfo {author} {\bibfnamefont {G.}~\bibnamefont {Dong}}, \bibinfo
  {author} {\bibfnamefont {R.}~\bibnamefont {Du}}, \bibinfo {author}
  {\bibfnamefont {C.~E.}\ \bibnamefont {Fiore}}, \bibinfo {author}
  {\bibfnamefont {A.~L.~M.}\ \bibnamefont {Vilela}},\ and\ \bibinfo {author}
  {\bibfnamefont {H.~E.}\ \bibnamefont {Stanley}},\ }\bibfield  {title}
  {\bibinfo {title} {Entropy production on cooperative opinion dynamics},\
  }\href {https://doi.org/10.1016/j.chaos.2024.114694} {\bibfield  {journal}
  {\bibinfo  {journal} {Chaos, Solitons \& Fractals}\ }\textbf {\bibinfo
  {volume} {181}},\ \bibinfo {pages} {114694} (\bibinfo {year}
  {2024})}\BibitemShut {NoStop}%
\bibitem [{\citenamefont {Boltzmann}(2023)}]{Boltzmann}%
  \BibitemOpen
  \bibfield  {author} {\bibinfo {author} {\bibfnamefont {L.}~\bibnamefont
  {Boltzmann}},\ }\href {https://doi.org/doi:10.1525/9780520327474} {\emph
  {\bibinfo {title} {Lectures on Gas Theory}}},\ edited by\ \bibinfo {editor}
  {\bibfnamefont {S.~G.}\ \bibnamefont {Brush}}\ (\bibinfo  {publisher}
  {University of California Press},\ \bibinfo {address} {Berkeley},\ \bibinfo
  {year} {2023})\BibitemShut {NoStop}%
\bibitem [{\citenamefont {Rold{\'a}n}\ \emph {et~al.}(2014)\citenamefont
  {Rold{\'a}n}, \citenamefont {Mart{\'\i}nez}, \citenamefont {Parrondo},\ and\
  \citenamefont {Petrov}}]{Roldan2014Jun}%
  \BibitemOpen
  \bibfield  {author} {\bibinfo {author} {\bibfnamefont {E.}~\bibnamefont
  {Rold{\'a}n}}, \bibinfo {author} {\bibfnamefont {I.~A.}\ \bibnamefont
  {Mart{\'\i}nez}}, \bibinfo {author} {\bibfnamefont {J.~M.~R.}\ \bibnamefont
  {Parrondo}},\ and\ \bibinfo {author} {\bibfnamefont {D.}~\bibnamefont
  {Petrov}},\ }\bibfield  {title} {\bibinfo {title} {{Universal features in the
  energetics of symmetry breaking}},\ }\href
  {https://doi.org/10.1038/nphys2940} {\bibfield  {journal} {\bibinfo
  {journal} {Nat. Phys.}\ }\textbf {\bibinfo {volume} {10}},\ \bibinfo {pages}
  {457} (\bibinfo {year} {2014})}\BibitemShut {NoStop}%
\bibitem [{\citenamefont {Castellano}\ \emph
  {et~al.}(2009{\natexlab{b}})\citenamefont {Castellano}, \citenamefont
  {Mu{\~n}oz},\ and\ \citenamefont
  {{Pastor-Satorras}}}]{castellanoNonlinear$q$voterModel2009}%
  \BibitemOpen
  \bibfield  {author} {\bibinfo {author} {\bibfnamefont {C.}~\bibnamefont
  {Castellano}}, \bibinfo {author} {\bibfnamefont {M.~A.}\ \bibnamefont
  {Mu{\~n}oz}},\ and\ \bibinfo {author} {\bibfnamefont {R.}~\bibnamefont
  {{Pastor-Satorras}}},\ }\bibfield  {title} {\bibinfo {title} {Nonlinear
  $q$-voter model},\ }\href {https://doi.org/10.1103/PhysRevE.80.041129}
  {\bibfield  {journal} {\bibinfo  {journal} {Physical Review E}\ }\textbf
  {\bibinfo {volume} {80}},\ \bibinfo {pages} {041129} (\bibinfo {year}
  {2009}{\natexlab{b}})}\BibitemShut {NoStop}%
\bibitem [{\citenamefont {Holley}\ and\ \citenamefont
  {Liggett}(1975)}]{Holley1975}%
  \BibitemOpen
  \bibfield  {author} {\bibinfo {author} {\bibfnamefont {R.~A.}\ \bibnamefont
  {Holley}}\ and\ \bibinfo {author} {\bibfnamefont {T.~M.}\ \bibnamefont
  {Liggett}},\ }\bibfield  {title} {\bibinfo {title} {Ergodic theorems for
  weakly interacting infinite systems and the voter model},\ }\href
  {http://www.jstor.org/stable/2959329} {\bibfield  {journal} {\bibinfo
  {journal} {The Annals of Probability}\ }\textbf {\bibinfo {volume} {3}},\
  \bibinfo {pages} {643} (\bibinfo {year} {1975})}\BibitemShut {NoStop}%
\bibitem [{\citenamefont {Clifford}\ and\ \citenamefont
  {Sudbury}(1973)}]{Clifford1973}%
  \BibitemOpen
  \bibfield  {author} {\bibinfo {author} {\bibfnamefont {P.}~\bibnamefont
  {Clifford}}\ and\ \bibinfo {author} {\bibfnamefont {A.}~\bibnamefont
  {Sudbury}},\ }\bibfield  {title} {\bibinfo {title} {A model for spatial
  conflict},\ }\href {http://www.jstor.org/stable/2335008} {\bibfield
  {journal} {\bibinfo  {journal} {Biometrika}\ }\textbf {\bibinfo {volume}
  {60}},\ \bibinfo {pages} {581} (\bibinfo {year} {1973})}\BibitemShut
  {NoStop}%
\bibitem [{\citenamefont {Nyczka}\ \emph {et~al.}(2012)\citenamefont {Nyczka},
  \citenamefont {{Sznajd-Weron}},\ and\ \citenamefont
  {Cis{\l}o}}]{nyczkaPhaseTransitions$q$voter2012}%
  \BibitemOpen
  \bibfield  {author} {\bibinfo {author} {\bibfnamefont {P.}~\bibnamefont
  {Nyczka}}, \bibinfo {author} {\bibfnamefont {K.}~\bibnamefont
  {{Sznajd-Weron}}},\ and\ \bibinfo {author} {\bibfnamefont {J.}~\bibnamefont
  {Cis{\l}o}},\ }\bibfield  {title} {\bibinfo {title} {Phase transitions in the
  $q$-voter model with two types of stochastic driving},\ }\href
  {https://doi.org/10.1103/PhysRevE.86.011105} {\bibfield  {journal} {\bibinfo
  {journal} {Physical Review E}\ }\textbf {\bibinfo {volume} {86}},\ \bibinfo
  {pages} {011105} (\bibinfo {year} {2012})}\BibitemShut {NoStop}%
\bibitem [{\citenamefont {Nyczka}\ and\ \citenamefont
  {{Sznajd-Weron}}(2013)}]{nyczka2013}%
  \BibitemOpen
  \bibfield  {author} {\bibinfo {author} {\bibfnamefont {P.}~\bibnamefont
  {Nyczka}}\ and\ \bibinfo {author} {\bibfnamefont {K.}~\bibnamefont
  {{Sznajd-Weron}}},\ }\bibfield  {title} {\bibinfo {title} {Anticonformity or
  {{Independence}}?---{{Insights}} from {{Statistical Physics}}},\ }\href
  {https://doi.org/10.1007/s10955-013-0701-4} {\bibfield  {journal} {\bibinfo
  {journal} {Journal of Statistical Physics}\ }\textbf {\bibinfo {volume}
  {151}},\ \bibinfo {pages} {174} (\bibinfo {year} {2013})}\BibitemShut
  {NoStop}%
\bibitem [{\citenamefont {Kirman}(1993)}]{Kirman93}%
  \BibitemOpen
  \bibfield  {author} {\bibinfo {author} {\bibfnamefont {A.}~\bibnamefont
  {Kirman}},\ }\bibfield  {title} {\bibinfo {title} {Ants, rationality, and
  recruitment},\ }\href {https://doi.org/10.2307/2118498} {\bibfield  {journal}
  {\bibinfo  {journal} {The Quarterly Journal of Economics}\ }\textbf {\bibinfo
  {volume} {108}},\ \bibinfo {pages} {137} (\bibinfo {year}
  {1993})}\BibitemShut {NoStop}%
\bibitem [{\citenamefont {Granovsky}\ and\ \citenamefont
  {Madras}(1995)}]{Granovsky95}%
  \BibitemOpen
  \bibfield  {author} {\bibinfo {author} {\bibfnamefont {B.~L.}\ \bibnamefont
  {Granovsky}}\ and\ \bibinfo {author} {\bibfnamefont {N.}~\bibnamefont
  {Madras}},\ }\bibfield  {title} {\bibinfo {title} {The noisy voter model},\
  }\href {https://doi.org/https://doi.org/10.1016/0304-4149(94)00035-R}
  {\bibfield  {journal} {\bibinfo  {journal} {Stochastic Processes and their
  Applications}\ }\textbf {\bibinfo {volume} {55}},\ \bibinfo {pages} {23}
  (\bibinfo {year} {1995})}\BibitemShut {NoStop}%
\bibitem [{\citenamefont {Siedlecki}\ \emph {et~al.}(2016)\citenamefont
  {Siedlecki}, \citenamefont {Szwabi{\'n}ski},\ and\ \citenamefont
  {Weron}}]{siedleckiInterplayConformityAnticonformity2016}%
  \BibitemOpen
  \bibfield  {author} {\bibinfo {author} {\bibfnamefont {P.}~\bibnamefont
  {Siedlecki}}, \bibinfo {author} {\bibfnamefont {J.}~\bibnamefont
  {Szwabi{\'n}ski}},\ and\ \bibinfo {author} {\bibfnamefont {T.}~\bibnamefont
  {Weron}},\ }\bibfield  {title} {\bibinfo {title} {The {{Interplay Between
  Conformity}} and {{Anticonformity}} and its {{Polarizing Effect}} on
  {{Society}}},\ }\href {https://doi.org/10.18564/jasss.3203} {\bibfield
  {journal} {\bibinfo  {journal} {Journal of Artificial Societies and Social
  Simulation}\ }\textbf {\bibinfo {volume} {19}},\ \bibinfo {pages} {9}
  (\bibinfo {year} {2016})}\BibitemShut {NoStop}%
\bibitem [{\citenamefont {Peralta}\ \emph
  {et~al.}(2018{\natexlab{a}})\citenamefont {Peralta}, \citenamefont {Carro},
  \citenamefont {San~Miguel},\ and\ \citenamefont
  {Toral}}]{peraltaAnalyticalNumericalStudy2018a}%
  \BibitemOpen
  \bibfield  {author} {\bibinfo {author} {\bibfnamefont {A.~F.}\ \bibnamefont
  {Peralta}}, \bibinfo {author} {\bibfnamefont {A.}~\bibnamefont {Carro}},
  \bibinfo {author} {\bibfnamefont {M.}~\bibnamefont {San~Miguel}},\ and\
  \bibinfo {author} {\bibfnamefont {R.}~\bibnamefont {Toral}},\ }\bibfield
  {title} {\bibinfo {title} {Analytical and numerical study of the non-linear
  noisy voter model on complex networks},\ }\href
  {https://doi.org/10.1063/1.5030112} {\bibfield  {journal} {\bibinfo
  {journal} {Chaos: An Interdisciplinary Journal of Nonlinear Science}\
  }\textbf {\bibinfo {volume} {28}},\ \bibinfo {pages} {075516} (\bibinfo
  {year} {2018}{\natexlab{a}})}\BibitemShut {NoStop}%
\bibitem [{\citenamefont {Ramirez}\ \emph {et~al.}(2024)\citenamefont
  {Ramirez}, \citenamefont {Vazquez}, \citenamefont {San~Miguel},\ and\
  \citenamefont {Galla}}]{ramirezOrderingDynamicsNonlinear2024}%
  \BibitemOpen
  \bibfield  {author} {\bibinfo {author} {\bibfnamefont {L.~S.}\ \bibnamefont
  {Ramirez}}, \bibinfo {author} {\bibfnamefont {F.}~\bibnamefont {Vazquez}},
  \bibinfo {author} {\bibfnamefont {M.}~\bibnamefont {San~Miguel}},\ and\
  \bibinfo {author} {\bibfnamefont {T.}~\bibnamefont {Galla}},\ }\bibfield
  {title} {\bibinfo {title} {Ordering dynamics of nonlinear voter models},\
  }\href {https://doi.org/10.1103/PhysRevE.109.034307} {\bibfield  {journal}
  {\bibinfo  {journal} {Physical Review E}\ }\textbf {\bibinfo {volume}
  {109}},\ \bibinfo {pages} {034307} (\bibinfo {year} {2024})}\BibitemShut
  {NoStop}%
\bibitem [{\citenamefont {J\ifmmode~\mbox{\k{e}}\else \k{e}\fi{}drzejewski}\
  and\ \citenamefont {Mendes}(2025)}]{Mendes2025}%
  \BibitemOpen
  \bibfield  {author} {\bibinfo {author} {\bibfnamefont {A.}~\bibnamefont
  {J\ifmmode~\mbox{\k{e}}\else \k{e}\fi{}drzejewski}}\ and\ \bibinfo {author}
  {\bibfnamefont {J.~F.~F.}\ \bibnamefont {Mendes}},\ }\bibfield  {title}
  {\bibinfo {title} {When does population diversity matter? a unified framework
  for binary-choice dynamics},\ }\href {https://doi.org/10.1103/4db2-7dpd}
  {\bibfield  {journal} {\bibinfo  {journal} {Phys. Rev. Lett.}\ }\textbf
  {\bibinfo {volume} {135}},\ \bibinfo {pages} {217401} (\bibinfo {year}
  {2025})}\BibitemShut {NoStop}%
\bibitem [{\citenamefont {van Kampen}(1992)}]{Kampen92}%
  \BibitemOpen
  \bibfield  {author} {\bibinfo {author} {\bibfnamefont {N.}~\bibnamefont {van
  Kampen}},\ }\href@noop {} {\emph {\bibinfo {title} {{S}tochastic {P}rocesses
  in {P}hysics and {C}hemistry}}}\ (\bibinfo  {publisher} {Elsevier Science
  Publishers, Amsterdam},\ \bibinfo {year} {1992})\BibitemShut {NoStop}%
\bibitem [{\citenamefont {Toral}\ and\ \citenamefont
  {Colet}(2014)}]{Toral2014StochasticNM}%
  \BibitemOpen
  \bibfield  {author} {\bibinfo {author} {\bibfnamefont {R.}~\bibnamefont
  {Toral}}\ and\ \bibinfo {author} {\bibfnamefont {P.}~\bibnamefont {Colet}},\
  }\href@noop {} {\emph {\bibinfo {title} {Stochastic Numerical Methods: An
  Introduction for Students and Scientists}}}\ (\bibinfo  {publisher}
  {Wiley-VCH},\ \bibinfo {year} {2014})\BibitemShut {NoStop}%
\bibitem [{\citenamefont {Esposito}\ \emph {et~al.}(2010)\citenamefont
  {Esposito}, \citenamefont {Lindenberg},\ and\ \citenamefont {Van~den
  Broeck}}]{Esposito10}%
  \BibitemOpen
  \bibfield  {author} {\bibinfo {author} {\bibfnamefont {M.}~\bibnamefont
  {Esposito}}, \bibinfo {author} {\bibfnamefont {K.}~\bibnamefont
  {Lindenberg}},\ and\ \bibinfo {author} {\bibfnamefont {C.}~\bibnamefont
  {Van~den Broeck}},\ }\bibfield  {title} {\bibinfo {title} {Entropy production
  as correlation between system and reservoir},\ }\href
  {https://doi.org/10.1088/1367-2630/12/1/013013} {\bibfield  {journal}
  {\bibinfo  {journal} {New Journal of Physics}\ }\textbf {\bibinfo {volume}
  {12}},\ \bibinfo {pages} {013013} (\bibinfo {year} {2010})}\BibitemShut
  {NoStop}%
\bibitem [{\citenamefont {Manzano}\ \emph {et~al.}(2018)\citenamefont
  {Manzano}, \citenamefont {Horowitz},\ and\ \citenamefont
  {Parrondo}}]{Manzano18}%
  \BibitemOpen
  \bibfield  {author} {\bibinfo {author} {\bibfnamefont {G.}~\bibnamefont
  {Manzano}}, \bibinfo {author} {\bibfnamefont {J.~M.}\ \bibnamefont
  {Horowitz}},\ and\ \bibinfo {author} {\bibfnamefont {J.~M.~R.}\ \bibnamefont
  {Parrondo}},\ }\bibfield  {title} {\bibinfo {title} {Quantum fluctuation
  theorems for arbitrary environments: Adiabatic and nonadiabatic entropy
  production},\ }\href {https://doi.org/10.1103/PhysRevX.8.031037} {\bibfield
  {journal} {\bibinfo  {journal} {Phys. Rev. X}\ }\textbf {\bibinfo {volume}
  {8}},\ \bibinfo {pages} {031037} (\bibinfo {year} {2018})}\BibitemShut
  {NoStop}%
\bibitem [{\citenamefont {Bauer}\ and\ \citenamefont {Cornu}(2014)}]{Bauer15}%
  \BibitemOpen
  \bibfield  {author} {\bibinfo {author} {\bibfnamefont {M.}~\bibnamefont
  {Bauer}}\ and\ \bibinfo {author} {\bibfnamefont {F.}~\bibnamefont {Cornu}},\
  }\bibfield  {title} {\bibinfo {title} {Local detailed balance: a microscopic
  derivation},\ }\href {https://doi.org/10.1088/1751-8113/48/1/015008}
  {\bibfield  {journal} {\bibinfo  {journal} {Journal of Physics A:
  Mathematical and Theoretical}\ }\textbf {\bibinfo {volume} {48}},\ \bibinfo
  {pages} {015008} (\bibinfo {year} {2014})}\BibitemShut {NoStop}%
\bibitem [{\citenamefont {Maes}(2021)}]{Maes21}%
  \BibitemOpen
  \bibfield  {author} {\bibinfo {author} {\bibfnamefont {C.}~\bibnamefont
  {Maes}},\ }\bibfield  {title} {\bibinfo {title} {{Local detailed balance}},\
  }\href {https://doi.org/10.21468/SciPostPhysLectNotes.32} {\bibfield
  {journal} {\bibinfo  {journal} {SciPost Phys. Lect. Notes}\ ,\ \bibinfo
  {pages} {32}} (\bibinfo {year} {2021})}\BibitemShut {NoStop}%
\bibitem [{\citenamefont {Falasco}\ and\ \citenamefont
  {Esposito}(2021)}]{Falasco21}%
  \BibitemOpen
  \bibfield  {author} {\bibinfo {author} {\bibfnamefont {G.}~\bibnamefont
  {Falasco}}\ and\ \bibinfo {author} {\bibfnamefont {M.}~\bibnamefont
  {Esposito}},\ }\bibfield  {title} {\bibinfo {title} {Local detailed balance
  across scales: From diffusions to jump processes and beyond},\ }\href
  {https://doi.org/10.1103/PhysRevE.103.042114} {\bibfield  {journal} {\bibinfo
   {journal} {Phys. Rev. E}\ }\textbf {\bibinfo {volume} {103}},\ \bibinfo
  {pages} {042114} (\bibinfo {year} {2021})}\BibitemShut {NoStop}%
\bibitem [{\citenamefont {Nowak}\ and\ \citenamefont
  {Sznajd-Weron}(2020)}]{Nowak2020}%
  \BibitemOpen
  \bibfield  {author} {\bibinfo {author} {\bibfnamefont {B.}~\bibnamefont
  {Nowak}}\ and\ \bibinfo {author} {\bibfnamefont {K.}~\bibnamefont
  {Sznajd-Weron}},\ }\bibfield  {title} {\bibinfo {title} {Symmetrical
  threshold model with independence on random graphs},\ }\href
  {https://doi.org/10.1103/PhysRevE.101.052316} {\bibfield  {journal} {\bibinfo
   {journal} {Phys. Rev. E}\ }\textbf {\bibinfo {volume} {101}},\ \bibinfo
  {pages} {052316} (\bibinfo {year} {2020})}\BibitemShut {NoStop}%
\bibitem [{\citenamefont {Nowak}\ \emph {et~al.}(2022)\citenamefont {Nowak},
  \citenamefont {Grabisch},\ and\ \citenamefont {Sznajd-Weron}}]{Nowak2022}%
  \BibitemOpen
  \bibfield  {author} {\bibinfo {author} {\bibfnamefont {B.}~\bibnamefont
  {Nowak}}, \bibinfo {author} {\bibfnamefont {M.}~\bibnamefont {Grabisch}},\
  and\ \bibinfo {author} {\bibfnamefont {K.}~\bibnamefont {Sznajd-Weron}},\
  }\bibfield  {title} {\bibinfo {title} {Threshold model with anticonformity
  under random sequential updating},\ }\href
  {https://doi.org/10.1103/PhysRevE.105.054314} {\bibfield  {journal} {\bibinfo
   {journal} {Phys. Rev. E}\ }\textbf {\bibinfo {volume} {105}},\ \bibinfo
  {pages} {054314} (\bibinfo {year} {2022})}\BibitemShut {NoStop}%
\bibitem [{\citenamefont {Castell\'o}\ \emph {et~al.}(2006)\citenamefont
  {Castell\'o}, \citenamefont {Egu\'{\i}luz},\ and\ \citenamefont {{San
  Miguel}}}]{Castello_2006}%
  \BibitemOpen
  \bibfield  {author} {\bibinfo {author} {\bibfnamefont {X.}~\bibnamefont
  {Castell\'o}}, \bibinfo {author} {\bibfnamefont {V.~M.}\ \bibnamefont
  {Egu\'{\i}luz}},\ and\ \bibinfo {author} {\bibfnamefont {M.}~\bibnamefont
  {{San Miguel}}},\ }\bibfield  {title} {\bibinfo {title} {Ordering dynamics
  with two non-excluding options: bilingualism in language competition},\
  }\href {https://doi.org/10.1088/1367-2630/8/12/308} {\bibfield  {journal}
  {\bibinfo  {journal} {New Journal of Physics}\ }\textbf {\bibinfo {volume}
  {8}},\ \bibinfo {pages} {308} (\bibinfo {year} {2006})}\BibitemShut {NoStop}%
\bibitem [{\citenamefont {Galam}(2002)}]{Galam2002}%
  \BibitemOpen
  \bibfield  {author} {\bibinfo {author} {\bibfnamefont {S.}~\bibnamefont
  {Galam}},\ }\bibfield  {title} {\bibinfo {title} {Minority opinion spreading
  in random geometry},\ }\href {https://doi.org/10.1140/epjb/e20020045}
  {\bibfield  {journal} {\bibinfo  {journal} {The European Physical Journal B -
  Condensed Matter and Complex Systems}\ }\textbf {\bibinfo {volume} {25}},\
  \bibinfo {pages} {403} (\bibinfo {year} {2002})}\BibitemShut {NoStop}%
\bibitem [{\citenamefont {de~Oliveira}(1992)}]{Oliveira1992}%
  \BibitemOpen
  \bibfield  {author} {\bibinfo {author} {\bibfnamefont {M.~J.}\ \bibnamefont
  {de~Oliveira}},\ }\bibfield  {title} {\bibinfo {title} {Isotropic
  majority-vote model on a square lattice},\ }\href
  {https://doi.org/10.1007/BF01060069} {\bibfield  {journal} {\bibinfo
  {journal} {Journal of Statistical Physics}\ }\textbf {\bibinfo {volume}
  {66}},\ \bibinfo {pages} {273} (\bibinfo {year} {1992})}\BibitemShut
  {NoStop}%
\bibitem [{\citenamefont {Krause}\ \emph {et~al.}(2012)\citenamefont {Krause},
  \citenamefont {B\"ottcher},\ and\ \citenamefont {Bornholdt}}]{Krause2012}%
  \BibitemOpen
  \bibfield  {author} {\bibinfo {author} {\bibfnamefont {S.~M.}\ \bibnamefont
  {Krause}}, \bibinfo {author} {\bibfnamefont {P.}~\bibnamefont {B\"ottcher}},\
  and\ \bibinfo {author} {\bibfnamefont {S.}~\bibnamefont {Bornholdt}},\
  }\bibfield  {title} {\bibinfo {title} {Mean-field-like behavior of the
  generalized voter-model-class kinetic {Ising} model},\ }\href
  {https://doi.org/10.1103/PhysRevE.85.031126} {\bibfield  {journal} {\bibinfo
  {journal} {Phys. Rev. E}\ }\textbf {\bibinfo {volume} {85}},\ \bibinfo
  {pages} {031126} (\bibinfo {year} {2012})}\BibitemShut {NoStop}%
\bibitem [{Note1()}]{Note1}%
  \BibitemOpen
  \bibinfo {note} {The introduction of the Boltzmann constant $k$ in the
  definition of entropy is unnecessary when dealing with systems which do not
  have a clear thermodynamic interpretation and hence we take $k=1$
  throughout.}\BibitemShut {Stop}%
\bibitem [{\citenamefont {Guryanova}\ \emph {et~al.}(2016)\citenamefont
  {Guryanova}, \citenamefont {Popescu}, \citenamefont {Short}, \citenamefont
  {Silva},\ and\ \citenamefont {Skrzypczyk}}]{Guryanova16}%
  \BibitemOpen
  \bibfield  {author} {\bibinfo {author} {\bibfnamefont {Y.}~\bibnamefont
  {Guryanova}}, \bibinfo {author} {\bibfnamefont {S.}~\bibnamefont {Popescu}},
  \bibinfo {author} {\bibfnamefont {A.~J.}\ \bibnamefont {Short}}, \bibinfo
  {author} {\bibfnamefont {R.}~\bibnamefont {Silva}},\ and\ \bibinfo {author}
  {\bibfnamefont {P.}~\bibnamefont {Skrzypczyk}},\ }\bibfield  {title}
  {\bibinfo {title} {Thermodynamics of quantum systems with multiple conserved
  quantities},\ }\href@noop {} {\bibfield  {journal} {\bibinfo  {journal}
  {Nature communications}\ }\textbf {\bibinfo {volume} {7}},\ \bibinfo {pages}
  {12049} (\bibinfo {year} {2016})}\BibitemShut {NoStop}%
\bibitem [{\citenamefont {Rao}\ and\ \citenamefont {Esposito}(2018)}]{Rao18}%
  \BibitemOpen
  \bibfield  {author} {\bibinfo {author} {\bibfnamefont {R.}~\bibnamefont
  {Rao}}\ and\ \bibinfo {author} {\bibfnamefont {M.}~\bibnamefont {Esposito}},\
  }\bibfield  {title} {\bibinfo {title} {Conservation laws shape dissipation},\
  }\href {https://doi.org/10.1088/1367-2630/aaa15f} {\bibfield  {journal}
  {\bibinfo  {journal} {New Journal of Physics}\ }\textbf {\bibinfo {volume}
  {20}},\ \bibinfo {pages} {023007} (\bibinfo {year} {2018})}\BibitemShut
  {NoStop}%
\bibitem [{Note2()}]{Note2}%
  \BibitemOpen
  \bibinfo {note} {Technically speaking, when the two reactions act
  simultaneously, the differences in generalized chemical potentials make the
  second term in the exponent to induce a non-conservative force, no longer
  derivable from a potential~\cite {Rao18}.}\BibitemShut {Stop}%
\bibitem [{\citenamefont {Maes}(2020)}]{Maes20}%
  \BibitemOpen
  \bibfield  {author} {\bibinfo {author} {\bibfnamefont {C.}~\bibnamefont
  {Maes}},\ }\bibfield  {title} {\bibinfo {title} {Frenesy: Time-symmetric
  dynamical activity in nonequilibria},\ }\href
  {https://doi.org/https://doi.org/10.1016/j.physrep.2020.01.002} {\bibfield
  {journal} {\bibinfo  {journal} {Physics Reports}\ }\textbf {\bibinfo {volume}
  {850}},\ \bibinfo {pages} {1} (\bibinfo {year} {2020})}\BibitemShut {NoStop}%
\bibitem [{Note3()}]{Note3}%
  \BibitemOpen
  \bibinfo {note} {We will generically set the time-scales of the dynamics by
  setting $\omega = 1$}\BibitemShut {NoStop}%
\bibitem [{Note4()}]{Note4}%
  \BibitemOpen
  \bibinfo {note} {When the number of individuals is finite, the use of the
  term \protect \textit {phase transition} is clearly an abuse of notation, as
  a truly symmetry-breaking phase transition can only occur in the limit $N \to
  \infty $, where the distribution becomes sharply peaked, approaching a sum of
  delta functions centered at its maxima~\cite {Toral:2011}.}\BibitemShut
  {Stop}%
\bibitem [{Note5()}]{Note5}%
  \BibitemOpen
  \bibinfo {note} {We will use this approximation for most of our analytical
  inquiry, for which we avoid carrying the symbols $\simeq $ to ease the
  notation.}\BibitemShut {Stop}%
\bibitem [{Note6()}]{Note6}%
  \BibitemOpen
  \bibinfo {note} {For simplicity we often refer to the agent's attribute as an
  \protect \textit {opinion}, but we keep in mind that the attribute can
  correspond to a generic social or cultural trait of the agents}\BibitemShut
  {NoStop}%
\bibitem [{Note7()}]{Note7}%
  \BibitemOpen
  \bibinfo {note} {An arbitrary number of reactions or jumps involving more
  than one agent change in opinion can be naturally incorporated in the
  framework.}\BibitemShut {Stop}%
\bibitem [{\citenamefont {Manzano}\ and\ \citenamefont
  {Zambrini}(2022)}]{Manzano22}%
  \BibitemOpen
  \bibfield  {author} {\bibinfo {author} {\bibfnamefont {G.}~\bibnamefont
  {Manzano}}\ and\ \bibinfo {author} {\bibfnamefont {R.}~\bibnamefont
  {Zambrini}},\ }\bibfield  {title} {\bibinfo {title} {Quantum thermodynamics
  under continuous monitoring: A general framework},\ }\href
  {https://doi.org/10.1116/5.0079886} {\bibfield  {journal} {\bibinfo
  {journal} {AVS Quantum Science}\ }\textbf {\bibinfo {volume} {4}},\ \bibinfo
  {pages} {025302} (\bibinfo {year} {2022})}\BibitemShut {NoStop}%
\bibitem [{\citenamefont {Hiura}\ and\ \citenamefont {Sasa}(2021)}]{Hiura21}%
  \BibitemOpen
  \bibfield  {author} {\bibinfo {author} {\bibfnamefont {K.}~\bibnamefont
  {Hiura}}\ and\ \bibinfo {author} {\bibfnamefont {S.-i.}\ \bibnamefont
  {Sasa}},\ }\bibfield  {title} {\bibinfo {title} {Kinetic uncertainty relation
  on first-passage time for accumulated current},\ }\href
  {https://doi.org/10.1103/PhysRevE.103.L050103} {\bibfield  {journal}
  {\bibinfo  {journal} {Phys. Rev. E}\ }\textbf {\bibinfo {volume} {103}},\
  \bibinfo {pages} {L050103} (\bibinfo {year} {2021})}\BibitemShut {NoStop}%
\bibitem [{\citenamefont {Mart{\'\i}nez}\ \emph {et~al.}(2019)\citenamefont
  {Mart{\'\i}nez}, \citenamefont {Bisker}, \citenamefont {Horowitz},\ and\
  \citenamefont {Parrondo}}]{Martinez19}%
  \BibitemOpen
  \bibfield  {author} {\bibinfo {author} {\bibfnamefont {I.~A.}\ \bibnamefont
  {Mart{\'\i}nez}}, \bibinfo {author} {\bibfnamefont {G.}~\bibnamefont
  {Bisker}}, \bibinfo {author} {\bibfnamefont {J.~M.}\ \bibnamefont
  {Horowitz}},\ and\ \bibinfo {author} {\bibfnamefont {J.~M.}\ \bibnamefont
  {Parrondo}},\ }\bibfield  {title} {\bibinfo {title} {Inferring broken
  detailed balance in the absence of observable currents},\ }\href@noop {}
  {\bibfield  {journal} {\bibinfo  {journal} {Nature communications}\ }\textbf
  {\bibinfo {volume} {10}},\ \bibinfo {pages} {3542} (\bibinfo {year}
  {2019})}\BibitemShut {NoStop}%
\bibitem [{\citenamefont {van~der Meer}\ \emph {et~al.}(2022)\citenamefont
  {van~der Meer}, \citenamefont {Ertel},\ and\ \citenamefont
  {Seifert}}]{VdMeer22}%
  \BibitemOpen
  \bibfield  {author} {\bibinfo {author} {\bibfnamefont {J.}~\bibnamefont
  {van~der Meer}}, \bibinfo {author} {\bibfnamefont {B.}~\bibnamefont
  {Ertel}},\ and\ \bibinfo {author} {\bibfnamefont {U.}~\bibnamefont
  {Seifert}},\ }\bibfield  {title} {\bibinfo {title} {Thermodynamic inference
  in partially accessible markov networks: A unifying perspective from
  transition-based waiting time distributions},\ }\href
  {https://doi.org/10.1103/PhysRevX.12.031025} {\bibfield  {journal} {\bibinfo
  {journal} {Phys. Rev. X}\ }\textbf {\bibinfo {volume} {12}},\ \bibinfo
  {pages} {031025} (\bibinfo {year} {2022})}\BibitemShut {NoStop}%
\bibitem [{\citenamefont {Harunari}\ \emph {et~al.}(2022)\citenamefont
  {Harunari}, \citenamefont {Dutta}, \citenamefont {Polettini},\ and\
  \citenamefont {Rold\'an}}]{Harunari22}%
  \BibitemOpen
  \bibfield  {author} {\bibinfo {author} {\bibfnamefont {P.~E.}\ \bibnamefont
  {Harunari}}, \bibinfo {author} {\bibfnamefont {A.}~\bibnamefont {Dutta}},
  \bibinfo {author} {\bibfnamefont {M.}~\bibnamefont {Polettini}},\ and\
  \bibinfo {author} {\bibfnamefont {E.}~\bibnamefont {Rold\'an}},\ }\bibfield
  {title} {\bibinfo {title} {What to learn from a few visible transitions'
  statistics?},\ }\href {https://doi.org/10.1103/PhysRevX.12.041026} {\bibfield
   {journal} {\bibinfo  {journal} {Phys. Rev. X}\ }\textbf {\bibinfo {volume}
  {12}},\ \bibinfo {pages} {041026} (\bibinfo {year} {2022})}\BibitemShut
  {NoStop}%
\bibitem [{\citenamefont {Crooks}(1999)}]{Crooks99}%
  \BibitemOpen
  \bibfield  {author} {\bibinfo {author} {\bibfnamefont {G.~E.}\ \bibnamefont
  {Crooks}},\ }\bibfield  {title} {\bibinfo {title} {Entropy production
  fluctuation theorem and the nonequilibrium work relation for free energy
  differences},\ }\href {https://doi.org/10.1103/PhysRevE.60.2721} {\bibfield
  {journal} {\bibinfo  {journal} {Phys. Rev. E}\ }\textbf {\bibinfo {volume}
  {60}},\ \bibinfo {pages} {2721} (\bibinfo {year} {1999})}\BibitemShut
  {NoStop}%
\bibitem [{\citenamefont {{Van den Broeck}}\ and\ \citenamefont
  {Esposito}(2015)}]{vandenbroeckEnsembleTrajectoryThermodynamics2015}%
  \BibitemOpen
  \bibfield  {author} {\bibinfo {author} {\bibfnamefont {C.}~\bibnamefont {{Van
  den Broeck}}}\ and\ \bibinfo {author} {\bibfnamefont {M.}~\bibnamefont
  {Esposito}},\ }\bibfield  {title} {\bibinfo {title} {Ensemble and trajectory
  thermodynamics: {{A}} brief introduction},\ }\href
  {https://doi.org/10.1016/j.physa.2014.04.035} {\bibfield  {journal} {\bibinfo
   {journal} {Physica A: Statistical Mechanics and its Applications}\ }\bibinfo
  {series} {Proceedings of the 13th {{International Summer School}} on
  {{Fundamental Problems}} in {{Statistical Physics}}},\ \textbf {\bibinfo
  {volume} {418}},\ \bibinfo {pages} {6} (\bibinfo {year} {2015})}\BibitemShut
  {NoStop}%
\bibitem [{\citenamefont {Seifert}(2005)}]{Seifert05}%
  \BibitemOpen
  \bibfield  {author} {\bibinfo {author} {\bibfnamefont {U.}~\bibnamefont
  {Seifert}},\ }\bibfield  {title} {\bibinfo {title} {Entropy production along
  a stochastic trajectory and an integral fluctuation theorem},\ }\href
  {https://doi.org/10.1103/PhysRevLett.95.040602} {\bibfield  {journal}
  {\bibinfo  {journal} {Phys. Rev. Lett.}\ }\textbf {\bibinfo {volume} {95}},\
  \bibinfo {pages} {040602} (\bibinfo {year} {2005})}\BibitemShut {NoStop}%
\bibitem [{\citenamefont {Herpich}\ \emph {et~al.}(2020)\citenamefont
  {Herpich}, \citenamefont {Cossetto}, \citenamefont {Falasco},\ and\
  \citenamefont {Esposito}}]{Herpich2020Jun}%
  \BibitemOpen
  \bibfield  {author} {\bibinfo {author} {\bibfnamefont {T.}~\bibnamefont
  {Herpich}}, \bibinfo {author} {\bibfnamefont {T.}~\bibnamefont {Cossetto}},
  \bibinfo {author} {\bibfnamefont {G.}~\bibnamefont {Falasco}},\ and\ \bibinfo
  {author} {\bibfnamefont {M.}~\bibnamefont {Esposito}},\ }\bibfield  {title}
  {\bibinfo {title} {{Stochastic thermodynamics of all-to-all interacting
  many-body systems}},\ }\href {https://doi.org/10.1088/1367-2630/ab882f}
  {\bibfield  {journal} {\bibinfo  {journal} {New J. Phys.}\ }\textbf {\bibinfo
  {volume} {22}},\ \bibinfo {pages} {063005} (\bibinfo {year}
  {2020})}\BibitemShut {NoStop}%
\bibitem [{\citenamefont {Seifert}(2019)}]{Seifert19}%
  \BibitemOpen
  \bibfield  {author} {\bibinfo {author} {\bibfnamefont {U.}~\bibnamefont
  {Seifert}},\ }\bibfield  {title} {\bibinfo {title} {From stochastic
  thermodynamics to thermodynamic inference},\ }\href
  {https://doi.org/https://doi.org/10.1146/annurev-conmatphys-031218-013554}
  {\bibfield  {journal} {\bibinfo  {journal} {Annual Review of Condensed Matter
  Physics}\ }\textbf {\bibinfo {volume} {10}},\ \bibinfo {pages} {171}
  (\bibinfo {year} {2019})}\BibitemShut {NoStop}%
\bibitem [{\citenamefont {Cover}\ and\ \citenamefont
  {Thomas}(2006)}]{Cover2006}%
  \BibitemOpen
  \bibfield  {author} {\bibinfo {author} {\bibfnamefont {T.~M.}\ \bibnamefont
  {Cover}}\ and\ \bibinfo {author} {\bibfnamefont {J.~A.}\ \bibnamefont
  {Thomas}},\ }\href@noop {} {\emph {\bibinfo {title} {Elements of Information
  Theory 2nd Edition (Wiley Series in Telecommunications and Signal
  Processing)}}}\ (\bibinfo  {publisher} {Wiley-Interscience},\ \bibinfo {year}
  {2006})\BibitemShut {NoStop}%
\bibitem [{\citenamefont {{Maes}}\ and\ \citenamefont
  {{Neto{\v{c}}n{\'y}}}(2003)}]{Maes03}%
  \BibitemOpen
  \bibfield  {author} {\bibinfo {author} {\bibfnamefont {C.}~\bibnamefont
  {{Maes}}}\ and\ \bibinfo {author} {\bibfnamefont {K.}~\bibnamefont
  {{Neto{\v{c}}n{\'y}}}},\ }\bibfield  {title} {\bibinfo {title}
  {{Time-Reversal and Entropy}},\ }\href
  {https://doi.org/10.1023/A:1021026930129} {\bibfield  {journal} {\bibinfo
  {journal} {Journal of Statistical Physics}\ }\textbf {\bibinfo {volume}
  {110}},\ \bibinfo {pages} {269} (\bibinfo {year} {2003})}\BibitemShut
  {NoStop}%
\bibitem [{\citenamefont {Kawai}\ \emph {et~al.}(2007)\citenamefont {Kawai},
  \citenamefont {Parrondo},\ and\ \citenamefont {den Broeck}}]{Kawai07}%
  \BibitemOpen
  \bibfield  {author} {\bibinfo {author} {\bibfnamefont {R.}~\bibnamefont
  {Kawai}}, \bibinfo {author} {\bibfnamefont {J.~M.~R.}\ \bibnamefont
  {Parrondo}},\ and\ \bibinfo {author} {\bibfnamefont {C.~V.}\ \bibnamefont
  {den Broeck}},\ }\bibfield  {title} {\bibinfo {title} {Dissipation: The
  phase-space perspective},\ }\href
  {https://doi.org/10.1103/PhysRevLett.98.080602} {\bibfield  {journal}
  {\bibinfo  {journal} {Phys. Rev. Lett.}\ }\textbf {\bibinfo {volume} {98}},\
  \bibinfo {pages} {080602} (\bibinfo {year} {2007})}\BibitemShut {NoStop}%
\bibitem [{\citenamefont {Gomez-Marin}\ \emph {et~al.}(2008)\citenamefont
  {Gomez-Marin}, \citenamefont {Parrondo},\ and\ \citenamefont {Van~den
  Broeck}}]{Gomez-Marin08}%
  \BibitemOpen
  \bibfield  {author} {\bibinfo {author} {\bibfnamefont {A.}~\bibnamefont
  {Gomez-Marin}}, \bibinfo {author} {\bibfnamefont {J.~M.~R.}\ \bibnamefont
  {Parrondo}},\ and\ \bibinfo {author} {\bibfnamefont {C.}~\bibnamefont
  {Van~den Broeck}},\ }\bibfield  {title} {\bibinfo {title} {The
  “footprints” of irreversibility},\ }\href
  {https://doi.org/10.1209/0295-5075/82/50002} {\bibfield  {journal} {\bibinfo
  {journal} {Europhysics Letters}\ }\textbf {\bibinfo {volume} {82}},\ \bibinfo
  {pages} {50002} (\bibinfo {year} {2008})}\BibitemShut {NoStop}%
\bibitem [{\citenamefont {Rold\'an}\ and\ \citenamefont
  {Parrondo}(2010)}]{Roldan10}%
  \BibitemOpen
  \bibfield  {author} {\bibinfo {author} {\bibfnamefont {E.}~\bibnamefont
  {Rold\'an}}\ and\ \bibinfo {author} {\bibfnamefont {J.~M.~R.}\ \bibnamefont
  {Parrondo}},\ }\bibfield  {title} {\bibinfo {title} {Estimating dissipation
  from single stationary trajectories},\ }\href
  {https://doi.org/10.1103/PhysRevLett.105.150607} {\bibfield  {journal}
  {\bibinfo  {journal} {Phys. Rev. Lett.}\ }\textbf {\bibinfo {volume} {105}},\
  \bibinfo {pages} {150607} (\bibinfo {year} {2010})}\BibitemShut {NoStop}%
\bibitem [{\citenamefont {Schnakenberg}(1976)}]{Schnakenberg1976Oct}%
  \BibitemOpen
  \bibfield  {author} {\bibinfo {author} {\bibfnamefont {J.}~\bibnamefont
  {Schnakenberg}},\ }\bibfield  {title} {\bibinfo {title} {{Network theory of
  microscopic and macroscopic behavior of master equation systems}},\ }\href
  {https://doi.org/10.1103/RevModPhys.48.571} {\bibfield  {journal} {\bibinfo
  {journal} {Rev. Mod. Phys.}\ }\textbf {\bibinfo {volume} {48}},\ \bibinfo
  {pages} {571} (\bibinfo {year} {1976})}\BibitemShut {NoStop}%
\bibitem [{\citenamefont {Proesmans}\ and\ \citenamefont
  {Horowitz}(2019)}]{Proesmans19}%
  \BibitemOpen
  \bibfield  {author} {\bibinfo {author} {\bibfnamefont {K.}~\bibnamefont
  {Proesmans}}\ and\ \bibinfo {author} {\bibfnamefont {J.~M.}\ \bibnamefont
  {Horowitz}},\ }\bibfield  {title} {\bibinfo {title} {Hysteretic thermodynamic
  uncertainty relation for systems with broken time-reversal symmetry},\ }\href
  {https://doi.org/10.1088/1742-5468/ab14da} {\bibfield  {journal} {\bibinfo
  {journal} {Journal of Statistical Mechanics: Theory and Experiment}\ }\textbf
  {\bibinfo {volume} {2019}},\ \bibinfo {pages} {054005} (\bibinfo {year}
  {2019})}\BibitemShut {NoStop}%
\bibitem [{\citenamefont {Koyuk}\ and\ \citenamefont
  {Seifert}(2020)}]{Timur20}%
  \BibitemOpen
  \bibfield  {author} {\bibinfo {author} {\bibfnamefont {T.}~\bibnamefont
  {Koyuk}}\ and\ \bibinfo {author} {\bibfnamefont {U.}~\bibnamefont
  {Seifert}},\ }\bibfield  {title} {\bibinfo {title} {Thermodynamic uncertainty
  relation for time-dependent driving},\ }\href
  {https://doi.org/10.1103/PhysRevLett.125.260604} {\bibfield  {journal}
  {\bibinfo  {journal} {Phys. Rev. Lett.}\ }\textbf {\bibinfo {volume} {125}},\
  \bibinfo {pages} {260604} (\bibinfo {year} {2020})}\BibitemShut {NoStop}%
\bibitem [{\citenamefont {Guarnieri}\ \emph {et~al.}(2019)\citenamefont
  {Guarnieri}, \citenamefont {Landi}, \citenamefont {Clark},\ and\
  \citenamefont {Goold}}]{Guarnieri19}%
  \BibitemOpen
  \bibfield  {author} {\bibinfo {author} {\bibfnamefont {G.}~\bibnamefont
  {Guarnieri}}, \bibinfo {author} {\bibfnamefont {G.~T.}\ \bibnamefont
  {Landi}}, \bibinfo {author} {\bibfnamefont {S.~R.}\ \bibnamefont {Clark}},\
  and\ \bibinfo {author} {\bibfnamefont {J.}~\bibnamefont {Goold}},\ }\bibfield
   {title} {\bibinfo {title} {Thermodynamics of precision in quantum
  nonequilibrium steady states},\ }\href
  {https://doi.org/10.1103/PhysRevResearch.1.033021} {\bibfield  {journal}
  {\bibinfo  {journal} {Phys. Rev. Res.}\ }\textbf {\bibinfo {volume} {1}},\
  \bibinfo {pages} {033021} (\bibinfo {year} {2019})}\BibitemShut {NoStop}%
\bibitem [{\citenamefont {Carollo}\ \emph {et~al.}(2019)\citenamefont
  {Carollo}, \citenamefont {Jack},\ and\ \citenamefont {Garrahan}}]{Carollo19}%
  \BibitemOpen
  \bibfield  {author} {\bibinfo {author} {\bibfnamefont {F.}~\bibnamefont
  {Carollo}}, \bibinfo {author} {\bibfnamefont {R.~L.}\ \bibnamefont {Jack}},\
  and\ \bibinfo {author} {\bibfnamefont {J.~P.}\ \bibnamefont {Garrahan}},\
  }\bibfield  {title} {\bibinfo {title} {Unraveling the large deviation
  statistics of markovian open quantum systems},\ }\href
  {https://doi.org/10.1103/PhysRevLett.122.130605} {\bibfield  {journal}
  {\bibinfo  {journal} {Phys. Rev. Lett.}\ }\textbf {\bibinfo {volume} {122}},\
  \bibinfo {pages} {130605} (\bibinfo {year} {2019})}\BibitemShut {NoStop}%
\bibitem [{\citenamefont {Esposito}\ \emph {et~al.}(2007)\citenamefont
  {Esposito}, \citenamefont {Harbola},\ and\ \citenamefont
  {Mukamel}}]{Esposito2007Apr}%
  \BibitemOpen
  \bibfield  {author} {\bibinfo {author} {\bibfnamefont {M.}~\bibnamefont
  {Esposito}}, \bibinfo {author} {\bibfnamefont {U.}~\bibnamefont {Harbola}},\
  and\ \bibinfo {author} {\bibfnamefont {S.}~\bibnamefont {Mukamel}},\
  }\bibfield  {title} {\bibinfo {title} {{Fluctuation theorem for counting
  statistics in electron transport through quantum junctions}},\ }\href
  {https://doi.org/10.1103/PhysRevB.75.155316} {\bibfield  {journal} {\bibinfo
  {journal} {Phys. Rev. B}\ }\textbf {\bibinfo {volume} {75}},\ \bibinfo
  {pages} {155316} (\bibinfo {year} {2007})}\BibitemShut {NoStop}%
\bibitem [{\citenamefont {Flindt}\ \emph {et~al.}(2008)\citenamefont {Flindt},
  \citenamefont {Novotn{\ifmmode\acute{y}\else\'{y}\fi}}, \citenamefont
  {Braggio}, \citenamefont {Sassetti},\ and\ \citenamefont
  {Jauho}}]{Flindt2008Apr}%
  \BibitemOpen
  \bibfield  {author} {\bibinfo {author} {\bibfnamefont {C.}~\bibnamefont
  {Flindt}}, \bibinfo {author} {\bibfnamefont {T.}~\bibnamefont
  {Novotn{\ifmmode\acute{y}\else\'{y}\fi}}}, \bibinfo {author} {\bibfnamefont
  {A.}~\bibnamefont {Braggio}}, \bibinfo {author} {\bibfnamefont
  {M.}~\bibnamefont {Sassetti}},\ and\ \bibinfo {author} {\bibfnamefont
  {A.-P.}\ \bibnamefont {Jauho}},\ }\bibfield  {title} {\bibinfo {title}
  {{Counting Statistics of Non-Markovian Quantum Stochastic Processes}},\
  }\href {https://doi.org/10.1103/PhysRevLett.100.150601} {\bibfield  {journal}
  {\bibinfo  {journal} {Phys. Rev. Lett.}\ }\textbf {\bibinfo {volume} {100}},\
  \bibinfo {pages} {150601} (\bibinfo {year} {2008})}\BibitemShut {NoStop}%
\bibitem [{\citenamefont {Walldorf}\ \emph {et~al.}(2020)\citenamefont
  {Walldorf}, \citenamefont {Brange}, \citenamefont {Padurariu},\ and\
  \citenamefont {Flindt}}]{Walldorf2020May}%
  \BibitemOpen
  \bibfield  {author} {\bibinfo {author} {\bibfnamefont {N.}~\bibnamefont
  {Walldorf}}, \bibinfo {author} {\bibfnamefont {F.}~\bibnamefont {Brange}},
  \bibinfo {author} {\bibfnamefont {C.}~\bibnamefont {Padurariu}},\ and\
  \bibinfo {author} {\bibfnamefont {C.}~\bibnamefont {Flindt}},\ }\bibfield
  {title} {\bibinfo {title} {{Noise and full counting statistics of a Cooper
  pair splitter}},\ }\href {https://doi.org/10.1103/PhysRevB.101.205422}
  {\bibfield  {journal} {\bibinfo  {journal} {Phys. Rev. B}\ }\textbf {\bibinfo
  {volume} {101}},\ \bibinfo {pages} {205422} (\bibinfo {year}
  {2020})}\BibitemShut {NoStop}%
\bibitem [{\citenamefont {Landi}\ \emph {et~al.}(2024)\citenamefont {Landi},
  \citenamefont {Kewming}, \citenamefont {Mitchison},\ and\ \citenamefont
  {Potts}}]{Landi2024Apr}%
  \BibitemOpen
  \bibfield  {author} {\bibinfo {author} {\bibfnamefont {G.~T.}\ \bibnamefont
  {Landi}}, \bibinfo {author} {\bibfnamefont {M.~J.}\ \bibnamefont {Kewming}},
  \bibinfo {author} {\bibfnamefont {M.~T.}\ \bibnamefont {Mitchison}},\ and\
  \bibinfo {author} {\bibfnamefont {P.~P.}\ \bibnamefont {Potts}},\ }\bibfield
  {title} {\bibinfo {title} {{Current Fluctuations in Open Quantum Systems:
  Bridging the Gap Between Quantum Continuous Measurements and Full Counting
  Statistics}},\ }\href {https://doi.org/10.1103/PRXQuantum.5.020201}
  {\bibfield  {journal} {\bibinfo  {journal} {PRX Quantum}\ }\textbf {\bibinfo
  {volume} {5}},\ \bibinfo {pages} {020201} (\bibinfo {year}
  {2024})}\BibitemShut {NoStop}%
\bibitem [{\citenamefont {Ledermann}\ and\ \citenamefont {Reuter
  G.~E.}(1954)}]{Ledermann1954Jan}%
  \BibitemOpen
  \bibfield  {author} {\bibinfo {author} {\bibfnamefont {W.}~\bibnamefont
  {Ledermann}}\ and\ \bibinfo {author} {\bibfnamefont {H.}~\bibnamefont {Reuter
  G.~E.}},\ }\bibfield  {title} {\bibinfo {title} {Spectral theory for the
  differential equations of simple birth and death processes},\ }\href
  {https://doi.org/10.1098/rsta.1954.0001} {\bibfield  {journal} {\bibinfo
  {journal} {Philosophical Transactions of the Royal Society of London, Series
  A, Mathematical and Physical Sciences}\ }\textbf {\bibinfo {volume} {246}},\
  \bibinfo {pages} {321} (\bibinfo {year} {1954})}\BibitemShut {NoStop}%
\bibitem [{\citenamefont {Andrieux}\ and\ \citenamefont
  {Gaspard}(2006)}]{Andrieux06}%
  \BibitemOpen
  \bibfield  {author} {\bibinfo {author} {\bibfnamefont {D.}~\bibnamefont
  {Andrieux}}\ and\ \bibinfo {author} {\bibfnamefont {P.}~\bibnamefont
  {Gaspard}},\ }\bibfield  {title} {\bibinfo {title} {Fluctuation theorems and
  the nonequilibrium thermodynamics of molecular motors},\ }\href
  {https://doi.org/10.1103/PhysRevE.74.011906} {\bibfield  {journal} {\bibinfo
  {journal} {Phys. Rev. E}\ }\textbf {\bibinfo {volume} {74}},\ \bibinfo
  {pages} {011906} (\bibinfo {year} {2006})}\BibitemShut {NoStop}%
\bibitem [{\citenamefont {Hayashi}\ \emph {et~al.}(2010)\citenamefont
  {Hayashi}, \citenamefont {Ueno}, \citenamefont {Iino},\ and\ \citenamefont
  {Noji}}]{Hayasi10}%
  \BibitemOpen
  \bibfield  {author} {\bibinfo {author} {\bibfnamefont {K.}~\bibnamefont
  {Hayashi}}, \bibinfo {author} {\bibfnamefont {H.}~\bibnamefont {Ueno}},
  \bibinfo {author} {\bibfnamefont {R.}~\bibnamefont {Iino}},\ and\ \bibinfo
  {author} {\bibfnamefont {H.}~\bibnamefont {Noji}},\ }\bibfield  {title}
  {\bibinfo {title} {Fluctuation theorem applied to
  ${\mathbf{f}}_{1}$-atpase},\ }\href
  {https://doi.org/10.1103/PhysRevLett.104.218103} {\bibfield  {journal}
  {\bibinfo  {journal} {Phys. Rev. Lett.}\ }\textbf {\bibinfo {volume} {104}},\
  \bibinfo {pages} {218103} (\bibinfo {year} {2010})}\BibitemShut {NoStop}%
\bibitem [{\citenamefont {Llabr\'es}\ \emph {et~al.}(2026)\citenamefont
  {Llabr\'es}, \citenamefont {San~Miguel},\ and\ \citenamefont
  {Toral}}]{Llabres25}%
  \BibitemOpen
  \bibfield  {author} {\bibinfo {author} {\bibfnamefont {J.}~\bibnamefont
  {Llabr\'es}}, \bibinfo {author} {\bibfnamefont {M.}~\bibnamefont
  {San~Miguel}},\ and\ \bibinfo {author} {\bibfnamefont {R.}~\bibnamefont
  {Toral}},\ }\bibfield  {title} {\bibinfo {title} {Universality of
  noise-induced transitions in nonlinear voter models},\ }\href
  {https://doi.org/10.1103/6xdk-2cjy} {\bibfield  {journal} {\bibinfo
  {journal} {Phys. Rev. Res.}\ }\textbf {\bibinfo {volume} {8}},\ \bibinfo
  {pages} {013015} (\bibinfo {year} {2026})}\BibitemShut {NoStop}%
\bibitem [{\citenamefont {Parrondo}(2001)}]{Parrondo01}%
  \BibitemOpen
  \bibfield  {author} {\bibinfo {author} {\bibfnamefont {J.~M.~R.}\
  \bibnamefont {Parrondo}},\ }\bibfield  {title} {\bibinfo {title} {The
  {S}zilard engine revisited: Entropy, macroscopic randomness, and symmetry
  breaking phase transitions},\ }\href {https://doi.org/10.1063/1.1388006}
  {\bibfield  {journal} {\bibinfo  {journal} {Chaos: An Interdisciplinary
  Journal of Nonlinear Science}\ }\textbf {\bibinfo {volume} {11}},\ \bibinfo
  {pages} {725} (\bibinfo {year} {2001})}\BibitemShut {NoStop}%
\bibitem [{Note8()}]{Note8}%
  \BibitemOpen
  \bibinfo {note} {Note however that the standard second law in Eq.~\protect
  \eqref {eq:second-law} with averages over the entire phase space is always
  verified in any case.}\BibitemShut {Stop}%
\bibitem [{\citenamefont {Gleeson}(2013)}]{Gleeson2013}%
  \BibitemOpen
  \bibfield  {author} {\bibinfo {author} {\bibfnamefont {J.~P.}\ \bibnamefont
  {Gleeson}},\ }\bibfield  {title} {\bibinfo {title} {Binary-state dynamics on
  complex networks: Pair approximation and beyond},\ }\href
  {https://doi.org/10.1103/PhysRevX.3.021004} {\bibfield  {journal} {\bibinfo
  {journal} {Phys. Rev. X}\ }\textbf {\bibinfo {volume} {3}},\ \bibinfo {pages}
  {021004} (\bibinfo {year} {2013})}\BibitemShut {NoStop}%
\bibitem [{\citenamefont {Peralta}\ and\ \citenamefont
  {Toral}(2020)}]{Peralta2020}%
  \BibitemOpen
  \bibfield  {author} {\bibinfo {author} {\bibfnamefont {A.~F.}\ \bibnamefont
  {Peralta}}\ and\ \bibinfo {author} {\bibfnamefont {R.}~\bibnamefont
  {Toral}},\ }\bibfield  {title} {\bibinfo {title} {Binary-state dynamics on
  complex networks: Stochastic pair approximation and beyond},\ }\href
  {https://doi.org/10.1103/PhysRevResearch.2.043370} {\bibfield  {journal}
  {\bibinfo  {journal} {Phys. Rev. Res.}\ }\textbf {\bibinfo {volume} {2}},\
  \bibinfo {pages} {043370} (\bibinfo {year} {2020})}\BibitemShut {NoStop}%
\bibitem [{\citenamefont {Peralta}\ \emph
  {et~al.}(2018{\natexlab{b}})\citenamefont {Peralta}, \citenamefont {Carro},
  \citenamefont {Miguel},\ and\ \citenamefont {Toral}}]{Peralta_2018}%
  \BibitemOpen
  \bibfield  {author} {\bibinfo {author} {\bibfnamefont {A.~F.}\ \bibnamefont
  {Peralta}}, \bibinfo {author} {\bibfnamefont {A.}~\bibnamefont {Carro}},
  \bibinfo {author} {\bibfnamefont {M.~S.}\ \bibnamefont {Miguel}},\ and\
  \bibinfo {author} {\bibfnamefont {R.}~\bibnamefont {Toral}},\ }\bibfield
  {title} {\bibinfo {title} {Stochastic pair approximation treatment of the
  noisy voter model},\ }\href {https://doi.org/10.1088/1367-2630/aae7f5}
  {\bibfield  {journal} {\bibinfo  {journal} {New Journal of Physics}\ }\textbf
  {\bibinfo {volume} {20}},\ \bibinfo {pages} {103045} (\bibinfo {year}
  {2018}{\natexlab{b}})}\BibitemShut {NoStop}%
\bibitem [{\citenamefont {Sood}\ and\ \citenamefont {Redner}(2005)}]{Sood2005}%
  \BibitemOpen
  \bibfield  {author} {\bibinfo {author} {\bibfnamefont {V.}~\bibnamefont
  {Sood}}\ and\ \bibinfo {author} {\bibfnamefont {S.}~\bibnamefont {Redner}},\
  }\bibfield  {title} {\bibinfo {title} {Voter model on heterogeneous graphs},\
  }\href {https://doi.org/10.1103/PhysRevLett.94.178701} {\bibfield  {journal}
  {\bibinfo  {journal} {Phys. Rev. Lett.}\ }\textbf {\bibinfo {volume} {94}},\
  \bibinfo {pages} {178701} (\bibinfo {year} {2005})}\BibitemShut {NoStop}%
\bibitem [{\citenamefont {Cui}\ \emph {et~al.}(2022)\citenamefont {Cui},
  \citenamefont {KhudaBukhsh},\ and\ \citenamefont {Koeppl}}]{Cui2022}%
  \BibitemOpen
  \bibfield  {author} {\bibinfo {author} {\bibfnamefont {K.}~\bibnamefont
  {Cui}}, \bibinfo {author} {\bibfnamefont {W.~R.}\ \bibnamefont
  {KhudaBukhsh}},\ and\ \bibinfo {author} {\bibfnamefont {H.}~\bibnamefont
  {Koeppl}},\ }\bibfield  {title} {\bibinfo {title} {Motif-based mean-field
  approximation of interacting particles on clustered networks},\ }\href
  {https://doi.org/10.1103/PhysRevE.105.L042301} {\bibfield  {journal}
  {\bibinfo  {journal} {Phys. Rev. E}\ }\textbf {\bibinfo {volume} {105}},\
  \bibinfo {pages} {L042301} (\bibinfo {year} {2022})}\BibitemShut {NoStop}%
\bibitem [{\citenamefont {Demirel}\ \emph {et~al.}(2014)\citenamefont
  {Demirel}, \citenamefont {Vazquez}, \citenamefont {Böhme},\ and\
  \citenamefont {Gross}}]{DEMIREL201468}%
  \BibitemOpen
  \bibfield  {author} {\bibinfo {author} {\bibfnamefont {G.}~\bibnamefont
  {Demirel}}, \bibinfo {author} {\bibfnamefont {F.}~\bibnamefont {Vazquez}},
  \bibinfo {author} {\bibfnamefont {G.}~\bibnamefont {Böhme}},\ and\ \bibinfo
  {author} {\bibfnamefont {T.}~\bibnamefont {Gross}},\ }\bibfield  {title}
  {\bibinfo {title} {Moment-closure approximations for discrete adaptive
  networks},\ }\href
  {https://doi.org/https://doi.org/10.1016/j.physd.2013.07.003} {\bibfield
  {journal} {\bibinfo  {journal} {Physica D: Nonlinear Phenomena}\ }\textbf
  {\bibinfo {volume} {267}},\ \bibinfo {pages} {68} (\bibinfo {year} {2014})},\
  \bibinfo {note} {evolving Dynamical Networks}\BibitemShut {NoStop}%
\bibitem [{\citenamefont {Zimmaro}\ \emph {et~al.}(2023)\citenamefont
  {Zimmaro}, \citenamefont {Contucci},\ and\ \citenamefont
  {Kertész}}]{e25060838}%
  \BibitemOpen
  \bibfield  {author} {\bibinfo {author} {\bibfnamefont {F.}~\bibnamefont
  {Zimmaro}}, \bibinfo {author} {\bibfnamefont {P.}~\bibnamefont {Contucci}},\
  and\ \bibinfo {author} {\bibfnamefont {J.}~\bibnamefont {Kertész}},\
  }\bibfield  {title} {\bibinfo {title} {Voter-like dynamics with conflicting
  preferences on modular networks},\ }\bibfield  {journal} {\bibinfo  {journal}
  {Entropy}\ }\textbf {\bibinfo {volume} {25}},\ \href
  {https://doi.org/10.3390/e25060838} {10.3390/e25060838} (\bibinfo {year}
  {2023})\BibitemShut {NoStop}%
\bibitem [{\citenamefont {Oestereich}\ \emph {et~al.}(2019)\citenamefont
  {Oestereich}, \citenamefont {Pires},\ and\ \citenamefont
  {Crokidakis}}]{Oestereich2019}%
  \BibitemOpen
  \bibfield  {author} {\bibinfo {author} {\bibfnamefont {A.~L.}\ \bibnamefont
  {Oestereich}}, \bibinfo {author} {\bibfnamefont {M.~A.}\ \bibnamefont
  {Pires}},\ and\ \bibinfo {author} {\bibfnamefont {N.}~\bibnamefont
  {Crokidakis}},\ }\bibfield  {title} {\bibinfo {title} {Three-state opinion
  dynamics in modular networks},\ }\href
  {https://doi.org/10.1103/PhysRevE.100.032312} {\bibfield  {journal} {\bibinfo
   {journal} {Phys. Rev. E}\ }\textbf {\bibinfo {volume} {100}},\ \bibinfo
  {pages} {032312} (\bibinfo {year} {2019})}\BibitemShut {NoStop}%
\bibitem [{\citenamefont {Stark}\ \emph {et~al.}(2008)\citenamefont {Stark},
  \citenamefont {Tessone},\ and\ \citenamefont {Schweitzer}}]{Stark2008}%
  \BibitemOpen
  \bibfield  {author} {\bibinfo {author} {\bibfnamefont {H.-U.}\ \bibnamefont
  {Stark}}, \bibinfo {author} {\bibfnamefont {C.~J.}\ \bibnamefont {Tessone}},\
  and\ \bibinfo {author} {\bibfnamefont {F.}~\bibnamefont {Schweitzer}},\
  }\bibfield  {title} {\bibinfo {title} {Decelerating microdynamics can
  accelerate macrodynamics in the voter model},\ }\href
  {https://doi.org/10.1103/PhysRevLett.101.018701} {\bibfield  {journal}
  {\bibinfo  {journal} {Phys. Rev. Lett.}\ }\textbf {\bibinfo {volume} {101}},\
  \bibinfo {pages} {018701} (\bibinfo {year} {2008})}\BibitemShut {NoStop}%
\bibitem [{\citenamefont {Artime}\ \emph {et~al.}(2018)\citenamefont {Artime},
  \citenamefont {Peralta}, \citenamefont {Toral}, \citenamefont {Ramasco},\
  and\ \citenamefont {San~Miguel}}]{Artime2018}%
  \BibitemOpen
  \bibfield  {author} {\bibinfo {author} {\bibfnamefont {O.}~\bibnamefont
  {Artime}}, \bibinfo {author} {\bibfnamefont {A.~F.}\ \bibnamefont {Peralta}},
  \bibinfo {author} {\bibfnamefont {R.}~\bibnamefont {Toral}}, \bibinfo
  {author} {\bibfnamefont {J.~J.}\ \bibnamefont {Ramasco}},\ and\ \bibinfo
  {author} {\bibfnamefont {M.}~\bibnamefont {San~Miguel}},\ }\bibfield  {title}
  {\bibinfo {title} {Aging-induced continuous phase transition},\ }\href
  {https://doi.org/10.1103/PhysRevE.98.032104} {\bibfield  {journal} {\bibinfo
  {journal} {Phys. Rev. E}\ }\textbf {\bibinfo {volume} {98}},\ \bibinfo
  {pages} {032104} (\bibinfo {year} {2018})}\BibitemShut {NoStop}%
\bibitem [{\citenamefont {Mobilia}(2003)}]{Mobilia2003}%
  \BibitemOpen
  \bibfield  {author} {\bibinfo {author} {\bibfnamefont {M.}~\bibnamefont
  {Mobilia}},\ }\bibfield  {title} {\bibinfo {title} {Does a single zealot
  affect an infinite group of voters?},\ }\href
  {https://doi.org/10.1103/PhysRevLett.91.028701} {\bibfield  {journal}
  {\bibinfo  {journal} {Phys. Rev. Lett.}\ }\textbf {\bibinfo {volume} {91}},\
  \bibinfo {pages} {028701} (\bibinfo {year} {2003})}\BibitemShut {NoStop}%
\bibitem [{\citenamefont {Khalil}\ \emph {et~al.}(2018)\citenamefont {Khalil},
  \citenamefont {San~Miguel},\ and\ \citenamefont {Toral}}]{Khalil2018}%
  \BibitemOpen
  \bibfield  {author} {\bibinfo {author} {\bibfnamefont {N.}~\bibnamefont
  {Khalil}}, \bibinfo {author} {\bibfnamefont {M.}~\bibnamefont {San~Miguel}},\
  and\ \bibinfo {author} {\bibfnamefont {R.}~\bibnamefont {Toral}},\ }\bibfield
   {title} {\bibinfo {title} {Zealots in the mean-field noisy voter model},\
  }\href {https://doi.org/10.1103/PhysRevE.97.012310} {\bibfield  {journal}
  {\bibinfo  {journal} {Phys. Rev. E}\ }\textbf {\bibinfo {volume} {97}},\
  \bibinfo {pages} {012310} (\bibinfo {year} {2018})}\BibitemShut {NoStop}%
\bibitem [{\citenamefont {Rahav}\ and\ \citenamefont
  {Harbola}(2014)}]{Rahav14}%
  \BibitemOpen
  \bibfield  {author} {\bibinfo {author} {\bibfnamefont {S.}~\bibnamefont
  {Rahav}}\ and\ \bibinfo {author} {\bibfnamefont {U.}~\bibnamefont
  {Harbola}},\ }\bibfield  {title} {\bibinfo {title} {An integral fluctuation
  theorem for systems with unidirectional transitions},\ }\href
  {https://doi.org/10.1088/1742-5468/2014/10/P10044} {\bibfield  {journal}
  {\bibinfo  {journal} {Journal of Statistical Mechanics: Theory and
  Experiment}\ }\textbf {\bibinfo {volume} {2014}},\ \bibinfo {pages} {P10044}
  (\bibinfo {year} {2014})}\BibitemShut {NoStop}%
\bibitem [{\citenamefont {Pal}\ \emph {et~al.}(2021)\citenamefont {Pal},
  \citenamefont {Reuveni},\ and\ \citenamefont {Rahav}}]{Rahav21}%
  \BibitemOpen
  \bibfield  {author} {\bibinfo {author} {\bibfnamefont {A.}~\bibnamefont
  {Pal}}, \bibinfo {author} {\bibfnamefont {S.}~\bibnamefont {Reuveni}},\ and\
  \bibinfo {author} {\bibfnamefont {S.}~\bibnamefont {Rahav}},\ }\bibfield
  {title} {\bibinfo {title} {Thermodynamic uncertainty relation for systems
  with unidirectional transitions},\ }\href
  {https://doi.org/10.1103/PhysRevResearch.3.013273} {\bibfield  {journal}
  {\bibinfo  {journal} {Phys. Rev. Res.}\ }\textbf {\bibinfo {volume} {3}},\
  \bibinfo {pages} {013273} (\bibinfo {year} {2021})}\BibitemShut {NoStop}%
\bibitem [{\citenamefont {Busiello}\ \emph {et~al.}(2020)\citenamefont
  {Busiello}, \citenamefont {Gupta},\ and\ \citenamefont
  {Maritan}}]{Busiello20}%
  \BibitemOpen
  \bibfield  {author} {\bibinfo {author} {\bibfnamefont {D.~M.}\ \bibnamefont
  {Busiello}}, \bibinfo {author} {\bibfnamefont {D.}~\bibnamefont {Gupta}},\
  and\ \bibinfo {author} {\bibfnamefont {A.}~\bibnamefont {Maritan}},\
  }\bibfield  {title} {\bibinfo {title} {Entropy production in systems with
  unidirectional transitions},\ }\href
  {https://doi.org/10.1103/PhysRevResearch.2.023011} {\bibfield  {journal}
  {\bibinfo  {journal} {Phys. Rev. Res.}\ }\textbf {\bibinfo {volume} {2}},\
  \bibinfo {pages} {023011} (\bibinfo {year} {2020})}\BibitemShut {NoStop}%
\bibitem [{\citenamefont {Bailey}(1975)}]{Bailey}%
  \BibitemOpen
  \bibfield  {author} {\bibinfo {author} {\bibfnamefont {N.~T.~J.}\
  \bibnamefont {Bailey}},\ }\href@noop {} {\emph {\bibinfo {title} {{The
  Mathematical Theory of Infectious Diseases}}}},\ \bibinfo {edition} {2nd}\
  ed.\ (\bibinfo  {publisher} {Griffin, London},\ \bibinfo {year}
  {1975})\BibitemShut {NoStop}%
\bibitem [{\citenamefont {Borba}\ \emph {et~al.}(2025)\citenamefont {Borba},
  \citenamefont {Gonçalves},\ and\ \citenamefont {Anteneodo}}]{Borba2025}%
  \BibitemOpen
  \bibfield  {author} {\bibinfo {author} {\bibfnamefont {J.~S.}\ \bibnamefont
  {Borba}}, \bibinfo {author} {\bibfnamefont {S.}~\bibnamefont {Gonçalves}},\
  and\ \bibinfo {author} {\bibfnamefont {C.}~\bibnamefont {Anteneodo}},\
  }\bibfield  {title} {\bibinfo {title} {Inequality in a model of capitalist
  economy},\ }\href
  {https://doi.org/https://doi.org/10.1016/j.physa.2025.130457} {\bibfield
  {journal} {\bibinfo  {journal} {Physica A: Statistical Mechanics and its
  Applications}\ }\textbf {\bibinfo {volume} {664}},\ \bibinfo {pages} {130457}
  (\bibinfo {year} {2025})}\BibitemShut {NoStop}%
\bibitem [{\citenamefont {Pastor-Satorras}\ and\ \citenamefont
  {Vespignani}(2001)}]{Vespignani01}%
  \BibitemOpen
  \bibfield  {author} {\bibinfo {author} {\bibfnamefont {R.}~\bibnamefont
  {Pastor-Satorras}}\ and\ \bibinfo {author} {\bibfnamefont {A.}~\bibnamefont
  {Vespignani}},\ }\bibfield  {title} {\bibinfo {title} {Epidemic spreading in
  scale-free networks},\ }\href {https://doi.org/10.1103/PhysRevLett.86.3200}
  {\bibfield  {journal} {\bibinfo  {journal} {Phys. Rev. Lett.}\ }\textbf
  {\bibinfo {volume} {86}},\ \bibinfo {pages} {3200} (\bibinfo {year}
  {2001})}\BibitemShut {NoStop}%
\bibitem [{\citenamefont {Suchecki}\ \emph {et~al.}(2005)\citenamefont
  {Suchecki}, \citenamefont {Egu\'{\i}luz},\ and\ \citenamefont
  {San~Miguel}}]{Suchecki05}%
  \BibitemOpen
  \bibfield  {author} {\bibinfo {author} {\bibfnamefont {K.}~\bibnamefont
  {Suchecki}}, \bibinfo {author} {\bibfnamefont {V.~M.}\ \bibnamefont
  {Egu\'{\i}luz}},\ and\ \bibinfo {author} {\bibfnamefont {M.}~\bibnamefont
  {San~Miguel}},\ }\bibfield  {title} {\bibinfo {title} {Voter model dynamics
  in complex networks: Role of dimensionality, disorder, and degree
  distribution},\ }\href {https://doi.org/10.1103/PhysRevE.72.036132}
  {\bibfield  {journal} {\bibinfo  {journal} {Phys. Rev. E}\ }\textbf {\bibinfo
  {volume} {72}},\ \bibinfo {pages} {036132} (\bibinfo {year}
  {2005})}\BibitemShut {NoStop}%
\bibitem [{\citenamefont {Lambiotte}(2007)}]{Lambiotte07}%
  \BibitemOpen
  \bibfield  {author} {\bibinfo {author} {\bibfnamefont {R.}~\bibnamefont
  {Lambiotte}},\ }\bibfield  {title} {\bibinfo {title} {How does degree
  heterogeneity affect an order-disorder transition?},\ }\href
  {https://doi.org/10.1209/0295-5075/78/68002} {\bibfield  {journal} {\bibinfo
  {journal} {Europhysics Letters}\ }\textbf {\bibinfo {volume} {78}},\ \bibinfo
  {pages} {68002} (\bibinfo {year} {2007})}\BibitemShut {NoStop}%
\bibitem [{\citenamefont {Gleeson}(2011)}]{Gleeson11}%
  \BibitemOpen
  \bibfield  {author} {\bibinfo {author} {\bibfnamefont {J.~P.}\ \bibnamefont
  {Gleeson}},\ }\bibfield  {title} {\bibinfo {title} {High-accuracy
  approximation of binary-state dynamics on networks},\ }\href
  {https://doi.org/10.1103/PhysRevLett.107.068701} {\bibfield  {journal}
  {\bibinfo  {journal} {Phys. Rev. Lett.}\ }\textbf {\bibinfo {volume} {107}},\
  \bibinfo {pages} {068701} (\bibinfo {year} {2011})}\BibitemShut {NoStop}%
\bibitem [{\citenamefont {Carro}\ \emph {et~al.}(2016)\citenamefont {Carro},
  \citenamefont {Toral},\ and\ \citenamefont {San~Miguel}}]{Carro16}%
  \BibitemOpen
  \bibfield  {author} {\bibinfo {author} {\bibfnamefont {A.}~\bibnamefont
  {Carro}}, \bibinfo {author} {\bibfnamefont {R.}~\bibnamefont {Toral}},\ and\
  \bibinfo {author} {\bibfnamefont {M.}~\bibnamefont {San~Miguel}},\ }\bibfield
   {title} {\bibinfo {title} {The noisy voter model on complex networks},\
  }\href@noop {} {\bibfield  {journal} {\bibinfo  {journal} {Scientific
  reports}\ }\textbf {\bibinfo {volume} {6}},\ \bibinfo {pages} {24775}
  (\bibinfo {year} {2016})}\BibitemShut {NoStop}%
\bibitem [{\citenamefont {van Kampen}(2007)}]{vanKampen:2007}%
  \BibitemOpen
  \bibfield  {author} {\bibinfo {author} {\bibfnamefont {N.}~\bibnamefont {van
  Kampen}},\ }\href@noop {} {\emph {\bibinfo {title} {{Stochastic Processes in
  Physics and Chemistry}}}},\ \bibinfo {edition} {3rd}\ ed.\ (\bibinfo
  {publisher} {North-Holland},\ \bibinfo {address} {Amsterdam},\ \bibinfo
  {year} {2007})\BibitemShut {NoStop}%
\bibitem [{Note9()}]{Note9}%
  \BibitemOpen
  \bibinfo {note} {Note that $\omega $ only fixes the relaxation time scale and
  does not affect the stationary distribution.}\BibitemShut {Stop}%
\bibitem [{Note10()}]{Note10}%
  \BibitemOpen
  \bibinfo {note} {Since $f(0) = \chi \geq 0$ and $f(1) = -\theta \leq 0$,
  there is at least one solution $x_{\protect \mathrm {st}} \in [0, 1]$
  provided $\chi \protect \neq 0$ and $\theta \protect \neq 0$.}\BibitemShut
  {Stop}%
\bibitem [{\citenamefont {Esposito}\ \emph {et~al.}(2009)\citenamefont
  {Esposito}, \citenamefont {Harbola},\ and\ \citenamefont
  {Mukamel}}]{Esposito2009Dec}%
  \BibitemOpen
  \bibfield  {author} {\bibinfo {author} {\bibfnamefont {M.}~\bibnamefont
  {Esposito}}, \bibinfo {author} {\bibfnamefont {U.}~\bibnamefont {Harbola}},\
  and\ \bibinfo {author} {\bibfnamefont {S.}~\bibnamefont {Mukamel}},\
  }\bibfield  {title} {\bibinfo {title} {{Nonequilibrium fluctuations,
  fluctuation theorems, and counting statistics in quantum systems}},\ }\href
  {https://doi.org/10.1103/RevModPhys.81.1665} {\bibfield  {journal} {\bibinfo
  {journal} {Rev. Mod. Phys.}\ }\textbf {\bibinfo {volume} {81}},\ \bibinfo
  {pages} {1665} (\bibinfo {year} {2009})}\BibitemShut {NoStop}%
\bibitem [{\citenamefont {Touchette}(2009)}]{Touchette2009Jul}%
  \BibitemOpen
  \bibfield  {author} {\bibinfo {author} {\bibfnamefont {H.}~\bibnamefont
  {Touchette}},\ }\bibfield  {title} {\bibinfo {title} {{The large deviation
  approach to statistical mechanics}},\ }\href
  {https://doi.org/10.1016/j.physrep.2009.05.002} {\bibfield  {journal}
  {\bibinfo  {journal} {Phys. Rep.}\ }\textbf {\bibinfo {volume} {478}},\
  \bibinfo {pages} {1} (\bibinfo {year} {2009})}\BibitemShut {NoStop}%
\bibitem [{\citenamefont {Toral}(2011)}]{Toral:2011}%
  \BibitemOpen
  \bibfield  {author} {\bibinfo {author} {\bibfnamefont {R.}~\bibnamefont
  {Toral}},\ }\bibfield  {title} {\bibinfo {title} {Noise-induced transitions
  vs. noise-induced phase transitions},\ }in\ \href
  {https://doi.org/10.1063/1.3577618} {\emph {\bibinfo {booktitle} {AIP
  Conference Proceedings}}},\ Vol.\ \bibinfo {volume} {1332}\ (\bibinfo {year}
  {2011})\ pp.\ \bibinfo {pages} {145--154}\BibitemShut {NoStop}%
\end{thebibliography}%

\end{document}